\documentclass[twocolumn]{aastex63}
\pdfoutput=1
\usepackage[latin1]{inputenc}
\usepackage{amsmath}
\usepackage{amsfonts}
\usepackage{amssymb}
\usepackage{chngcntr}

\usepackage{graphicx,booktabs}
\usepackage[varg]{txfonts}
\usepackage{mwe} 
\usepackage{url}
\usepackage{longtable}
\usepackage{float}

\hypersetup{pdfauthor={BASS}}

\newcommand{\todo}{\ifmmode \text{\color{red}\Huge{\(\bullet\)}} \else {\color{red}{\Huge$\bullet$}}\fi}
\newcommand{\tido}{\ifmmode {{\color{red}\bullet}} \else {\color{red}$\bullet$}\fi}
\newcommand{\tudo}{\ifmmode {{\color{magenta}\bullet}} \else {\color{magenta}$\bullet$}\fi}

\newcommand{\tomike}{\ifmmode \text{\color{red}\Huge{\(\bullet\)}MIKE} \else {\color{red}{\Huge$\bullet$}MIKE}\fi}

\newcommand{\Nbonusraw}{207} 
\newcommand{\Nbonusgood}{177} 
\newcommand{\Nhabonusgood}{140} 
\newcommand{\Nhbbonusgood}{109} 
\newcommand{\Nmgbonusgood}{17} 
\newcommand{\Ncivbonusgood}{5} 

\newcommand{\Nfullgood}{689} 
\newcommand{\Nhafullgood}{597} 
\newcommand{\Nhbfullgood}{490} 
\newcommand{\Nmgfullgood}{45} 
\newcommand{\Ncivfullgood}{27} 

\newcommand{\Nha}{457} 

\newcommand{\Nhai}{341} 
\newcommand{\Nhaii}{77} 
\newcommand{\Nlargeerrha}{39} 
\newcommand{\Nhabad}{29} 

\newcommand{\Nhagoodnb}{434} 
\newcommand{\Nhbgoodnb}{343} 
\newcommand{\Nmggoodnb}{3} 
\newcommand{\Ncivgoodnb}{0} 
\newcommand{\Nhgoodnb}{445} 

\newcommand{\Ncivgoodb}{22} 
\newcommand{\Nhagoodb}{23} 
\newcommand{\Nhbgoodb}{38} 
\newcommand{\Nmggoodb}{25} 

\newcommand{\Nhadouble}{29} 
\newcommand{\Nhbdouble}{20} 

\newcommand{\Ncivdouble}{0} 
\newcommand{\Nmgdouble}{0} 

\newcommand{\Ntotdouble}{29} 

\newcommand{\Ntotdoublefull}{49} 

\newcommand{\Nhatotal}{515} 

\newcommand{\Nhb}{381}
\newcommand{\Nhbi}{245}
\newcommand{\Nhbii}{118}
\newcommand{\Nlargeerrhb}{18}
\newcommand{\Nhbbad}{32} 

\newcommand{\Nhbtotal}{433} %
\newcommand{\Nhahb}{348}
\newcommand{\Nmg}{28}
\newcommand{\Nmgi}{15}  
\newcommand{\Nmgii}{13} 

\newcommand{\Nmgbad}{13}

\newcommand{\Nciv}{22}
\newcommand{\Ncivi}{13} 
\newcommand{\Ncivii}{9} 
\newcommand{\Nmgciv}{6} 
\newcommand{\Nlargeerrciv}{0}
\newcommand{\Ncivbad}{10}

\newcommand{\Ncivtotal}{32} 
\newcommand{\Nmgtotal}{41} 

\newcommand{\NBZQlowz}{22} 
\newcommand{\NBZQgood}{67} 

\newcommand{\Nlargeerr}{43} 

\newcommand{\NbothM}{75} 
\newcommand{\NbothMinter}{50} 

\newcommand{\Ntoth}{485} 
\newcommand{\Ntotz}{43}
\newcommand{\Nuniquez}{27}

\newcommand{\Nblr}{559} 
\newcommand{\NblrGood}{512} 

\newcommand{\Nhasnr}{15}
\newcommand{\Nhbsnr}{18}
\newcommand{\Ncivsnr}{4}
\newcommand{\Nmgsnr}{8}
\newcommand{\Ntotalsnr}{45}

\newcommand{\Nhatellu}{7}
\newcommand{\Nhbtellu}{0}
\newcommand{\Nmgtellu}{1}
\newcommand{\Ncivtellu}{0}
\newcommand{\Ntotaltellu}{8}

\newcommand{\Nhaunclear}{0}
\newcommand{\Nhbunclear}{1}
\newcommand{\Ncivunclear}{0}
\newcommand{\Nmgunclear}{0}
\newcommand{\Ntotalunclear}{1}

\newcommand{\Nhareduc}{2}
\newcommand{\Nhbreduc}{9}
\newcommand{\Ncivreduc}{2}
\newcommand{\Nmgreduc}{4}
\newcommand{\Ntotalreduc}{17} 

\newcommand{\Nhaincom}{3}
\newcommand{\Nhbincom}{1}
\newcommand{\Ncivincom}{2}
\newcommand{\Nmgincom}{0}
\newcommand{\Ntotalincom}{6} 

\newcommand{\Nhaabs}{0}
\newcommand{\Nhbabs}{1}
\newcommand{\Ncivabs}{2}
\newcommand{\Nmgabs}{0}
\newcommand{\Ntotalabs}{3} 

\newcommand{\Nhanlr}{2}
\newcommand{\Nhbnlr}{0}
\newcommand{\Ncivnlr}{0}
\newcommand{\Nmgnlr}{0}
\newcommand{\Ntotalnlr}{2}

\newcommand{\Nhafe}{0}
\newcommand{\Nhbfe}{2}
\newcommand{\Ncivfe}{0}
\newcommand{\Nmgfe}{0}
\newcommand{\Ntotalfe}{2}

\newcommand{\Nhabelowinst}{26}
\newcommand{\Nhbbelowinst}{69}

\newcommand{\FQ  }{\ifmmode f_{\rm Q} \else $f_{\rm Q}$\fi}

\newcommand{\RBLR  }{\ifmmode R_{\rm BLR}\else $R_{\rm BLR}$\fi}
\newcommand{\Mbh   }{\ifmmode M_{\rm BH} \else $M_{\rm BH}$\fi}	
\newcommand{\logNH }{\ifmmode \log (N_{\rm H}/{\rm cm}^{-2}) \else $\log (N_{\rm H}/{\rm cm}^{-2})$\fi}	
\newcommand{\NH }{\ifmmode N_{\rm H} \else $N_{\rm H}$\fi}	
\newcommand{\logNHo}{\ifmmode \log N_{\rm H} \else $\log N_{\rm H}$\fi}	
\newcommand{\LLedd }{\ifmmode L/L_{\rm Edd} \else $L/L_{\rm Edd}$\fi}
\newcommand{\lledd }{\ifmmode L/L_{\rm Edd} \else $L/L_{\rm Edd}$\fi}
\newcommand{\Lbol }{\ifmmode L_{\rm bol} \else $L_{\rm bol}$\fi}
\newcommand{ \sigmas}{\ifmmode \sigma_{\star} \else $\sigma_{\star}$\fi}
\newcommand{  \sigs }{\sigmas}

\newcommand{  \Halpha   }{\ifmmode {\text{H} }\alpha \else H$\alpha$\fi}
\newcommand{  \Hbeta    }{\ifmmode {\text{H} }\beta \else H$\beta$\fi}
\newcommand{  \ha   	}{\Halpha}
\newcommand{  \hb   	}{\Hbeta}
\newcommand{  \CIV      }{\ifmmode {\rm C}\,\textsc{iv}\,\lambda1549 \else C\,\textsc{iv}\,$\lambda1549$\fi}
\newcommand{  \civ      }{\ifmmode {\rm C}\,\textsc{iv}  \else C\,\textsc{iv}\fi}
\newcommand{  \MgII     }{\ifmmode {\rm Mg}\,\textsc{ii}\,\lambda2798 \else Mg\,\textsc{ii}\,$\lambda2798$\fi}
\newcommand{  \mgii     }{\ifmmode {\rm Mg}\,\textsc{ii} \else Mg\,\textsc{ii}\fi}
\newcommand{\OIII}{\ifmmode \left[{\rm O}\,\textsc{iii}\right]\,\lambda5007 \else [O\,{\sc iii}]\,$\lambda5007$\fi}
\newcommand{\oiii}{\ifmmode \left[{\rm O}\,\textsc{iii}\right] \else [O\,{\sc iii}]\fi}
\newcommand{  \NII   }{\ifmmode \left[{\rm N}\,\textsc{ii}\right]\,\lambda6584 \else [N\,\textsc{ii}]\,$\lambda6584$\fi}
\newcommand{  \nii      }{\ifmmode \left[{\rm N}\,\textsc{ii}\right]  \else [N\,\textsc{ii}]\fi}
\newcommand{  \SII   }{\ifmmode \left[{\rm S}\,\textsc{ii}\right]\,\lambda6731 \else [S\,\textsc{ii}]\,$\lambda6731$\fi}
\newcommand{  \sii      }{\ifmmode \left[{\rm S}\,\textsc{ii}\right]  \else
[S\,\textsc{ii}]\fi}
\newcommand{\caii       }{\ifmmode {\rm Ca}\,\textsc{ii}   \else Ca\,\textsc{ii}\fi}
\newcommand{\Mgb        }{\ifmmode {\rm Mg}\,\textsc{i}\,\lambda5175 \else Mg\,{\sc i}\,$\lambda5175$\fi}
\newcommand{\Cahk       }{\ifmmode {\rm Ca H+K}\,\textsc{ii}\,\lambda\lambda3935,3968 \else Ca\,\textsc{ii} H+K$\,\lambda\lambda3935,3968$\fi}

\newcommand{\ergs}	{\ifmmode {\text{erg\,s}}^{-1} \else erg s$^{-1}$\fi}
\newcommand{\ergscm}	{\ifmmode {\text{erg\,s}}^{-1}\,{\text cm}^{-2} \else erg s$^{-1}$cm$^{-2}$\fi}
\newcommand{\ergsA}	{\ifmmode {\rm erg}^{-2}\,{\rm s}^{-1}\,{\rm\AA}^{-1} \else erg\,s$^{-1}$\,\AA$^{-1}$\fi}
\newcommand{\ergcmsA}	{\ifmmode {\rm erg\,cm}^{-2}\,{\rm s}^{-1}\,{\rm\AA}^{-1} \else erg\,cm$^{-2}$\,s$^{-1}$\,\AA$^{-1}$\fi}
\newcommand{\kms}	{\ifmmode {\rm km\,s}^{-1} \else km\,s$^{-1}$\fi}
\newcommand{\kev}	{\ifmmode {\text{keV}} \else keV\fi}
\newcommand{\mic}	{\ifmmode {\rm \mu m} \else $\mu$m\fi}
\newcommand{\Msun}{\ifmmode M_{\odot} \else $M_{\odot}$\fi}
\newcommand{\Lsun}{\ifmmode L_{\odot} \else $L_{\odot}$\fi}
\newcommand{\mpyr}{\ifmmode \Msun\,{\rm yr}^{-1} \else $\Msun\,{\rm yr}^{-1}$\fi}

\newcommand{\Msol}{\Msun}

\newcommand{  \Luv      }{\ifmmode L_{1450} \else $L_{1450}$\fi}
\newcommand{  \Fuv      }{\ifmmode F_{1450} \else $F_{1450}$\fi}
\newcommand{  \Lop      }{\ifmmode L_{5100} \else $L_{5100}$\fi}
\newcommand{  \Fop      }{\ifmmode F_{5100} \else $F_{5100}$\fi}
\newcommand{  \Lsix     }{\ifmmode L_{6200} \else $L_{6200}$\fi}
\newcommand{  \Fsix     }{\ifmmode F_{6200} \else $F_{6200}$\fi}
\newcommand{  \Lthree   }{\ifmmode L_{3000} \else $L_{3000}$\fi}
\newcommand{  \Fthree   }{\ifmmode F_{3000} \else $F_{3000}$\fi}
\newcommand{  \lamLlam  }{\ifmmode \lambda L_{\lambda} \else $\lambda L_{\lambda}$\fi}
\newcommand{  \lamFlam  }{\ifmmode \lambda F_{\lambda} \else $\lambda F_{\lambda}$\fi}
\newcommand{ \Lhard   }{\ifmmode L_{2-10\,\kev} \else $L_{2-10\,\kev}$\fi}
\newcommand{ \Luhard  }{\ifmmode L_{14-150\,\kev} \else $L_{14-150\,\kev}$\fi}
\newcommand{ \Lx  }{\Luhard}

\newcommand{\Lir}{\ifmmode L_{12\,\mic} \else $L_{12\,\mic}$ \fi}

\newcommand{\Lbha}{\ifmmode L\left({\rm b}\ha\right) \else $L\left({\rm b}\ha\right)$\fi}
\newcommand{\Fbha}{\ifmmode L\left({\rm b}\ha\right) \else $F\left({\rm b}\ha\right)$\fi}

\newcommand{\Lbhb}{\ifmmode L\left({\rm b}\hb\right) \else $L\left({\rm b}\hb\right)$\fi}
\newcommand{\Fbhb}{\ifmmode F\left({\rm b}\hb\right) \else $F\left({\rm b}\hb\right)$\fi}

\newcommand{ \fwha  }{\ifmmode {\text{FWHM}}(\ha) \else FWHM(\ha)\fi}

\newcommand{ \Lmg   }{\ifmmode L\left(\mgii\right) \else $L\left(\mgii\right)$\fi}
\newcommand{ \Fmg   }{\ifmmode F\left(\mgii\right) \else $F\left(\mgii\right)$\fi}
\newcommand{ \fwmg  }{\ifmmode {\text{FWHM}}\left(\mgii\right) \else FWHM(\mgii)\fi}
\newcommand{ \Lciv  }{\ifmmode L\left(\civ\right) \else $L\left(\civ\right)$\fi}
\newcommand{ \Fciv  }{\ifmmode F\left(\civ\right) \else $F\left(\civ\right)$\fi}

\newcommand{ \fwciv }{\ifmmode {\text{FWHM} }\left(\civ\right) \else FWHM(\civ)\fi}

\newcommand{ \fwhm  }{\ifmmode {\text{FWHM} } \else \text{FWHM}\fi} 
\newcommand{ \fwhb  }{\ifmmode {\text{FWHM} }\left(\Hbeta\right) \else FWHM(\Hbeta)\fi}

\newcommand{ \Lbhax  }{\ifmmode \Lbha/\Luhard \else $\Lbha/\Luhard$\fi}
\newcommand{ \Lbhbx  }{\ifmmode \Lbhb/\Luhard \else $\Lbhb/\Luhard$\fi}

\newcommand{\Lbhair}{\ifmmode \Lbha/\Lir \else $\Lbha/\Lir$ \fi}

\newcommand{ \Lbhboiii  }{\ifmmode \Lbhb/L\left(\oiii \right) \else $\Lbhb/L\left(\oiii \right)$\fi}

\received{March 31, 2021}
\revised{\today}
\accepted{TBD}
\submitjournal{ApJS}

\shorttitle{BASS XXV: DR2 broad-line masses}
\shortauthors{Mej\'ia-Restrepo et al.}
\graphicspath{{./}{figures/}}

\begin{document}

\title{BASS XXV: DR2 Broad-line Based Black Hole Mass Estimates and Biases from Obscuration}


\correspondingauthor{Juli\'an E. Mej\'ia-Restrepo}
\email{julianmejia@gmail.com}

\author[0000-0001-8450-7463]{Julian E. Mej\'ia-Restrepo}
\affiliation{European Southern Observatory, Casilla 19001, Santiago 19, Chile}

\author[0000-0002-3683-7297]{Benny Trakhtenbrot}
\affiliation{School of Physics and Astronomy, Tel Aviv University, Tel Aviv 69978, Israel}

\author[0000-0002-7998-9581]{Michael J. Koss}
\affiliation{Eureka Scientific, 2452 Delmer Street Suite 100, Oakland, CA 94602-3017, USA}
\affiliation{Space Science Institute, 4750 Walnut Street, Suite 205, Boulder, Colorado 80301, USA}


\author[0000-0002-5037-951X]{Kyuseok Oh}
\affiliation{Korea Astronomy \& Space Science institute, 776, Daedeokdae-ro, Yuseong-gu, Daejeon 34055, Republic of Korea}
\affiliation{Department of Astronomy, Kyoto University, Kitashirakawa-Oiwake-cho, Sakyo-ku, Kyoto 606-8502, Japan}
\affiliation{JSPS Fellow}

\author[0000-0002-8760-6157]{Jakob den Brok}
\affiliation{Argelander-Institut fur Astronomie, Universit{\"a}t Bonn, Auf dem H{\"u}gel 71, D-53121 Bonn, Germany}

\author[0000-0003-2686-9241]{Daniel Stern}
\affiliation{Jet Propulsion Laboratory, California Institute of Technology, 4800 Oak Grove Drive, MS 169-224, Pasadena, CA 91109, USA}

\author[0000-0003-2284-8603]{Meredith C. Powell}
\affiliation{Kavli Institute of Particle Astrophysics and Cosmology, Stanford University, 452 Lomita Mall, Stanford, CA 94305, USA}

\author[0000-0001-5742-5980]{Federica Ricci}
\affiliation{Instituto de Astrof\'isica and Centro de Astroingenier\'ia, Facultad de F\'isica, Pontificia Universidad Cat\'olica de Chile, Casilla 306, Santiago 22, Chile}
\affiliation{Dipartimento di Fisica e Astronomia, Universit\'a di Bologna, via Piero
Gobetti 93/2, I-40129 Bologna, Italy}
\affiliation{INAF - Osservatorio di Astrofisica e Scienza dello Spazio di Bologna,
Via Gobetti, 93/3, I40129 Bologna, Italy.}

\author[0000-0002-9144-2255]{Turgay Caglar}
\affiliation{Leiden Observatory, PO Box 9513, 2300 RA, Leiden, The Netherlands}

\author[0000-0001-5231-2645]{Claudio Ricci}
\affiliation{N\'ucleo de Astronom\'ia de la Facultad de Ingenier\'ia, Universidad Diego Portales, Av. Ej\'ercito Libertador 441, Santiago 22, Chile}
\affiliation{Kavli Institute for Astronomy and Astrophysics, Peking University, Beijing 100871, People's Republic of China}

\author[0000-0002-8686-8737]{Franz E. Bauer}
\affiliation{Instituto de Astrof\'isica and Centro de Astroingenier\'ia, Facultad de F\'isica, Pontificia Universidad Cat\'olica de Chile, Casilla 306, Santiago 22, Chile}
\affiliation{Millennium Institute of Astrophysics (MAS), Nuncio Monse\~{n}or S\'otero Sanz 100, Providencia, Santiago, Chile}
\affiliation{Space Science Institute, 4750 Walnut Street, Suite 205, Boulder, CO 80301, USA}

\author[0000-0001-7568-6412]{Ezequiel Treister}
\affiliation{Instituto de Astrof{\'i}sica, Facultad de F{\'i}sica, Pontificia Universidad Cat{\'o}lica de Chile, Casilla 306, Santiago 22, Chile}
 
\author{Fiona A. Harrison}
\affiliation{Cahill Center for Astronomy and Astrophysics, California Institute of Technology, Pasadena, CA 91125, USA}

\author[0000-0002-0745-9792]{C. M. Urry}
\affiliation{Yale Center for Astronomy \& Astrophysics, Physics Department, Yale University, PO BOX 201820, New Haven, CT 06520-8120, USA}


\author[0000-0001-8211-3807]{Tonima Tasnim Ananna}
\affiliation{Department of Physics \& Astronomy, Dartmouth College, 6127 Wilder Laboratory, Hanover, NH 03755, USA}

\author[0000-0003-0220-2063]{Daniel Asmus}
\affiliation{Department of Physics \& Astronomy, University of Southampton, SO17 1BJ, Hampshire, Southampton, United Kingdom}

\author[0000-0002-9508-3667]{Roberto J. Assef}
\affiliation{N\'ucleo de Astronom\'ia de la Facultad de Ingenier\'ia y Ciencias, Universidad Diego Portales, Av. Ej\'ercito Libertador 441, Santiago 8370191, Chile}

\author[0000-0001-5481-8607]{Rudolf E. B\"{a}r}
\affiliation{Institute for Particle Physics and Astrophysics, ETH Z{\"u}rich, Wolfgang-Pauli-Strasse 27, CH-8093 Z{\"u}rich, Switzerland}

\author[0000-0002-0205-5940]{Patricia S. Bessiere}
\affiliation{Instituto de Astrof\'isica and Centro de Astroingenier\'ia, Facultad de F\'isica, Pontificia Universidad Cat\'olica de Chile, Casilla 306, Santiago 22, Chile}
\affiliation{Instituto de Astrof\'\i sica de Canarias (IAC), C/V\'i a L\'actea, s/n, E-38205, La Laguna, Tenerife, Spain}

\author[0000-0003-1014-043X]{Leonard Burtscher}
\affiliation{Leiden Observatory, PO Box 9513, 2300 RA, Leiden, the Netherlands}

\author[0000-0002-4377-903X]{Kohei Ichikawa}
\affiliation{Frontier Research Institute for Interdisciplinary Sciences, Tohoku University, Sendai 980-8578, Japan}

\author[0000-0002-2603-2639]{Darshan Kakkad}
\affiliation{European Southern Observatory, Casilla 19001, Santiago 19, Chile}
\affiliation{Department of Physics, University of Oxford, Denys Wilkinson Building, Keble Road, Oxford, OX1 3RH, UK}

\author[0000-0002-3233-2451]{Nikita Kamraj}
\affiliation{Cahill Center for Astronomy and Astrophysics, California Institute of Technology, Pasadena, CA 91125, USA}


\author[0000-0002-7962-5446]{Richard Mushotzky}
\affiliation{Department of Astronomy, University of Maryland, College Park, MD 20742, USA}

\author[0000-0003-3474-1125]{George C. Privon}
\affiliation{National Radio Astronomy Observatory, 520 Edgemont Rd, Charlottesville, VA 22903, USA}
\affiliation{Department of Astronomy, University of Florida, 211 Bryant Space Sciences Center, Gainesville, FL 32611, USA}

\author[0000-0003-0006-8681]{Alejandra F. Rojas}
\affiliation{Centro de Astronomia (CITEVA), Universidad de Antofagasta, Avenida Angamos 601, Antofagasta, Chile}

\author[0000-0002-3140-4070]{Eleonora Sani}
\affiliation{European Southern Observatory, Casilla 19001, Santiago 19, Chile}

\author[0000-0001-5464-0888]{Kevin Schawinski}
\affiliation{Modulos AG, Technoparkstrasse 1, CH-8005 Zurich, Switzerland}

\author[0000-0002-3158-6820]{Sylvain Veilleux}
\affiliation{Department of Astronomy, University of Maryland, College Park, MD 20742, USA}




\begin{abstract}
We present measurements of broad emission lines and virial estimates of supermassive black hole masses (\Mbh) for a large sample of ultra-hard X-ray selected active galactic nuclei (AGNs) as part of the second data release of the BAT AGN Spectroscopic Survey (BASS/DR2). 
Our catalog includes \Mbh\ estimates for a total \Nfullgood\ AGNs, determined from the \Halpha, \Hbeta, \MgII, and/or \CIV\ broad emission lines.
The core sample includes a total of \NblrGood\ AGNs drawn from the 70-month Swift/BAT all-sky catalog.
We also provide measurements for \Nbonusgood\ additional AGNs that are drawn from deeper Swift/BAT survey data. 
We study the links between \Mbh\ estimates and line-of-sight obscuration measured from X-ray spectral analysis. We find that broad H$\alpha$ emission lines in obscured AGNs ($\logNH> 22.0$) are on average a factor of $8.0_{-2.4}^{+4.1}$ weaker, relative to ultra-hard X-ray emission, and about $35_{-12}^{~+7}$\% narrower than in unobscured sources (i.e., $\logNH < 21.5$). This indicates that the innermost part of the broad-line region is preferentially absorbed. 
Consequently, current single-epoch \Mbh\ prescriptions result in severely underestimated ($>$1 dex) masses for Type 1.9 sources (AGNs with broad \Halpha\ but no broad \Hbeta) and/or sources with $\logNH\gtrsim 22.0$. 
We provide simple multiplicative corrections for the observed luminosity and width of the broad H$\alpha$ component ($L[{\rm b}\Halpha]$ and FWHM[b\Halpha]) in such sources to account for this  effect, and to (partially) remedy \Mbh\ estimates for Type 1.9 objects. 
As key ingredient of BASS/DR2, our work provides the community with  the data needed to further study powerful AGNs  in the low-redshift Universe.
\end{abstract}



\section{Introduction} 
\label{sec:intro}

Accurate estimates of super-massive black hole (SMBH) masses (\Mbh) in Active Galactic Nuclei (AGNs) are critical to understand SMBH demographics and growth, and their apparent co-evolution with their host galaxies \citep[e.g.,][]{Ferrarese2000,KormendyHo2013}. 
This requires large, highly complete surveys of AGNs (and SMBHs in general), as well as a detailed characterization of the different sources of uncertainties involved in the currently available methods to estimate \Mbh. 

In unobscured AGN, \Mbh\ is commonly determined through the so-called ``single epoch'' (SE), or ``virial'' black hole mass estimation method, which uses detailed spectral measurements probing the broad emission line region (BLR; see, e.g., works by \citealt{GreeneHo2005,Wang2009,TrakhtenbrotNetzer2012,ShenLiu2012,MejiaRestrepo2016a}, or reviews by \citealt{Shen2013_rev,Peterson2014_review}). 
This method is based on (1) the assumption of virialized motion of the BLR gas and (2) empirical relations between the accretion-related continuum luminosity and the BLR size. 
These latter relations are calibrated in reverberation mapping (RM) experiments, and take the general form $\RBLR  \propto  L_\lambda^{\alpha}$, where $L_\lambda$ is the monochromatic luminosity at a particular wavelength $\lambda$ 
\cite[e.g.,][]{Kaspi2000,Kaspi2005,Bentz2009,Park2012,Bentz2013}.
Under these assumptions, the width of the broad-emission-line profiles, such as the full width at half maximum (\fwhm), can be used as a proxy for the virial velocity of the BLR clouds, $v_{\rm BLR}$. 
\Mbh\ can thus be  expressed as:
\begin{equation}
 M_{\rm BH} = G^{-1} R_{\text{BLR} }\, v_{\text{BLR}}^{2} \propto f \, L_\lambda^{\alpha}\,\fwhm^{2}.
\end{equation}\label{eq:mbh_general}
Here $G$ is the gravitational constant and $f$ is a geometrical factor that accounts for the unknown structure and inclination of the BLR with respect to the line 
of sight. 
The $\RBLR\left(\Hbeta\right)-L_\lambda (5100\,{\rm \AA})$ relation is the only $\RBLR-L_\lambda$ relation that has been established for a large number of AGN covering a broad luminosity range, $10^{43} \lesssim \Lop/\ergs \lesssim 10^{47}$ \citep[][]{Bentz2013,BentzKatz2015}. 
Consequently, it has been used to calibrate several other SE  \Mbh\ prescriptions, using other emission lines and/or continuum bands \cite[e.g.,][]{MclureJarvis2002, GreeneHo2005,TrakhtenbrotNetzer2012}. 

Single-epoch mass prescriptions have allowed the estimation of \Mbh\ for tens of thousands of AGNs in large spectroscopic surveys \cite[e.g.][]{Shen2008,Kozlowski2017}, and have thus allowed probing of the evolution of the active SMBH population \cite[e.g.,][]{GreeneHo2007_BHMF,Vestergaard2009,Schulze2010_HES,TrakhtenbrotNetzer2012,KellyShen2013,Schulze2015} and of the links between SMBHs and their hosts \citep[e.g.,][]{Jahnke2009,Decarli2010,Merloni2010,Bongiorno2014,Suh2020} -- out to the highest accessible redshifts \cite[e.g.,][]{Mazzucchelli2017,Shen2019_hiz}.
Most of these studies, and particularly those dedicated to the largest samples and/or the highest-redshift ones (i.e., $z>3$), had to be based on surveys of luminous, unobscured and optically-selected AGN (e.g., SDSS quasars).

Despite these significant achievements, the SE approach should be used with care as it is subject to several significant (systematic) uncertainties that, in principle, may total to 0.4 dex in \Mbh, or even more \citep[e.g.,][]{Shen2013_rev,Pancoast2014a,Peterson2014_review}. 
Below, we briefly describe the most critical uncertainties relevant to the current work.
A major source of uncertainty stems from the need to assume a structural geometrical factor, $f$. 
The common approach is to deduce a universal $f$ by requiring that RM-based BH masses match those expected from the relation between \Mbh\ and the stellar velocity dispersion ($\sigma_{*}$) found in local galaxies \citep{Onken2004, Graham2016, Woo2015,Batiste2017}.
Some studies, however, have put forward the idea that the BLR may have a disk-like structure, at least in some AGN \cite[e.g.,][and references therein]{EracleousHalpern1994,Grier2013,Pancoast2014b,MejiaRestrepo2018a}. 
Such a distribution of gas would introduce a bias to the SE \Mbh\ determination, as the (unknown) inclination angle of the BLR disk with respect to the line of sight, for each AGN, limits the ability to measure the true virial velocity. 
In particular, BH masses would be overestimated (underestimated) at larger (smaller) inclination angles  \citep[e.g.][]{Collin2006,ShenHo2014,MejiaRestrepo2018a}.

Another important bias comes from the possible presence of winds, which could potentially affect the (observed) BLR gas dynamics. Indeed, several studies have shown that high ionization lines such as \CIV, commonly used to estimate \Mbh\ at $z\gtrsim2$, show highly blue-shifted profiles \cite[by up to 8000 \kms; e.g.,][and references therein]{Marziani2015,Vietri2020}, and thus their line widths are known to be poorer tracers of the {\it virial} velocity of the BLR gas, compared to other lines \citep[e.g.,][]{Richards2011,Coatman2016,MejiaRestrepo2018b}. 
In the case of the \MgII\ line, several studies have shown that the innermost, highest-velocity gas is affected by fountain-like winds and the global virial assumption is likely no longer valid for systems with $\fwhm \gtrsim 6000\,\kms$ \citep[e.g.,][]{TrakhtenbrotNetzer2012,Marziani2013,Popovic2019}.

Unlike the aforementioned biases, the partial obscuration of the broad line emitting region and its potential effect on \Mbh\ estimates remains poorly understood. 
\citet{Gaskell2018} proposed that compact, outflowing dusty clumps driven by radiation pressure may partially block the BLR emission. Such partial obscuration may explain the lack of correlation between disk and BLR line variabilities occasionally reported in RM campaigns  \cite[e.g.,][]{Goad1999,Cackett2015,Goad2016_STORM}.  
Preliminary observational evidence for this comes from the recent study of \citet{Caglar2020}, who identified a systematic offset of roughly $-0.6$ dex between the broad-\Halpha\ based \Mbh\ estimates and those based on the stellar velocity dispersion in 19 hard X-ray selected AGN drawn from the volume-complete LLAMA sample \citep{Davies2015}, including both unobscured and partially-obscured systems (as deduced from the relative strength of broad \Hbeta). 
Further support for the idea that this discrepancy could be (partially) attributed to dust obscuration comes from the fact that the discrepancy is found to be more dramatic in systems that completely lack broad \Hbeta\ emission \cite[Type 1.9 AGNs; see also][]{Goodrich1989,Goodrich1990,RicciFed2017_Msig}. 

One way to overcome these complications is to focus on the (rest-frame) near-infrared (NIR) regime, which is at least 10 times less sensitive to extinction than the optical. \citet[][]{RicciFed2017_Mvir} have provided SE \Mbh\ prescriptions that rely on several broad NIR lines (Pa$\alpha$, Pa$\beta$ and He\,\textsc{i} $\lambda1.083$ \micron), and on the hard X-ray continuum luminosity (in either the 2-10 keV and/or the 14-195 keV regime) as BLR probes. 
In addition to the advantages in overcoming obscuration, the use of the hard X-ray luminosity allows measurement of \Mbh\ even in low luminosity systems, where host contamination significantly affects optical AGN continuum estimates.
One further advantage of this method is that it can even be applied to some Type 2 AGNs -- the so-called hidden BLR Type 2s, where broad lines are detected in the NIR regime while the optical spectrum shows only narrow \Hbeta\ and/or \Halpha\ \citep[see e.g.,][]{Veilleux1997, Veilleux1997b,Veilleux1999,Riffel2006, Lamperti2017,Onori2017,denBrok_DR2_NIR}. 
However, calibrating this method requires  larger datasets of high signal-to-noise NIR spectra in order to improve the reliability of the method and better characterize the obscuration effects in Type 2 AGNs.

In order to further investigate all these issues, one has to obtain high quality optical-NIR spectroscopy of broad AGN emission lines and robust, independent line-of-sight obscuration measures for a large AGN sample that is unbiased with regard to obscuration. 
The BAT AGN Spectroscopic Survey (BASS) has been collecting and analyzing optical and NIR spectroscopy, X-ray spectral observations, and other multi-wavelength data for bright AGNs selected in the ultra-hard X-rays (14-195 keV) by the Swift/BAT mission.
The first data release of BASS \cite[DR1;][]{Koss2017,Lamperti2017,Ricci2017_Xray_cat} has already provided \Mbh\ estimates for hundreds of AGNs, over a wide range of obscuration, drawn from the 70-month catalog of Swift/BAT \citep{Baumgartner2013}. 
This highly complete and rich collection of multi-wavelength data has already been used in several studies that examined the links between AGN physics, structure and various emission components, and specifically to investigate topics where obscuration and/or orientation may play a key role \cite[e.g.,][]{Ricci2017N,Shimizu2018ApJ,Baer2019,Rojas2020}.

In this paper we present broad emission line measurements as part of the second data release of the BAT AGN Spectroscopic Survey (BASS/DR2), including the analysis of hundreds of new spectra and improved estimates of the black hole masses of hundreds of AGN, thus greatly improving and expanding on the first data release described in \citet{Koss2017}.
We then combine these new measurements with the rich BASS/DR2 multi-wavelength dataset to explore the effect of dust obscuration on single-epoch \Mbh\ estimates from optical broad emission lines. 
Other BASS/DR2 studies present extensive NIR spectroscopy and use it to address complementary issues \citep{denBrok_DR2_NIR,Ricci_DR2_NIR_Mbh}.
Throughout this work, we adopt $\Omega_{\rm M} = 0.3$, $\Omega_\Lambda = 0.7$, and $H_0 = 70 \,\kms\,{\rm Mpc}^{-1}$.

\section{Data Content and Analysis}

\subsection{Overview of Survey, Sample and Data Content}

The ultimate goal of BASS is to complement the largest available sample of Swift/BAT, hard X-ray selected AGNs with optical spectroscopy and ancillary multi-wavelength data using dedicated observations and archival data, to complete the first large survey (${\gtrsim}1000$ sources) of the most powerful accreting SMBHs in the low-redshift Universe. 
This work is part of a series of papers devoted to the 2nd Data Release (DR2) of BASS.
In particular, this paper presents detailed spectral measurements of the broad-line AGN, with either a broad  \Hbeta\ or \Halpha\ line (i.e., $\fwhm>1000\,\kms$), as well as a smaller subset of higher-redshift sources with \mgii\ and \civ\ broad emission lines. 
\cite{Koss_DR2_overview} provides an overview of BASS DR2, while \cite{Koss_DR2_catalog} provides a detailed account of the BASS/DR2 AGN catalog and main observational data, in particular the optical spectroscopy which is used here.
Other key BASS DR2 papers include \cite{Koss_DR2_sigs}, where we present the velocity dispersion measurements for (obscured) BASS sources; 
\cite{Oh_DR2_NLR}, where we focus on spectral measurements for narrow-line AGN and host light decomposition;
and \cite{denBrok_DR2_NIR} and \cite{Ricci_DR2_NIR_Mbh}, where we present extensive NIR spectroscopic observations, and analyze (broad) hydrogen and high-ionization (coronal) emission lines.
The broad line measurements and related \Mbh\ (and \lledd) estimates presented herein are used in \citealt{Ananna_DR2_XLF_BHMF_ERDF} to determine the BH mass and Eddington ratio distribution functions among essentially all BASS/DR2 AGNs.

BASS/DR1 used mostly archival telescope data \citep[see e.g. Fig 1 in][]{Koss2017} for 641 BAT AGNs, including $>$250 spectra from the SDSS and 6dF surveys.  
In this DR2 paper we provide a complete sample of black hole mass estimates from optical broad emission lines for \NblrGood\  AGNs with such lines in the 70-month Swift/BAT survey \citep{Baumgartner2013}. 
As part of our efforts towards DR2, we obtained new spectroscopy for many AGNs that did not have reliable black hole mass determination in DR1.
This includes 
(1) AGNs that did not have sufficient data (or data quality) to yield a black hole mass measurement, including cases where the DR1 archival spectra were not properly flux-calibrated \citep[e.g., 6dF/2dF spectra;][]{Jones2009}; 
(2) Type 1.9 AGNs that were lacking a sufficiently high quality spectrum to derive their BH masses, i.e. either a high-quality spectrum of their broad \Halpha\ lines or a spectrum that enables a robust velocity dispersion measurement;
and (3) any DR1 AGN with only a broad \Hbeta\ line measurement, where \Halpha\ coverage was missing.  

The new BASS/DR2 spectroscopic observations were carried out with a variety of facilities and instruments, as detailed in the main DR2 Catalog and Data paper \citep{Koss_DR2_catalog}. 
Here we note that the large majority of new spectra were obtained with either 
the Double Beam Spectrograph (DBSP) mounted on the Hale 5 m telescope at Palomar observatory (\citealt{OkeGunn1982_DBSP}; $>$400 AGNs, mainly northern targets); 
the X-Shooter spectrograph at the Very Large Telescope (\citealt{Vernet2011_XShooter}; $>$200 sources, mainly southern);
or the Goodman spectrograph mounted on the SOAR telescope at Cerro Pachon (\citealt{Clemens2004_Goodman}; $>$150 sources, also southern).
More details on the facilities used, the spectroscopic setups and spectral resolutions, the observations, and the reduction procedures, can be found in \cite{Koss_DR2_catalog}.

%

The DR2 also includes {\color{black} publicly available optical spectroscopy from the SDSS ($\sim$150 sources; drawn from SDSS DR16, \citealt{Ahumada2020_SDSS_DR16}), and} 
a small number of additional archival spectra, obtained as part of  follow-up observations of ROSAT sources \citep{Grupe1999a,Grupe1999b}. Finally, it includes spectra from recent studies of newly identified BAT AGNs \cite[see][]{Rojas2017}.

In a non-negligible number of cases, the extensive data collecting process resulted in multiple optical spectra of the same source.  
In such cases
we select for each AGN the single best spectrum for broad-line based \Mbh\ measurements, and use it in the present analysis (i.e., in Sections \ref{sec:cat} and \ref{sec:bHa_and_obsc}). 
This selection is done by considering the signal-to-noise ratio, spectral resolution and quality of our spectral fits (see Section \ref{sec:fitq} below).  
Following this selection, our dataset consists of a total of \Nblr\ unique AGNs with at least one useful optical spectrum. 
These \Nblr\ AGNs cover the redshift range $z\sim0-4$, with the vast majority ($>$90\%) being at $z\lesssim0.5$.

All BASS DR2 spectra used here have sufficient spectral resolution to robustly measure the broad emission lines that are at the heart of the present paper.
As noted, the main DR2 Catalog and Data paper \citep{Koss_DR2_catalog} provides ample details about the new, the previously-obtained (DR1), and the archival spectra used throughout BASS DR2.
%

As mentioned above, the BASS/DR2 sample is fundamentally based on AGNs identified through the 70-month Swift/BAT catalog. However, the broad line measurements described in the present study were also carried out on the optical spectra of \Nbonusraw\ additional AGNs, which were acquired as part of the on-going BASS efforts to follow up on the increasingly deeper (and larger) content of the Swift/BAT all-sky survey \cite[e.g.,][]{Oh2018_BAT_105m}.
While we provide these measurements, we stress that this ``bonus'' sample is neither complete nor final: it does {\it not} represent any sort of flux- or volume-complete subset of deeper BAT data, and it is possible that future BASS follow-up observations and analysis could reveal significant changes to the determinations of optical counterparts, their redshifts, their AGN nature, and/or any other property.
Apart from providing the relevant spectral measurements, we \textit{ignore} this bonus sample throughout the rest of this paper. In particular, we do \textit{not} include the bonus sample measurements when further discussing the BASS broad-line AGN statistics, measurements or implications for any of the analyses we present (unless explicitly mentioned otherwise).  {\color{black} A summary of the number of  \Mbh\ estimations from the \Halpha, \Hbeta, \mgii\ and \civ\ Broad Emission lines is presented in Table \ref{tab:totalfits}}

The data reduction and analysis for the DR2 has maintained the same uniform approach described in the initial DR1 paper \citep{Koss2017}. All the spectra were processed using standard tasks (in IRAF or comparable reduction frameworks) for cosmic ray removal, 1d spectral extraction, and wavelength and flux calibrations.  
The spectra were flux calibrated using standard stars, which were typically observed twice per night, whenever possible.  In the DR2, we have also implemented the use of the \texttt{molecfit} software \citep{Smette2015} to correct spectral regions affected by telluric absorption (e.g. H$_2$O, CO$_2$, CH$_4$ and O$_2$), based on nightly weather data \citep{Koss_DR2_overview,Koss_DR2_catalog}.

\subsection{Continuum and Line Emission Modeling}
\label{sec:modeling}

For each of the \Nblr\ broad-line emitting sources in BASS with optical spectroscopy of either one of the \Halpha, \Hbeta, \MgII\ and/or \CIV\ broad emission lines, we fitted the spectral complexes around these lines following the established and well-tested procedures initially presented in \citet{TrakhtenbrotNetzer2012} and further developed in  \citet{MejiaRestrepo2016a}, where more details can be found.

We note that the spectral modeling of \textit{narrow}-line (i.e., Type 2) AGNs in BASS/DR2, and generally host galaxy decomposition and narrow line emission (including beyond the spectral regions considered here), was carried out independently, using a different fitting procedure, and is described in a dedicated BASS/DR2 paper \citep{Oh_DR2_NLR}. 
In Section~\ref{sec:cat} we show a few basic properties of the BASS/DR2 Type 2 AGN population, based on this independent spectral analysis, which is however not used in any other part of the present study.

The broad line fitting procedures use the \texttt{PySpecKit} Python package \citep{GinsburgMirocha2011} to measure broad emission line properties. 
In brief, 
each spectrum is first corrected for Milky-Way (foreground) dust extinction, using the \cite{Schlegel1998} maps and the \cite{Cardelli1989} extinction law (with $R_{\rm V}=3.1$).
Next, the continuum emission is modeled with a (local) power law,  fitted to certain continuum-dominated bands around the emission line complex of interest (see Table~1 in \citealt{MejiaRestrepo2016a}). 
After subtracting the continuum emission we proceed with the emission line modeling as follows. 
Narrow line components, including the 
[O\,\textsc{iii}]\,$\lambda\lambda$4959,5007, 
[N\,\textsc{ii}]\,$\lambda\lambda$6548,6584, and [S\,\textsc{ii}]\,$\lambda\lambda$6717,6731 lines, as well as the narrow components of the \Halpha\ and \Hbeta\ lines, 
are modeled with a single Gaussian profile, each, except for rare cases where a visual inspection of the residuals motivated us to use an additional Gaussian. 
The widths and relative (velocity) shifts of these (primary) narrow profiles are tied to each other, to avoid over-fitting in heavily blended line complexes such as the \Halpha\ spectral complex. 
%
The profiles of the most prominent broad lines (\Halpha, \Hbeta, \mgii\ and \civ) are modeled using two broad Gaussian components (each), while weaker emission lines are modeled with a single broad Gaussian (including He\,\textsc{ii}\,$\lambda$1640, N\,\textsc{iv}\,$\lambda$1718, Si\,\textsc{iii}]\,$\lambda$1892). 
We emphasize that the two broad Gaussian components are used only in order to account for the total broad emission line profiles, and we do \textit{not} consider any physical interpretation to the two separate components. This choice is based on previous works, which showed that two broad Gaussian components provide a good compromise between the number of free parameters (i.e., 6) and the achieved goodness of fit  \citep[e.g.,]{Shang2007,TrakhtenbrotNetzer2012,MejiaRestrepo2016a}.
All broad emission features are restricted to have $\fwhm > 1000\,\kms$ and to be broader than the narrow emission features (including of the same transition; see above).
We allow the central wavelength of each Gaussian component to be shifted by up to 1500 \kms\ relative to the laboratory central wavelength of the transition. 
The blue-shift of the \civ\ and He\,\textsc{ii}\,$\lambda$1640 components is allowed to be larger, up to 5000\,\kms, in agreement with what is observed in other large AGN samples \cite[e.g.,][]{Shang2007,Runnoe2014,Coatman2017}. 
We verified that yet larger shifts are neither observed in our sample nor required in our modeling of the spectra.
When fitting the \Hbeta, \mgii\ and \civ\ spectral complexes, we also account for (heavily blended) Fe\,\textsc{ii} and Fe\,\textsc{iii}  using the iron template described in \citet{MejiaRestrepo2016a}, broadened and shifted separately for each source. 
Since the \mgii\ spectral complex was modeled using a relatively narrow, ``local'' part of the spectrum ($\sim$2600--3000~\AA), we did not include a designated model for the Balmer continuum. In our modeling approach, the Balmer continuum is assumed to be blended with the underlying disk continuum, forming the ``local'' continuum emission (see \citealt{MejiaRestrepo2016a} for a detailed discussion).

From each fitted emission line profile we extract 
(1) the shift of the line centroid, i.e. the flux-weighted average center of the line emission, and 
(2) the shift of the line peak -- providing two probes of $\Delta v$;
(3) the total line luminosity; and
(4) the line width, in terms of \fwhm. 
From the latter we subtract in quadrature the instrumental spectral resolution, according to the observational setup (i.e., in velocity space; see \citealt{Koss_DR2_overview} and  \citealt{Koss_DR2_catalog}).
Whenever the resolution-corrected FWHMs of narrow lines fall below the corresponding instrumental resolution, we regard the emission line FWHMs as upper limits, and report the (velocity-equivalent) instrumental resolution in our catalog (see descriptions of Tables~\ref{tab:HA}-\ref{tab:CIV}).
We note that even in these cases (affecting \Nhabelowinst\ sources with  \Hbeta\ measurements and \Nhbbelowinst\ sources with \Halpha\ measurements), the narrow lines still provide the best way to decompose the complex key broad emission line profiles, and tease out the \textit{broad} emission line widths, which are crucial for BH mass estimates.

Together with the line profile properties, we also computed the monochromatic continuum luminosities at several narrow wavelength bands, $L_{\lambda} \equiv \lambda\ L\left(\lambda \right)$, to be used for the estimation of \Mbh.
In particular, we measured  \Luv,  \Lthree, \Lop\ and \Lsix\ for \Mbh\ estimates using \civ, \mgii, \Hbeta\ and \Halpha, respectively \citep[see, e.g.,][]{GreeneHo2005,VestergaardPeterson2006,Wang2009,MejiaRestrepo2016a}. 

Uncertainties on all spectral measurements were derived by a resampling procedure. Each observed spectrum was used to generate 100 mock spectra, based on its noise (variance). 
Each of these mock spectra were then fit using our spectral decomposition procedures. 
For each measured quantity, of each AGN, the 16th and 84th percentiles of the corresponding distribution of measurements were then used to determine the corresponding uncertainty.

All spectral fits were visually inspected by at least three independent, experienced team members (J. M.-R., B. T., and M. K.).
In the cases where the fits were inadequate, we have adjusted some of the parameters of the fitting procedure and re-fitted the data. These manual adjustments typically involved the continuum placement and/or the limits to the widths and/or shifts of emission line components. We note that these minor numerical adjustments did \textit{not} contradict the physical motivation and/or meaning of the emission components (e.g., broad Gaussian components always remained broader than the narrow ones, etc.).

We ultimately visually inspected all the final (adjusted) spectral fits, used them to derive the spectral measurements we rely upon throughout the rest of the paper and catalog (as well as the related uncertainties, using our re-sampling procedure), and assigned them spectral fit quality flags, which we describe immediately below.\\

\subsection{Model Fit Quality}
\label{sec:fitq}

We visually inspected all the (final) spectral fits and assigned a quality flag (\FQ) representing the quality of each fit (i.e., each spectral complex for each source), ranging from 1 to 3. %
$\FQ=1$ marks good quality fits, with randomly distributed residuals, providing the most reliable line measurements we can hope to achieve within the scope of a large effort like BASS. 
$\FQ=2$ is used to mark good/acceptable fits, that may show slight systematic residuals and that could be \textit{slightly} improved with further, less-trivial manual adjustments; however $\FQ=2$ fits can still be used to provide reliable broad line measurements. 
In such cases we preferred not to further adjust the fits, as this may make our fitting procedure too heterogeneous. 
Finally, $\FQ=3$ marks those spectral fits that have failed, and/or data that exhibit very low signal-to-noise or otherwise severe issues. 
In such cases, our (reasonable) attempts to manually adjust the spectral fitting procedure could not result in an acceptable fit.
We exclude all such $\FQ=3$ fits from both the BASS/DR2 catalog and any of the analysis that follows.
We further discuss these problematic fits in Section \ref{sec:prob_fits} below.

Examples of spectra and best-fit models representing the three \FQ\ classes are shown in Appendix \ref{app:fitq} and Fig~\ref{fig:fitq}.  
By examining the results of the ``useful'' fits (i.e., with $\FQ=1$ and 2), we estimate that the minimum rest-frame equivalent width (rEW) needed to achieve such high-quality fits for the \Halpha, \Hbeta, \mgii\ and \civ\ lines are ${\rm EW}=10$, 3, 9, and 10 \AA, respectively.\footnote{\textcolor{black}{Given the highly heterogeneous nature of the BASS/DR2 spectroscopic observations and data, we prefer to report the lowest EWs (i.e. physically weakest lines) we can measure, instead of reporting a more generic $S/N$ criterion.}}

Out of the initial \Nhatotal\ and \Nhbtotal\ unique objects with available spectra of the broad \Halpha\ and/or \Hbeta\ lines (respectively), after retaining only fits with $\FQ=1$ and $2$, we end up with \Nha\ unique AGNs with useful measurements of the broad \Halpha\ line, of which \Nhai\ and \Nhaii\ unique objects have $\FQ=1$ and 2, (respectively); and \Nhb\ AGNs with useful measurements of the broad \Hbeta\ line, of which \Nhbi\ and \Nhbii\ unique objects have $\FQ=1$ and 2 (respectively). 
The remaining useful measurements come from sources with with acceptable fits but large errors ($\FQ=2.5$; see Section~\ref{sec:prob_fits} below).
Also, there are \Nhahb\ AGNs for which we have useful measurements of both the \Hbeta\ and \Halpha\ broad emission line. 
In total, we have \Ntoth\ unique AGNs with useful measurements of either the \Halpha\ and/or \Hbeta\ broad emission lines. 
For higher-redshift sources ($z\gtrsim0.7$), essentially all of which are classified as beamed AGNs, only the \mgii\ and/or the \civ\ lines are available in our optical spectroscopy. 
From the initial \Nmgtotal\ and \Ncivtotal\ spectra with broad \mgii\ and \civ\ lines (respectively), we obtained \Nmg\ and \Nciv\ useful measurements of the \mgii\ and \civ\ broad emission line complexes (respectively). These are further split to \Nmgi\ and \Ncivi\ fits with $\FQ=1$, and \Nmgii\ and \Ncivii\ fits with $\FQ=2$ (in each case, for \mgii\ and \civ, respectively). 
There are \Nmgciv\ objects with useful measurements in both \mgii\ and \civ. %
Our data-set thus consists of \Ntotz\ unique AGNs with either \mgii\ and/or \civ\ broad lines measurements {\color{black} out of which \Nuniquez\ do not have complementary \Halpha\ and/or \Hbeta\ broad line measurements. Therefore, we end up with a total of \NblrGood\ objects ( $\Ntoth$, from \Halpha\ and \Hbeta, plus  \Nuniquez, from \mgii\ and \civ) with black hole mass estimations from Broad emission lines}. 

\begin{deluxetable}{lrrrrr}
\label{tab:totalfits}
\tablecaption{{\color{black} Summary of good and acceptable \Mbh\  estimations from different Broad emission lines (see section \ref{sec:fitq}). Bonus 105m objects correspond to additional sources drawn from deeper-than-70-month DR2 Swift/BAT survey data.   }}
\tablewidth{0pt}
\tablehead{
\colhead{Subset} & \colhead{\Halpha} & \colhead{\Hbeta} & \colhead{\mgii} & \colhead{\civ} & \colhead{Total Objects$^{b}$}}
\startdata
70m non beamed            &   \Nhagoodnb     &   \Nhbgoodnb     &  \Nmggoodnb     & \Ncivgoodnb    & \Nhgoodnb \\ 
70m beamed  &   \Nhagoodb     &   \Nhbgoodb     &  \Nmggoodb    & \Ncivgoodb    &  \NBZQgood   \\
\hline
Total  70m               & \Nha      &   \Nhb   & \Nmg   &  \Nciv   &   \NblrGood  \\ 
Bonus 105m$^{a}$            &   \Nhabonusgood   &   \Nhbbonusgood  &  \Nmgbonusgood   & \Ncivbonusgood  &  \Nbonusgood \\
\hline
Total Objects               & \Nhafullgood      &   \Nhbfullgood   & \Nmgfullgood   &  \Ncivfullgood   &   \Nfullgood \\ 
\hline
\enddata
\tablenotetext{a}{{\color{black} Apart from providing the relevant spectral measurements, we \textit{ignore} this bonus sample throughout the rest of this paper.} }
\tablenotetext{b}{\color{black} Since there are objects with simultaneous \Mbh\ estimations from different broad emission lines (e.g. \Halpha-\Hbeta, \Hbeta-\mgii\ and \mgii-\civ), the total number objects is smaller than the sum of available \Mbh\ measurements from the different emission lines.} 
\end{deluxetable}

\subsubsection{Problematic Fits}
\label{sec:prob_fits}

After our visual inspection of the spectral fits (including those that required minor adjustments), we have a total of \Nhabad\ objects with failed fits ($\FQ=3$) of the broad \Halpha\ line, \Nhbbad\ in the case of \Hbeta, \Nmgbad\ in the case of \mgii, and \Ncivbad\ in the case of \civ. 
There is a variety of reasons for such failed fits, including a low signal-to-noise ratio ($S/N$), imperfect correction of telluric features associated with certain redshift ranges, incomplete profiles due to the specific source redshift and observation setup, and/or difficulties in data reduction. 
Table \ref{tab:failedfits} summarizes the breakdown of the failed fits according to these (and other) categories, and below we further discuss some of the main ones.

Among the \Nhasnr\ objects with low $S/N$ over the \Halpha\ complex, 4 are classified as Type 1.9 AGNs, that is sources which show broad \Halpha\ but \textit{no} broad \Hbeta\ line emission (see Section~\ref{subsec:types} for more details). 
This 
may result from significant obscuration by dust, dimming even the \Halpha\ emission, and making even prominent emission lines like \Halpha\ harder to detect and model. 
There are also two beamed (``BZQ'') AGNs whose \Halpha\ line is difficult to detect due to their relatively high redshift. 
The remaining systems with low $S/N$ are bona-fide broad-line (Type 1) AGNs whose spectra have not been re-observed since DR1, despite their apparent low $S/N$ in that initial data release. 
The higher number of objects with low $S/N$ in the \Hbeta\ line, \Nhbsnr, is not surprising as \Hbeta\ is weaker than \Halpha\ by at least a factor of $\sim$3, and thus requires a higher overall $S/N$, particularly given the need to properly model the blended iron emission complex. 
These low $S/N$ sources will likely be re-observed in future BASS spectroscopy and analyzed as part of a future DR.


For one source (BAT ID 1204, a.k.a. RBS~2043), the \mgii\ emission line is considerably narrower ($\approx$1200 \kms) than what is seen in quasars, while the \Hbeta\ line is narrower still, consistent with an NLR origin. 
The \mgii\ width leads to a BH mass estimate (see Section~\ref{sec:mbh_lledd} below) that is much lower than what is deduced from stellar velocity dispersion ($\log\left[\Mbh/\Msol\right] \simeq 7.4$ vs. 9.6; \citealt{Koss_DR2_sigs}).
Our broad line catalog thus reports the basic measurements for the \mgii\ spectral complex, but not the associated BH mass.

{\color{black} There is a single object with unclear presence of BLR features (in \Hbeta; BAT ID 349, a.k.a. UGC~3601), that is with the data in hand we could not robustly determine whether there is a broad emission component.}
We note that, in principle, in extreme cases one could also expect a mis-identification of blended {\it narrow} \Halpha\ and [N\,\textsc{ii}] lines as a broad and weak \Halpha\ profile. 
{\color{black} We have not identified such questionable broad \Halpha\ profiles among our BASS/DR2 AGNs.}
At any rate, if such ambiguous cases indeed have broad Balmer emission lines (that is, they originate from the BLR), their measurement would require sufficiently high $S/N$ and line EW, as well as more detailed analysis 
(see, e.g., \citealt{Oh2015} for detailed examples and discussion).


In addition to the failed fits (i.e., $\FQ=3$), our inspection of the fitting results uncovered another subset of \Nlargeerr\ spectral fits for which the resampling technique resulted in exceptionally large (fractional) uncertainties on the 
\Halpha\ (\Nlargeerrha),
\Hbeta\ (\Nlargeerrhb), and 
\civ\ (\Nlargeerrciv) line widths, $\Delta[\fwha]/\fwha \geq 50\%$, {\color{black} and/or large systemic offsets of the narrow line}. However, our visual inspection suggests the fits are acceptable. 
Upon closer inspection, it seems that in these cases the contrast between the broad emission line and the adjacent continuum emission is relatively low, which led a significant fraction of the re-sampled (mock) spectra to be fitted by extremely broad profiles (i.e., FWHM[\Halpha] $>10,000\,\kms$). 
Since the best-fit parameters appear to represent the observed spectra well, we do not downgrade the quality flags of such fits to $\FQ=3$, and instead mark these \Nlargeerr\ cases with a dedicated flag, $\FQ=2.5$.

{\color{black} We finally note that any physical interpretation of the line (velocity) shifts reported in our catalog should be done with care, as these naturally depend on the precise redshifts used for our spectral analysis. Specifically, any interpretation of {\it narrow} emission line shifts should consider the fact that our redshifts are, themselves, based on narrow emission lines (i.e., \OIII; \citealt{Koss_DR2_catalog}).
The two sources in our catalogs that have extremely large shifts listed ($>$1000\,\kms; BAT IDs 334 \& 1332) are also marked as $\FQ=2.5$.}

To summarize, users of our catalog who prefer to have the largest possible sample of reliable fits of the broad \Halpha\ line, and of derived quantities, can use the default quality cut $\FQ<3$.
More cautious analyses may however prefer to impose the stricter $\FQ\leq2$ cut. 
We indeed adopt this stricter cut for all of the analyses presented below.


\begin{deluxetable}{lrrrrr}
\label{tab:failedfits}
\tablecaption{Breakdown of failed and/or problematic spectral fits}
\tablewidth{0pt}
\tablehead{
\colhead{Reason for failure} & \colhead{\Halpha} & \colhead{\Hbeta} & \colhead{\mgii} & \colhead{\civ} & \colhead{Total Spectra}}
\startdata
Low $S/N$             &   \Nhasnr     &   \Nhbsnr     &  \Nmgsnr     & \Ncivsnr     &  \Ntotalsnr  \\
Tellurics             &   \Nhatellu   &   \Nhbtellu   &  \Nmgtellu   & \Ncivtellu   &  \Ntotaltellu  \\
Unclear BLR           &   \Nhaunclear &   \Nhbunclear &  \Nmgunclear & \Ncivunclear &  \Ntotalunclear  \\
 Mismatched NLR       &   \Nhanlr     &   \Nhbnlr     &  \Nmgnlr     & \Ncivnlr     &  \Ntotalnlr  \\
Incomplete profile    &   \Nhaincom   &   \Nhbincom   &  \Nmgincom   & \Ncivincom   &  \Ntotalincom  \\
Reduction issue\tablenotemark{a}
 &    \Nhareduc    &   \Nhbreduc &  \Nmgreduc  &   \Ncivreduc     &       \Ntotalreduc  \\
FeII fit issue        &   \Nhafe      &    \Nhbfe   &  \Nmgfe    &   \Ncivfe     &     \Ntotalfe  \\
Absorption features\tablenotemark{b}   &   \Nhaabs      &  \Nhbabs      &  \Nmgabs   &  \Ncivabs   &     \Ntotalabs   \\ 
Double-peak broad lines & \Nhadouble & \Nhbdouble & \Nmgdouble & \Ncivdouble &  \Ntotdoublefull\\
\hline 
Total Failed fits                & 58      &   52   &   13 &  10   &   133  \\ 
\hline 
\enddata
\tablenotetext{a}{This includes fringing (or otherwise `wavy' spectral features), problems with flux calibration, and other artifacts.}
\tablenotetext{b}{Indicating a strong absorption feature superimposed on the (broad) emission line, which limits our ability to properly model the latter.} 
\end{deluxetable}

In addition to the failed fits described above, we also excluded from our main catalog and analysis objects with indications of double-peaked broad emission lines where the SE black hole mass estimation approach is not applicable, and their physical origin is still debated. 
The proposed origins include the accretion disk, dual BLR in a binary SMBH system, bipolar outflows, and/or flares or spiral arms in the accretion disk  \cite[see,e.g.,][and references therein]{VeilleuxZheng1991,Zheng1991,EracleousHalpern1994,Jovanovic2010,StorchiBergmann2017,RicciSteiner2019}. 
Out of our initial sample of  broad-lined Swift/BAT AGNs, we find a total of \Ntotdouble\ candidate double-peaked systems, comprising \Nhadouble\ and \Nhbdouble\ sources with double-peaked \Halpha\ and \Hbeta, respectively (20 sources have double-peaked profiles in both Balmer lines).
We provide basic information regarding these double-peaked sources in Table~\ref{tab:double_peaked}.

\subsection{Black hole Mass and Eddington Ratio Estimates}
\label{sec:mbh_lledd}

Black hole masses are estimated using the \Halpha, \Hbeta, \mgii, and \civ\ emission line measurements, according to their availability for each source. 
In all our estimates, we assume a common (and universal) virial factor of $f=1$.\footnote{This virial factor is appropriate for \Mbh\ prescriptions that depend on the FWHM of the respective emission line, and corresponds to  $f_\sigma = 5.5$ if one uses line velocity dispersion instead.}
This value is appropriate for \Mbh\ estimates that rely on the \fwhm\ of broad emission lines as the BLR velocity tracer, and is further motivated by the observationally-derived mean value from \cite{Woo2015}, where RM-based \Mbh\ estimates were matched, on average, to the corresponding expectations from the \Mbh-\sigmas\ relation. 
\cite{Woo2015} found an uncertainty on the mean value of $f$ of about 30\%. 

For the \Halpha\ line, we used the specific prescription calibrated in \cite{GreeneHo2005}, and in particular their Eq. 6. 
We note, however, that their calibration assumed $f=0.75$ (corresponding to a spherical BLR distribution). 
Our choice of $f=1$ therefore requires adjusting the \cite{GreeneHo2005} calibration by $\times4/3$ (or $+$0.125 dex).
Our \Halpha-based prescription for \Mbh\ is thus
\begin{multline}\label{eq:mbh_ha}
\Mbh(\Halpha) = \\ 
2.67 \times 10^6 \left(\frac{L[{\rm b}\Halpha]}{10^{42}\,\ergs}\right)^{0.55} \,
     \left(\frac{{\rm FWHM}[{\rm b}\Halpha]}{10^3\,\kms}\right)^{2.06}\,\Msol \,.
\end{multline}
%
We note that this prescription was calibrated to best match  \Hbeta-related (RM) measurements, and it does {\it not} strictly follow the virial assumption, in the sense that the exponent of the velocity term is 2.06 (and not 2.00). 
In the case of \Hbeta, we used the calibration presented in \citet[][]{TrakhtenbrotNetzer2012} while
in the case  of \mgii\ and \civ\ we followed \citet[][]{MejiaRestrepo2016a}. 
A summary of the specific calibrations that we adopted can be found in Table~\ref{tab:mass}. 

\begin{deluxetable}{lccl}
\label{tab:mass}
\tablecaption{Parameters of virial BH mass prescriptions}
\tablewidth{0.475\textwidth}
\tablehead{
\colhead{Observables} & \colhead{$\log{K}$} & \colhead{$\alpha$} & \colhead{$\beta$} }
\startdata
\fwha, \Lbha                & 7.526      &  0.55     &  2.06     \\
\fwhb, \Lop                 & 6.721      &  0.65     &  2.00     \\
\fwmg, $L_{\rm 3000}$       & 6.906      &  0.61     &  2.00     \\
\fwciv,$L_{\rm 1450}$       & 6.331      &  0.60     &  2.00     \\
\enddata
\tablecomments{The mass prescriptions are of the form $\Mbh = K\,\, (\lambda L_{\lambda} )^{\alpha}\,\, \fwhm^{\beta}$, with $\lambda L_{\lambda}$ in units of $10^{44}\,\ergs$, FWHM in units of 1000\,\kms, and \Mbh\ in units of \Msun.}
\end{deluxetable}

While not essential for the main findings of the present study, we use two key properties of broad-line AGNs to provide context in Section~\ref{sec:cat}: their bolometric luminosities, \Lbol, and Eddington ratios, $\lledd \equiv \Lbol/(1.5\times10^{38}\Mbh)$.{\color{black}  This scaling assumes \Lbol\ in units of \ergs, \Mbh\ in units of $M_{\odot}$, and  solar composition gas.} 
For \Lbol\ estimates, we rely on the intrinsic, absorption-corrected X-ray luminosities in the $2-10$ keV regime, $L_{\rm int}(2-10\,\kev)$, as determined through the elaborate spectral decomposition of the entire X-ray SEDs, discussed in detail in \cite{Ricci2017_Xray_cat}. 
We then assume a simple, universal bolometric correction of $\kappa_{2-10\,\kev}\equiv\Lbol/L_{\rm int}(2-10\,\kev)=20$, which is a typical value for luminous AGNs \cite[e.g.,][]{Marconi2004,Vasudevan2009}.
There are many other bolometric corrections suggested in the literature, including those defined at other spectral regimes \cite[e.g.,][]{2006ApJS..166..470R,Runnoe2012,TrakhtenbrotNetzer2012}, luminosity-dependent corrections \cite[e.g.,][]{Marconi2004,TrakhtenbrotNetzer2012,Duras2020_BC},  \lledd-dependent ones \cite[e.g.,][]{Vasudevan2009}, and/or those motivated by accretion disc models \cite[e.g.,][]{Netzer2019_BC}.
To exemplify how these various prescriptions may affect the simple \lledd\ estimates we use here, we note that given the range of $L_{\rm int}(2-10\,\kev)$ for our broad-line sources, the luminosity-dependent prescription of \cite{Marconi2004} would suggest $\kappa_{2-10\,\kev}\sim 10 - 70$.

The publicly available DR2 catalogs provide many measurements that can be used for alternative determinations of \Lbol\ (and thus of \lledd), and in particular rest-frame optical monochromatic continuum luminosities (\Lop) and broad \Halpha\ line luminosities (\Lbha; both included in this catalog\footnote{{\color{black}  The luminosities measured in the optical regime are corrected for galactic extinction.}}), as well as X-ray continuum luminosities (\Lhard\ and \Luhard). 
Indeed, other BASS (DR2) publications may use different choices for the determination of \Lbol\ and \lledd, as best suits their science goals.


We finally stress that neither the \Mbh\ nor the \Lbol\ prescriptions we use were calibrated using, are meant to be applied on, beamed AGNs. 
In such sources the continuum X-ray, UV and optical luminosities may be boosted, and thus both \Mbh\ and \Lbol\ may be significantly overestimated. 
We describe the identification of beamed sources among our sample of BASS/DR2 broad-line AGNs in Section \ref{subsec:types}.

\section{The BASS/DR2 Broad Line catalog}
\label{sec:cat}

In this Section we present the BASS/DR2 broad emission line catalog and some key characteristics of the broad line AGN demographics in BASS.

Our detailed spectral measurements, their uncertainties, and select derived quantities, are provided in electronic form herein, and online.\footnote{\url{http://www.bass-survey.com/data.html}}
Tables \ref{tab:HA}, \ref{tab:HB}, \ref{tab:MG} and \ref{tab:CIV} (in Appendix~\ref{app:blr_catalog}) describe the content of the BASS/DR2 broad-line catalogs for the \ha, \hb, \mgii, and \civ\ spectral regions (respectively), for the BASS/DR2 AGNs with adequate spectral fit quality ($\FQ<3$). 
We also provide, in a separate set of tables with identical format, spectral measurements for the \Nbonusgood\ AGNs from the ``bonus'' sample
(i.e., sources drawn from deeper-than-70-month Swift/BAT survey data), which had adequate spectral fit quality (i.e., $\FQ\leq2$).

In Appendix~\ref{sec:DR2vsDR1} we provide a detailed comparison of line width and BH mass measurements derived in BASS/DR2 and DR1 for those broad-line AGNs that are part of both DRs. Fig.~\ref{fig:MbhHADR12} summarizes these comparisons graphically, highlighting that our DR2 spectral measurements are, overall, in excellent agreement with DR1 measurements.

\subsection{Demographics of Optical AGN Emission Line Classes}
\label{subsec:types}

Here we further refine the classification of broad line AGNs in BASS/DR2, with coverage of both \Hbeta\ and \Halpha, based on the presence and (relative) strength of the broad components of these emission lines \cite[e.g.,][]{Osterbrock1981}.  
Specifically, we follow the quantitative approach outlined in \cite{Winkler1992} to classify our sources into AGN sub-classes (Type 1, 1.2, 1.5, 1.8, 1.9 and 2) using the observed flux ratio of the broad \Hbeta\ to the \oiii\ emission lines, \Lbhboiii, as follows:
\begin{itemize}
\item Sy1   if  $\Lbhboiii>5.0$;
\item Sy1.2 if  $2.0<\Lbhboiii<5.0$;
\item Sy1.5 if  $1/3<\Lbhboiii<2.0$;
\item Sy1.8 if  $\Lbhboiii <1/3$;
broad component visible in \Halpha\ and \Hbeta; 
\item Sy1.9 if there is a broad component visible in \Halpha\ but not in \Hbeta.
\item Sy2 if no broad components are visible.
\end{itemize}
Throughout the rest of this work, we refer to ``Type 1.x'' AGNs simply as ``Sy1.x'' sources. This ``Sy'' nomenclature is used here for the sake of simplicity and consistency with previous work, despite the fact that most of our BASS/DR2 AGNs may not be considered as ``Seyfert galaxies'' given their high (X-ray) luminosities.

We acknowledge that this classification scheme practically depends on source distance (or redshift), as it combines aperture-limited measurements of the compact, unresolved BLR (broad Balmer lines) and of the extended, host-wide \oiii\ emission. 
Thus, for any given slit width and/or angular extraction 
aperture, and a given (intrinsic) \Lbhboiii, the {\it measured} \oiii\ flux would increase, and thus the \Lbhboiii\ ratio would decrease, with increasing source distance (redshift). 
This may systematically shift the classification of sources towards weaker broad components, or more specifically shift the classification of a source from, e.g., Sy1 to Sy1.2 or from Sy1.2 to Sy1.5. 
We stress, however, that the present study focuses on the comparison between Type 1.9 AGNs (Sy1.9s) and the combined group of Type 1, 1.2 and 1.5 AGNs (Sy1-1.5s). 
Consequently, this caveat does not affect our key results. 
Additionally, whenever we present separate results for Sy1, Sy1.2 and Sy1.5 sources, we verify that the quoted statistics of each AGN Type sub-sample (i.e., medians and/or means) are {\color{black}  not} statistically different from each other. 
More generally, the interested reader is encouraged to use the tabulated slit/aperture widths of all BASS/DR2 optical spectra, and the distances to all AGNs, to address this caveat in any future study that relies on BASS data (details are available in \citealt{Koss_DR2_catalog}).
%

A previous BASS study by \cite{Shimizu2018ApJ} investigated Sy1.9 sources in BASS/DR1 and showed that Sy1.9s with high column densities, i.e.,  $\logNH>22$ and especially galaxies with $\logNH \gtrsim 23$, have optical spectra that may be contaminated by line emission from outflowing gas. 
Such systems have broad \Halpha\ lines that are relatively narrow, and that are blueshifted with respect to the NLR emission, as well as outflow signatures in their \oiii\ profiles.
%
That study was based on ad-hoc emission line diagnostics which were motivated by spatially-resolved (IFU) data for certain exemplary systems, and further noted that with higher resolution spectroscopy, these mis-classified outflowing systems would be easy to identify.
The superior BASS/DR2 data we use here, with hundreds of new VLT/X-Shooter spectra, indeed allows us to more directly rule out the possibility that outflows \textit{dominate} the key emission line complexes considered in the present study, and to be more confident in our classification of (high-\NH) Sy1.9s in BASS/DR2.
We thus proceed with our analysis of all BASS/DR2 AGNs, including Sy1.9s, according to the respective criteria listed above. 
We defer the identification of (weak) outflow signatures in such sources to a future study.

In addition to the AGN sub-classification, we also mark the \NBZQgood\ beamed AGNs, comprised of high-$z$ systems and ``candidate'' beamed sources, with a dedicated flag (\texttt{BZQ}). 
These are blazars or flat spectrum radio quasars, where Doppler boosting may significantly amplify the non-thermal emission, including the hard X-rays. 
This classification was done based on commonly-used techniques (e.g., intense radio emission, a flat radio spectrum, dramatic variability), combined with cross-matching to {\it Fermi} data products and multi-wavelength broad-band SED fitting \cite[e.g.,][]{Oh2018_BAT_105m,Paliya2019_BASS16_blazars}.
The BASS/DR2 Data \& Catalog paper \citep{Koss_DR2_catalog} provides further information on the classification of these sources, particularly those not identified as beamed AGNs in BASS/DR1 (see \citealt{Paliya2019_BASS16_blazars} {\color{black} and \citealt{Marcotulli22_BASS_Blazars}}).
%
This identification of beamed AGNs eventually included {\it all} AGNs with (reliable) measurements of the \mgii\ and/or \civ\ broad emission lines (i.e., $z\gtrsim0.43$ and $\gtrsim3.67$ respectively).
There are also \NBZQlowz\ (candidate) beamed systems among the lower-$z$ sources, where our optical spectra cover the spectral complexes of \Hbeta\ and/or \Halpha. 
Our catalog provides all the available spectral measurements, and derived properties (including AGN Type sub-classes when possible), regardless of any evidence for beaming.
We note that continuum and line measurements for such sources should be used with caution, while line ratios may be more robust. 

\begin{figure}
	\includegraphics[width=\columnwidth]{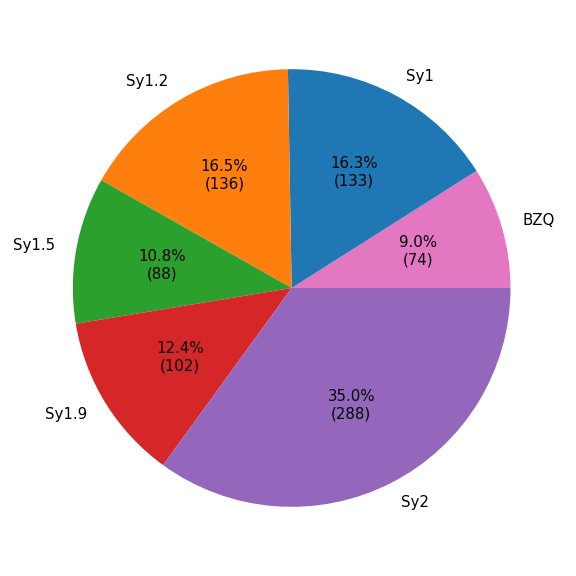}
    \caption{The distribution of BASS/DR2 AGNs in terms of various AGN Type sub-classes.
    We show the percentages and total numbers of Sy1 (blue), Sy1.2 (orange), Sy1.5 (green), Sy1.9 (red), Sy2 (purple) and BZQ (magenta) sources. 
    See text for classification details.}
    \label{fig:NSy}
\end{figure}

\begin{figure*}
	\includegraphics[width=\textwidth]{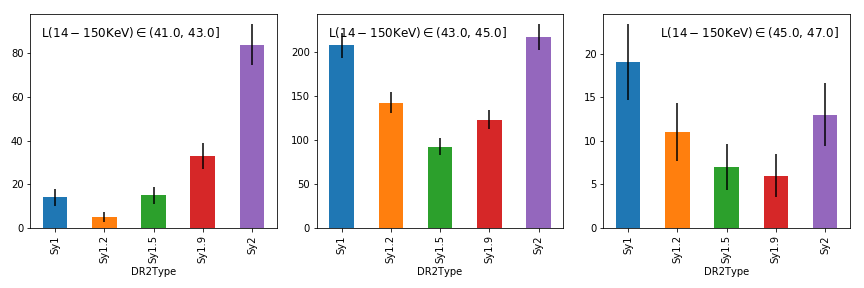}
    \caption{The distribution of BASS/DR2 AGNs among AGN Type sub-classes in bins of ultra-hard X-ray luminosity ($\log\Luhard$), as indicated in each panel. 
    Higher (lower) luminosity AGNs tend to preferentially belong to the Sy1-1.2 (Sy1.9-2) classes.}
    \label{fig:Lhist}
\end{figure*}

\begin{figure}
	\includegraphics[width=\columnwidth]{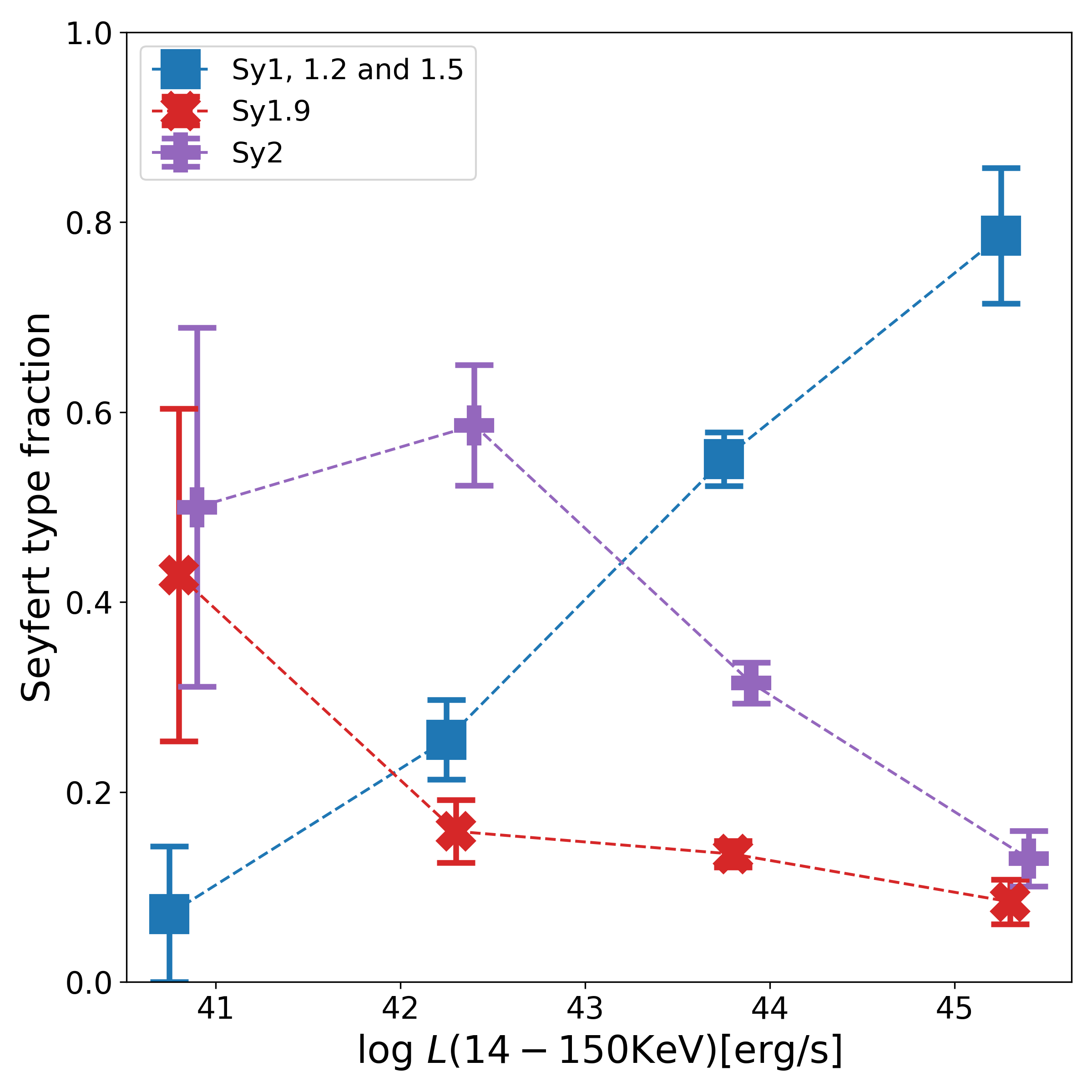}
    \caption{The fraction of BASS/DR2 AGNs of various AGN Type sub-classes as a function of ultra-hard X-ray luminosity ($\log \Luhard$).
    The various symbols and lines trace AGNs belonging to the Sy1-1.5 (blue), Sy1.9 (red), and Sy2 (purple) sub-classes.}
    \label{fig:NSyL}
\end{figure}

In Figure~\ref{fig:NSy} we show the composition of our BASS/DR2 sample of broad line AGNs in terms of the fraction and total number of sources belonging to each of the aforementioned AGN Type sub-classes (Sy1.0, 1.2, 1.5, 1.9, and BZQ). 
{\color{black} We do not identify any Sy1.8 sources among our BASS/DR2 AGNs.}
For completeness, we also include narrow-line BASS/DR2 AGNs (Sy2s), which are not part of the present catalog and are instead presented in other BASS/DR2 papers \citep{Koss_DR2_overview,Koss_DR2_catalog,Oh_DR2_NLR}.

In Figures~\ref{fig:Lhist} and \ref{fig:NSyL} we further illustrate how the fractions of AGNs in each of the AGN sub-classes varies with (ultra-hard) X-ray luminosity. 
At low luminosities ($L_{\rm 14-150\,\kev}<10^{43}\,\ergs$) the population is mostly dominated by Sy2 and Sy1.9 sources; 
however, as the X-ray luminosity increases the relative fraction of Sy1-1.5 sources increases, while the fraction of Sy1.9s and Sy2s decreases. 
This trend is in agreement with several previous studies \citep[e.g.,][]{Lawrence1991,Maiolino2007,Merloni2014,Oh2015,Ricci2017N,Ichikawa2019}, which suggest that the typical dust covering factor in AGNs decreases as the radiative power of the accretion disk increases. 
Earlier studies attributed this trend to the ``receding torus'' scenario, where the increasing (UV) disk emission sublimates dust at increasingly larger (inner) radii of the dusty torus.
A previous BASS study by \cite{Ricci2017N} conclusively showed that the underlying trend is in fact that the fraction of unobscured sources increases with increasing \lledd\ (and not $L$).
The dearth of high-\lledd, high-\NH\ AGNs is commonly interpreted as evidence for the amount of obscuring material, and indeed the degree of obscuration, to be driven by radiation pressure exerted by the central engine on the (inner) obscuring torodial structure \cite[e.g.,][and references therein]{Fabian2009,Ricci2017N,Ishibashi2018}.
Revisiting the distribution of BASS/DR2 AGNs in the $\lledd - \NH$ plane, and the relevant physical scenarios, is beyond the scope of the present study, and will be addressed in a forthcoming BASS publication (see however the results of the companion BASS/DR2 paper by \citealt{Ananna_DR2_XLF_BHMF_ERDF}).

We note that selection effects may also play a role in the trends seen in Figs.~\ref{fig:Lhist} and \ref{fig:NSyL}. 
For instance, as the accretion disk luminosity decreases, the contrast of AGN with respect to the host galaxy also decreases.
In the context of BASS optical spectroscopy, the $S/N$ required to robustly detect broad emission lines would become unrealistically large, and our analysis may thus favor the classification of low-luminosity AGNs as Sy2 and Sy1.9 sources, over the Sy1-1.5 sub-classes.




\subsection{Comparison of H$\alpha$ and H$\beta$ Line Width and $M_{\rm BH}$ Measurements}

\begin{figure*}
	\includegraphics[width=0.475\textwidth]{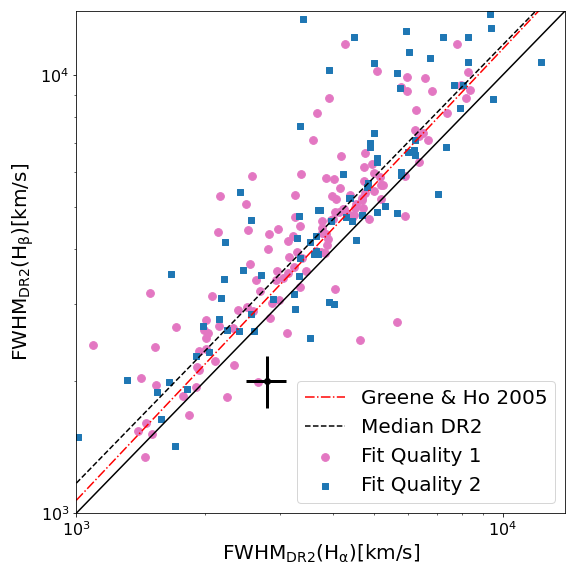}\hfill
	\includegraphics[width=0.475\textwidth]{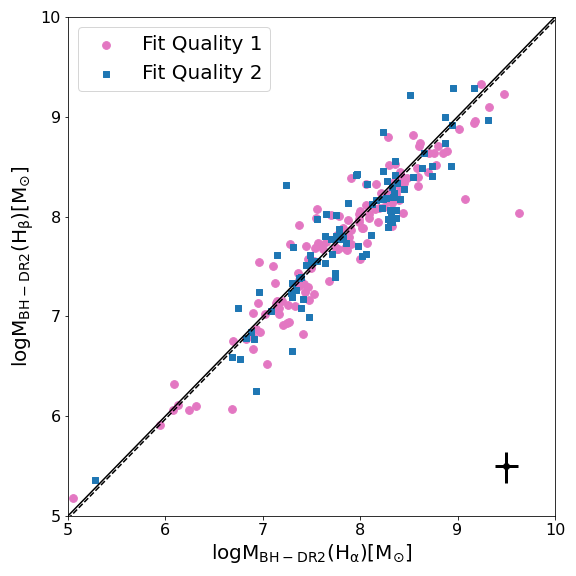}
    \caption{H$\beta$ versus H$\alpha$ FWHMs (left) and black hole masses (right) in BASS/DR2. Black solid crosses represent the median error bars associated to the \fwhm\ and \Mbh\ estimates. 
    The relation we find between \fwha\ and \fwhb\ (black dotted line) is broadly consistent with the relation derived in \citet[][red dotted line]{GreeneHo2005}, which supports our use of their \Halpha-based \Mbh\ prescription. The black solid line represents the 1:1 relation.}
    \label{fig:FWMbhDR2}
\end{figure*}

In Figure~\ref{fig:FWMbhDR2} we compare the FWHM measurements (left panel) and \Mbh\ estimates (right panel) associated with the broad \Hbeta\ and \Halpha\ emission lines. 
Our \fwha\ and \fwhb\ measurements are generally in good agreement with the relation derived by \citet[][see the red dash-dotted line]{GreeneHo2005}. 
%
A formal fit to our set of FWHM measurements, derived using the \texttt{emcee} Markov Chain Monte Carlo (MCMC) sampler \citep{emcee}, yields the best-fit relation
\begin{equation}\label{eq:fwhm_best_fit}
    \log\fwhb = (0.98\pm0.05)\log\fwha + (0.16^{+0.19}_{-0.17}) \, ,
\end{equation}
where FWHMs are given in \kms, and the quoted uncertainties represent 95\% confidence intervals. 
A fit using the BCES(Y$|$X) method \citep{Akritas1996_BCES} provides an indistinguishable best-fit relation. 
%

The right panel of Figure~\ref{fig:FWMbhDR2} shows that our \Hbeta- and \Halpha-based \Mbh\ estimates are indeed in excellent agreement, with a median offset of merely 0.03 dex, and a scatter of 0.25 dex. 
This scatter is mostly driven by the scatter between \fwha\ and \fwhb, which is found to be 0.11 dex, which is expected to yield a scatter in \Mbh\ of 0.23 dex.
This agreement between \Hbeta- and \Halpha-based \Mbh\ estimates further justifies our choice to use the not-strictly-virial \Halpha-based \Mbh\ prescription, derived by \cite{GreeneHo2005}.
%
We stress again that the excellent agreement between the two kinds of mass estimates is reached only after considering a virial factor of $f_{\fwhm}=1$ for both the \Halpha\ and \Hbeta\ mass prescriptions. 
This is justified as the two lines are expected to be formed in a similar circumnuclear region and consequently should have the same geometrical factor.

\subsection{Black hole Mass and Eddington Ratio Distributions}

\begin{figure*}
	\includegraphics[width=0.50\textwidth]{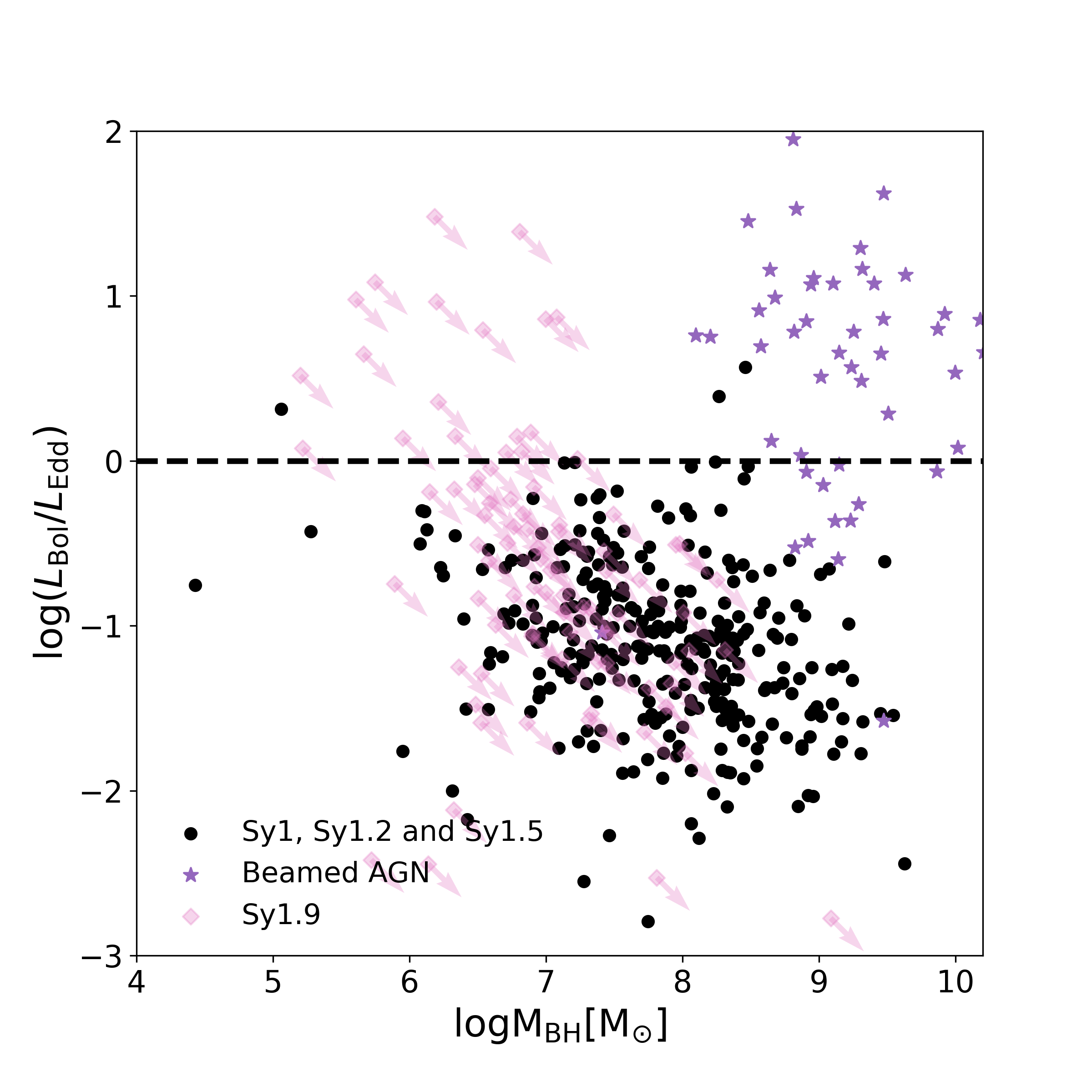}\hfill
	\includegraphics[width=0.50\textwidth]{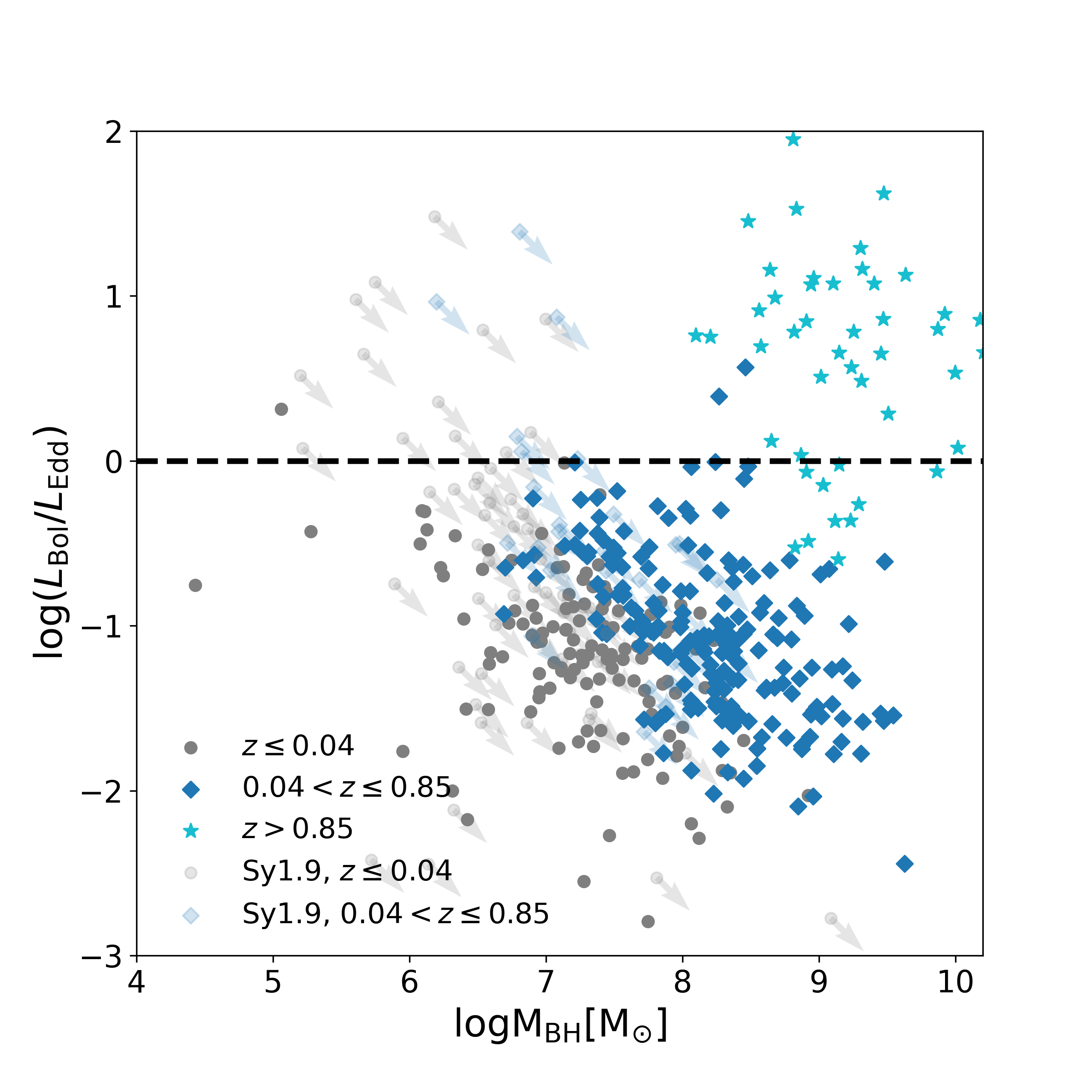}\\
	\includegraphics[width=0.33\textwidth]{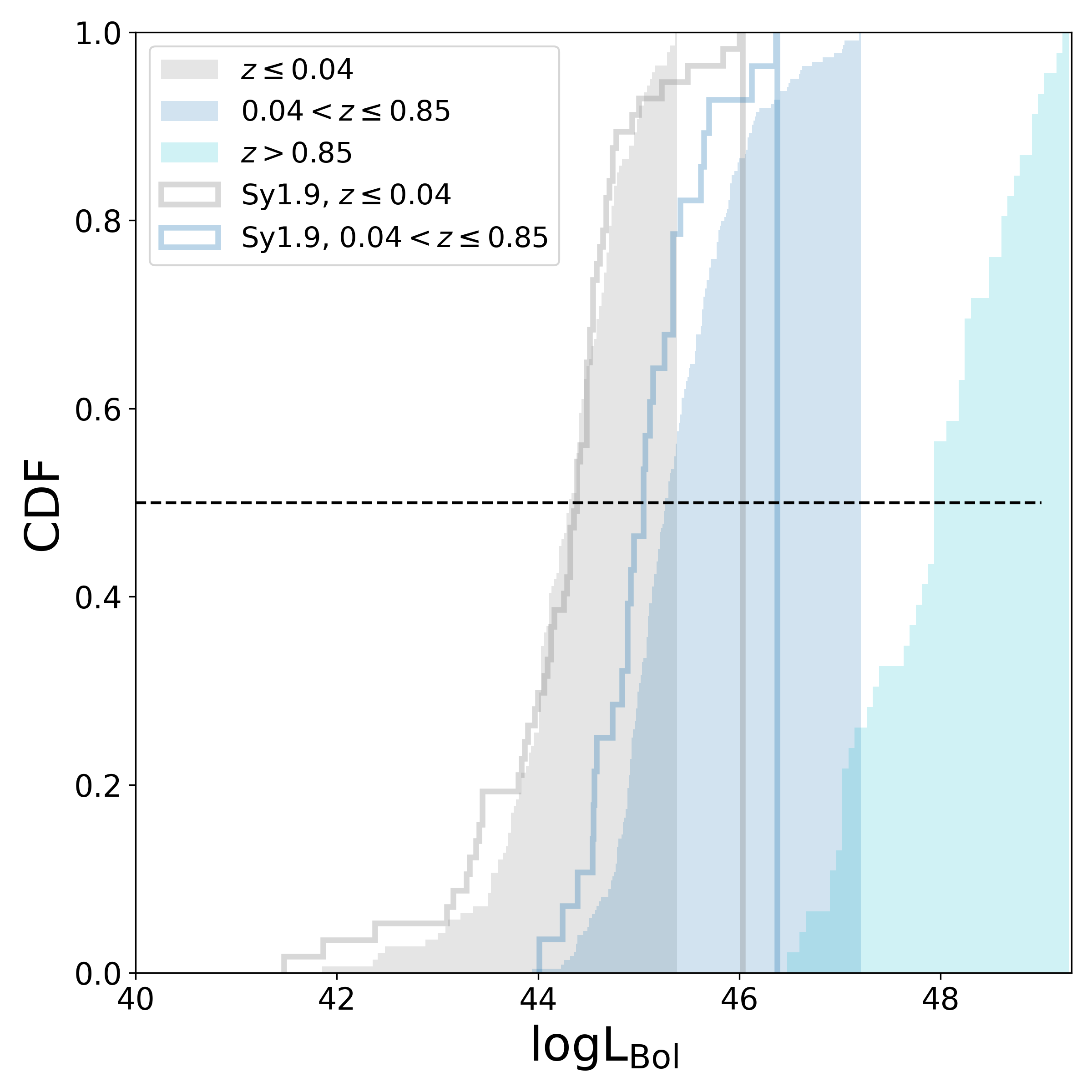}
	\includegraphics[width=0.33\textwidth]{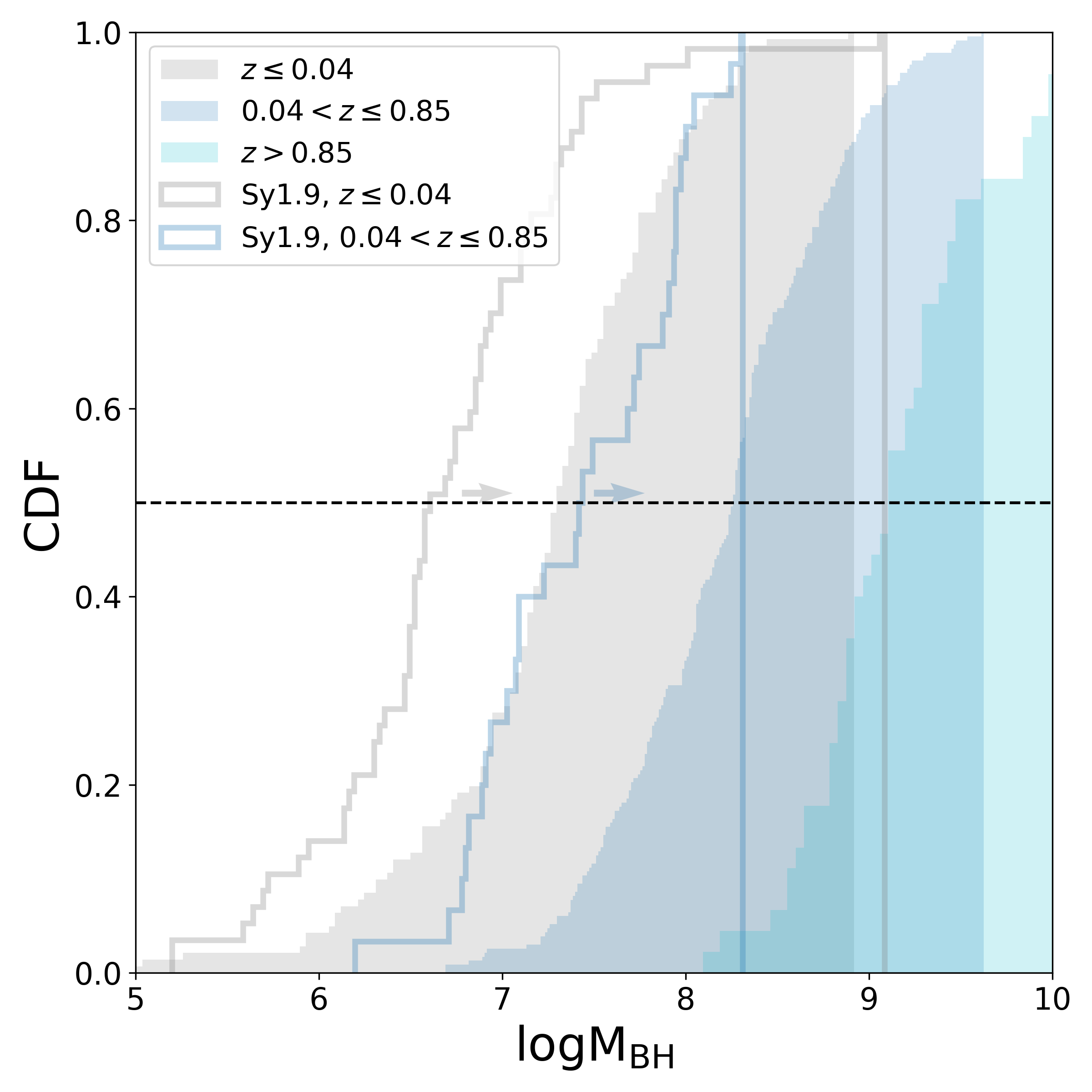}
	\includegraphics[width=0.33\textwidth]{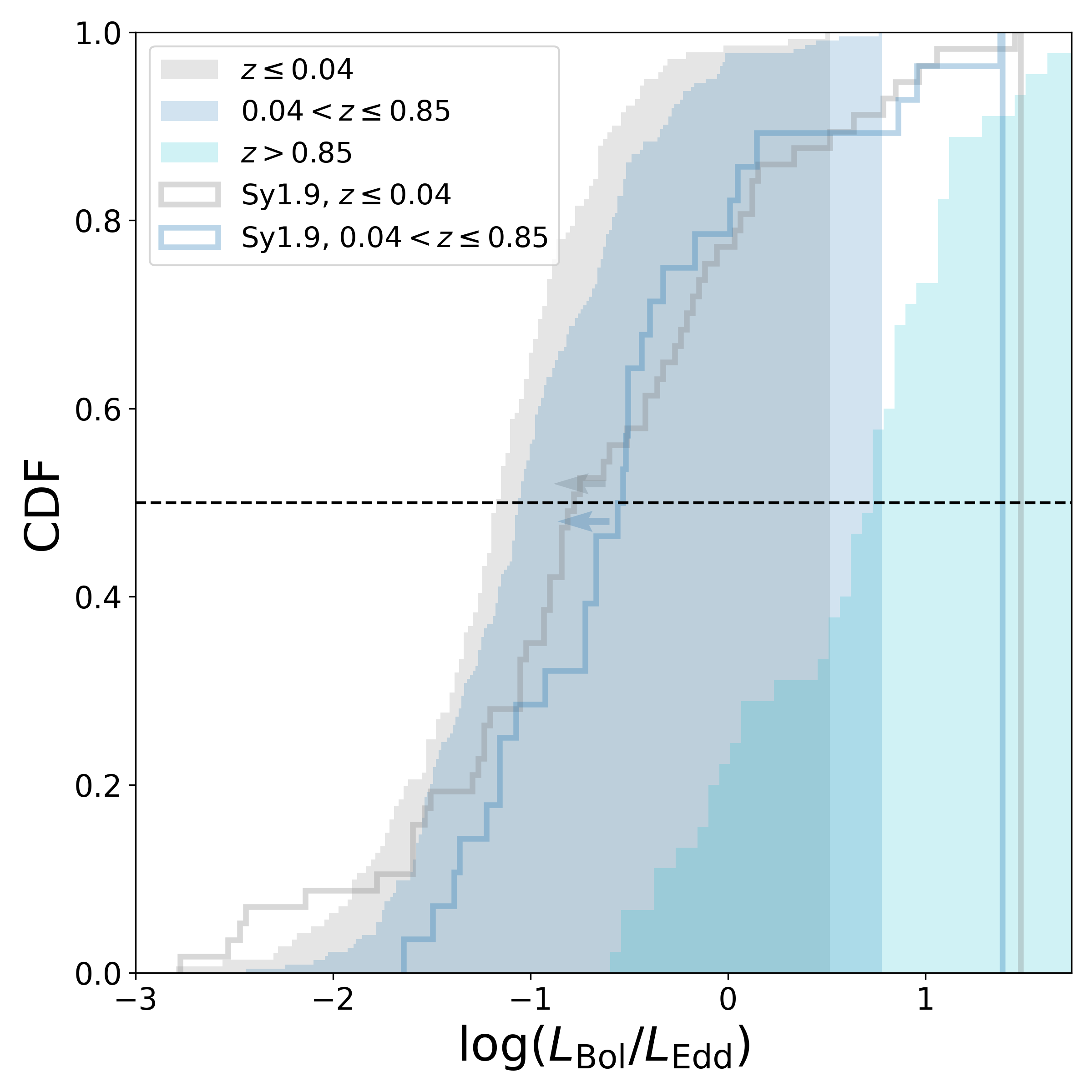}
    \caption{The luminosities, BH masses, and Eddington ratios of BASS/DR2 broad line AGNs.
    Top: The distribution of BASS/DR2 AGNs in the $\LLedd-\Mbh$ plane. 
    Our AGNs are further split by AGN Type ({\it left} panel) and by  redshift range ({\it right} panel). 
    Bottom: cumulative distributions of \Lbol\ (left), \Mbh\ (center) and  \LLedd\ (right).
    The \Mbh\ of Sy1.9 sources are likely underestimated, and their \LLedd\ are thus over-estimated (see text for discussion). 
    The small arrows in the top panels indicate the \textit{direction} of a simple, uniform correction towards the true \Mbh\ and \LLedd\ of Sy1.9s 
    (see \S\ref{sec:correcting_mbh} for more detailed discussion of mass corrections).
    The \Lbol\ and \lledd\ of beamed AGNs may be significantly over-estimated due to jet emission and/or relativistic boosting.
    }
    \label{fig:MbhLedd}
\end{figure*}

In Figure~\ref{fig:MbhLedd} we show the distribution of broad-line BASS/DR2 AGNs in the $\Mbh-\LLedd$ plane, with sources further divided either by AGN sub-class (top-left) or by redshift regime (top-right). 
We also show the cumulative distributions of \Lbol\ (bottom-left), \Mbh\ (bottom-center) and \LLedd\ (bottom-right).  
In this analysis we include all AGNs for which reliable estimates of \Mbh\ (and thus of \lledd) are derived from broad emission lines, that is Type 1-1.9 AGNs, and beamed sources (BZQ). 

Figure~\ref{fig:MbhLedd} clearly demonstrates the wide range in both \Mbh\ and \lledd\ that is sampled by BASS/DR2 AGNs.
First, unbeamed AGNs where both broad \Hbeta\ and \Halpha\ lines are robustly detected (Sy1-1.5s hereafter) cover $6 \lesssim  \log(\Mbh/\Msun) \lesssim 10$ and $-3 \lesssim \log \LLedd \lesssim 1$. 
This is comparable to the distribution reported in BASS/DR1 (see Fig.~16 in \citealt{Koss2017}).\footnote{For a few sources, BASS/DR1 measurements indicated exceptionally low \lledd\ (i.e., $\log\LLedd\lesssim -4$). 
The higher quality DR2 data and measurements have corrected these outliers.} 
Compared with other wide-field AGN surveys in the local Universe where SE mass estimates were used \cite[e.g.,][]{GreeneHo2007_BHMF,VestergaardOsmer2009,Schulze2010_HES}, BASS naturally includes the most luminous, rarest AGNs accessible, powered by the most massive and/or highest-\lledd\ BHs. 

Second, beamed AGNs in BASS/DR2, which preferentially reside at higher redshifts, appear to have higher \Lbol, \Mbh, and \LLedd, covering $8 \lesssim \log(\Mbh/\Msun) \lesssim  10$ and, importantly, $-1 \lesssim \log\LLedd \lesssim 2$ and $46 \lesssim \log(\Lbol/\ergs) \lesssim 49$. 
Although this may be partially attributed to the higher redshifts of the beamed sources (given the flux-limited nature of the Swift/BAT all-sky survey),
we stress again that in such systems \Luhard\ is most likely over-estimated, as their X-ray emission is affected by jets, and is boosted by relativistic effects. 
This propagates to an over-estimated \Lbol\ and thus \lledd.

Finally, a large fraction of (unbeamed) Sy1.9 sources show lower masses, $ \log(\Mbh/\Msun) \lesssim  7$, and higher Eddington ratios, $\log\LLedd \gtrsim -1$, compared to Sy1-1.5 sources, while covering a similar luminosity range.
This difference, however, likely highlights a bias among this class.
As we show in the next Section, 
%
the \Halpha-based masses of Sy1.9s are  underestimated, and their \lledd\ are thus overestimated, likely due to the suppression of broad \Halpha\ emission, which we argue is linked to (partial) obscuration of the BLR by dust.
The small arrows added to each Sy1.9 in the top panels of Fig.~\ref{fig:MbhLedd} demonstrate how a simple (uniform) correction for this bias would be reflected in the $\Mbh-\lledd$ plane, {\color{black} with increasing \Mbh\ and  accordingly decreasing $\lledd\propto L/\Mbh$.}
%
In Section~\ref{sec:correcting_mbh} we provide a set of simple \Mbh\ corrections for Sy1.9 sources.


\section{Reduced broad Balmer line emission and obscuration in BASS/DR2 AGNs}
\label{sec:bHa_and_obsc}

In what follows, we examine in detail the properties of the broad Balmer emission lines in our BASS/DR2 sample of broad-line AGNs. 
We particularly focus on those sources where only \Halpha, but not \Hbeta\ broad line emission is identified - i.e., Type 1.9 AGNs (Sy1.9s), and use the rich BASS data-set to better understand these  systems.\\

\subsection{Preliminaries: linking broad Balmer lines with X-ray measurements}
\label{subsec:balmer_to_xray}

As a first step, we look into the most basic links between the broad Balmer line measurements, and the key properties deduced from the X-ray analysis of the BASS AGNs \citep{Ricci2017_Xray_cat}. 
Namely, we examine the observed links between 
(i) the broad \Halpha\ and ultra-hard X-rays luminosities (\Lbha, \Luhard), and 
(ii) the broad Balmer decrement ($\Lbha/\Lbhb$, or simply $\Halpha/\Hbeta$ in what follows) and the line-of-sight column densities (\NH).

\begin{figure}
	\includegraphics[trim={0 0 0 1cm}, clip,width=0.475\textwidth]{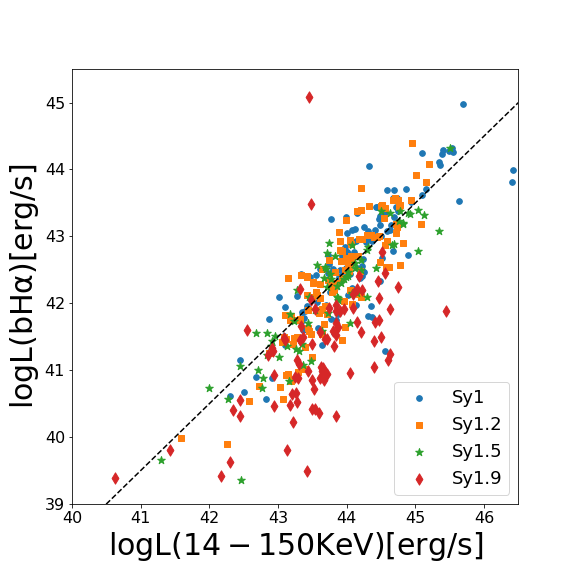}
    \caption{Broad \Halpha\ emission line luminosity (\Lbha) vs. ultra-hard X-ray luminosity (\Luhard) for BASS/DR2 AGNs. 
    Symbol colors mark AGNs of different (emission line) sub-classes, as indicated in the legend.
    The dashed line marks a scaling of  $\log(L[{\rm b}\Halpha]/\Luhard) = -1.5$, which describes the general trend seen in most of our bona-fide broad line AGNs. 
    The broad \Halpha\ emission in Type 1.9 AGNs, however, appears suppressed relative to \Luhard.
    }
    \label{fig:LbHa_Lbat}
\end{figure}

In Figure~\ref{fig:LbHa_Lbat} we show \Lbha\ vs. \Luhard\ for the \Nhagoodnb\ non-beamed AGNs with reliable broad \Halpha\ measurements in BASS/DR2, further highlighting the different AGN sub-classes. 
Unsurprisingly, the two independently-measured emission probes show a roughly-uniform scaling for the vast majority of AGNs.
However, for Sy1.9 sources, \Lbha\ deviates downwards from the general scaling, by roughly 0.8 dex.  
Thus, broad \Halpha\ emission seems to be suppressed in Sy1.9 sources relative to all other AGNs with detectable broad \Halpha\ emission (Sy1-1.5 sources) at any given \Luhard. 
We note that this apparent suppression is not limited to particularly high- or low-luminosity sources (in terms of \Lbha\ and/or \Luhard).
In the following section we further investigate this suppression and how it may be linked to other basic AGN observables and properties.

When considering the measured decrements between the broad \Halpha\ and \Hbeta\ emission lines, 
we first note that close to 30\% of the AGNs in our sample are Type 1.9 AGNs where the broad \Hbeta\ line cannot be detected, and thus formally have $\Lbhb = 0$ (and infinite \Halpha/\Hbeta).
Deducing a robust upper limit on \Lbhb\ (and thus a robust lower limit on \Halpha/\Hbeta) for such sources is challenging, and requires a full spectral decomposition of the (stellar) host emission.
In addition, about 40\% of those sources with detectable broad \Hbeta\ show $\Halpha/\Hbeta > 3$, which--if taken at face value--may indicate significant attenuation over the \Hbeta\ wavelength regime, perhaps by dusty BLR gas \cite[see, e.g.,][and references therein]{Dong2008,Baron2016}.

\begin{figure*}
    \includegraphics[trim={0 0 0 0}, clip,width=0.350\textwidth]{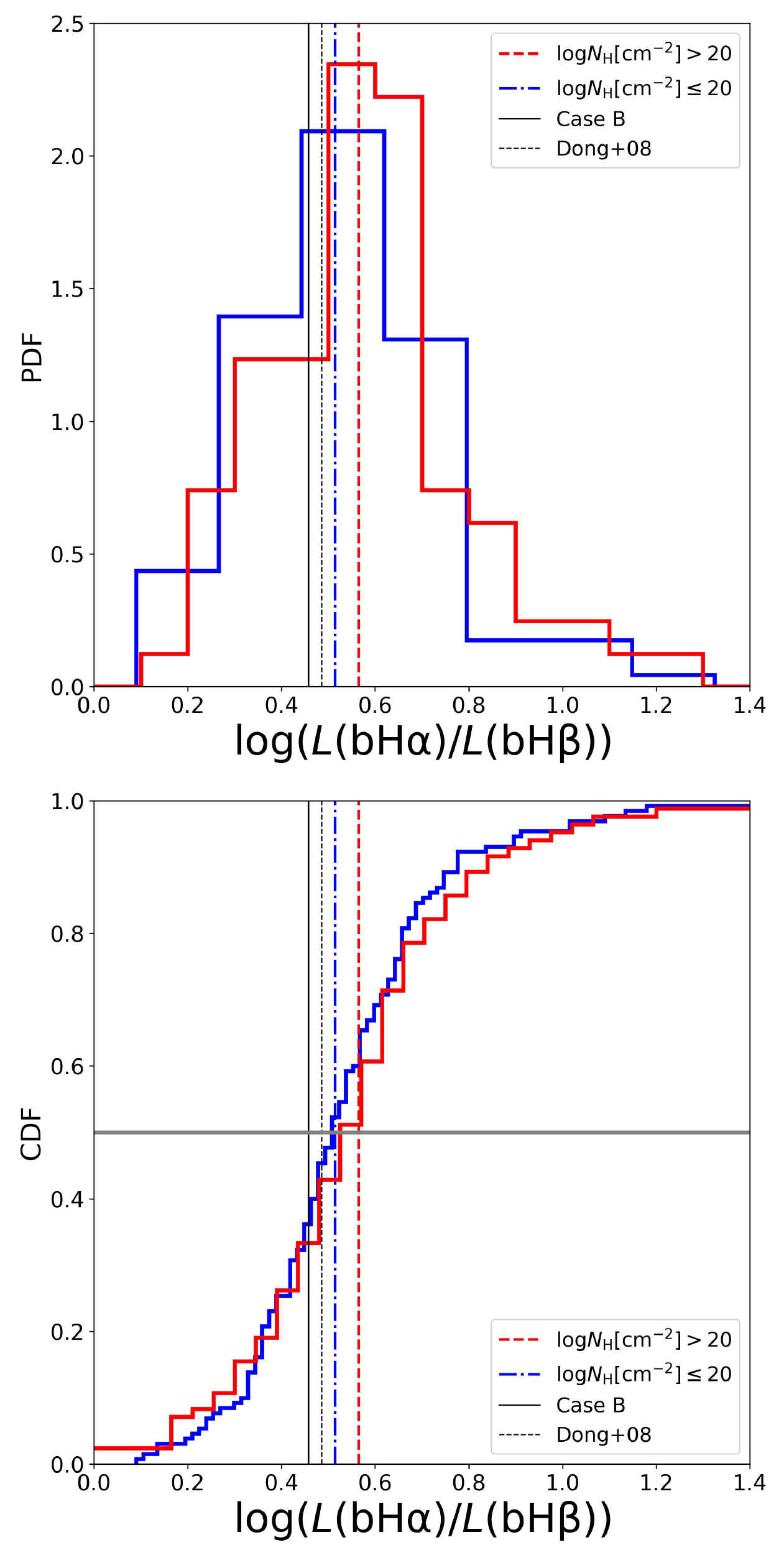} \hfill  
    \includegraphics[trim={0 0 1.75cm 1cm}, clip,width=0.595\textwidth]{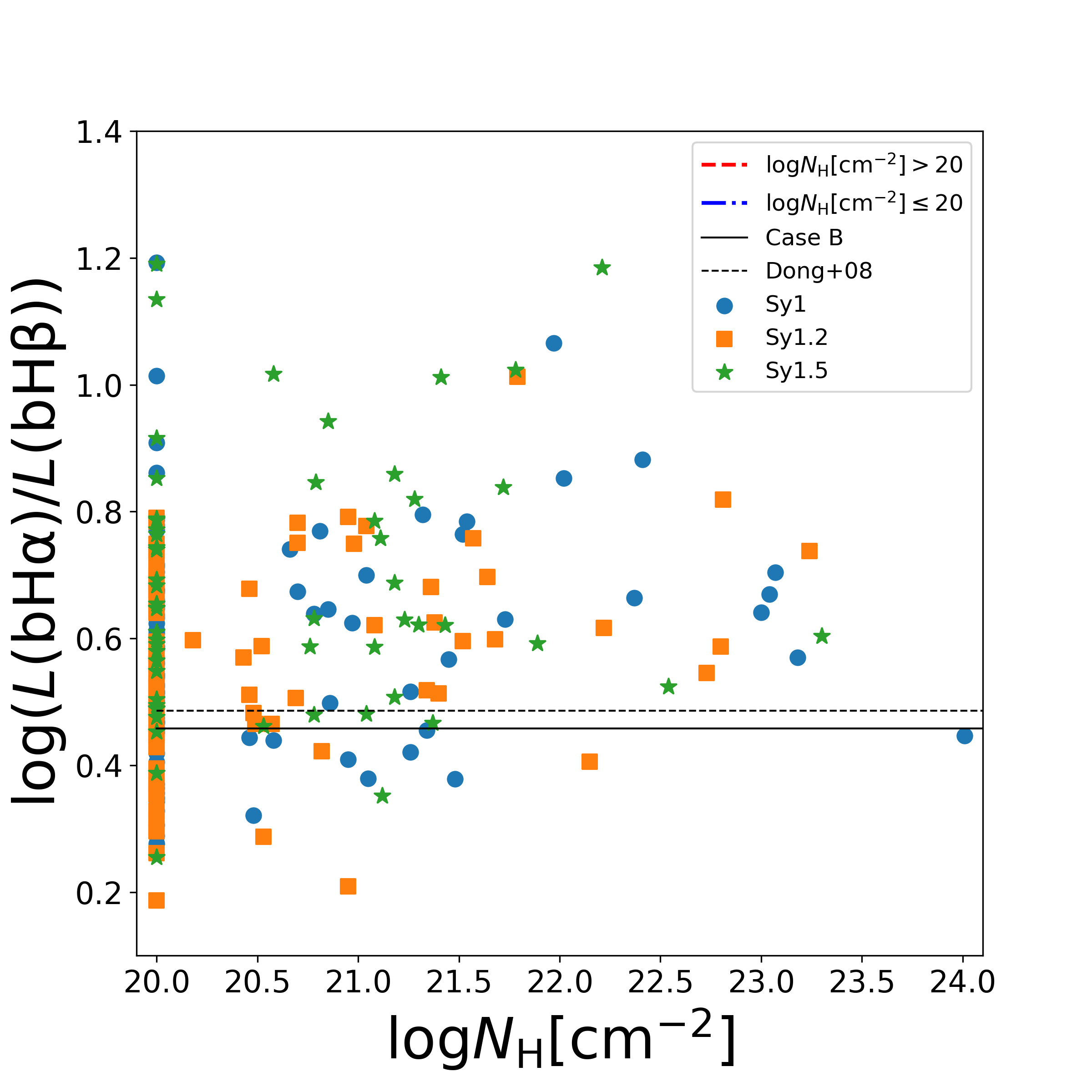} 
    \caption{
    The Balmer decrement ($\Lbha/\Lbhb$) for broad-line BASS/DR2 AGNs.
    \textit{Left:} 
    The distribution of $\log(\Lbha/\Lbhb)$ among unbeamed, broad-line BASS/DR2 AGNs, excluding Type 1.9 sources and further split to sources with $\logNH>20$ and $\logNH
    \leq20$ (top - probability distribution function; bottom - cumulative distribution function).
    \textit{Right:} 
    $\log(\Lbha/\Lbhb)$ vs. $\logNH$ for AGNs with $\logNH>20$, again excluding Type 1.9 sources. 
    Note the mild (yet statistically significant) positive correlation between the sources
    In all panels we mark the median values for the $\logNH\leq20$ (completely unobscured) and $\logNH>20$ sub-samples, as well as two reference values from the literature (see legends and text for details).
    } 
    \label{fig:balmer_dec}
\end{figure*}

In Figure~\ref{fig:balmer_dec} we show the available Balmer decrement measurements for our sample, and how it varies with \NH.
The left-hand-side panels show the distribution of $\Halpha/\Hbeta$ for all the BASS/DR2 AGNs for which these quantities are robustly measured, i.e.\ omitting Type 1.9 AGNs. 
We further split our sample to AGNs with $\logNH>20$ and ``completely unobscured'' AGNs, which formally have $\logNH=20$ in the \cite{Ricci2017_Xray_cat} catalog.
Note that this latter sub-sample includes sources with upper limits on \NH, so in practice it covers $\logNH\leq20$ (see \citealt{Ricci2017_Xray_cat} for details).
The right panel of Fig.~\ref{fig:balmer_dec} shows $\Halpha/\Hbeta$ vs. \NH\ for broad-line BASS/DR2 AGNs with $\logNH>20$, again excluding Type 1.9 AGNs.
All panels of Fig.~\ref{fig:balmer_dec} also mark the canonical value
of $\Halpha/\Hbeta=2.87$, derived for Case B recombination in H\,\textsc{ii} regions, as well as $\Halpha/\Hbeta=3.1$ which is  more relevant for AGNs. 
The latter is commonly adopted for the low-density NLR in AGN, and is also consistent with what is found for the broad Balmer lines (emitted from the high-density BLR) in large samples of optically-selected quasars \cite[see][and references therein]{Dong2008}.

The median Balmer decrements for our sub-samples of $\logNH\leq20$ and $\logNH>20$ AGNs are {\color{black} $\log\left(\Halpha/\Hbeta\right)=0.52$ and $0.58$ (respectively)}, and the scatter measures (standard deviations) in $\Halpha/\Hbeta$ for these two sub-samples are 0.24 and 0.41 dex, respectively.
The median Balmer decrements in our BASS/DR2 AGNs are in agreement with what is found for optically-selected SDSS quasars \cite[e.g.,][see reference lines in Fig.~\ref{fig:balmer_dec}]{Dong2008}. 
The scatter we find is higher than what is found for SDSS quasars \cite[i.e., $\sim$0.05 dex;][]{Dong2008}. 
This is expected given that SDSS quasars are pre-selected based on their blue continuum colors, tracing unobscured accretion disk emission \cite{Richards2002_SDSS_QSO_select}, while our Swift/BAT-selected broad-line AGNs indeed cover a wider range of (circumnuclear) obscuration.

Among the $\logNH>20$ BASS/DR2 broad-line AGNs, there is a (mild) trend of increasing Balmer decrement with increasing column density, with a significant amount of scatter (right panel of Fig.~\ref{fig:balmer_dec}). 
This trend seems to involve objects of all sub-classes (i.e., Sy1s, 1.2s and 1.5s).
A formal Spearman correlation test confirms that the correlation between $\Lbha/\Lbhb$ and \NH, for all Sy1-1.8 AGNs with $\logNH>20$, is statistically significant but rather weak ($r_{\rm s}=0.28$, $P_{\rm s}=0.01$). 
Given the large scatter and the limited strength of the correlation, we refrain from fitting a formal relation that links $\Lbha/\Lbhb$ and \NH.

The Balmer decrements we measure are far lower than what is expected from the corresponding column densities. 
For reference, for $\logNH=22$ one would expect a Balmer decrement of roughly $\Halpha/\Hbeta \simeq 17$, assuming a standard Galactic absorption scaling \cite[i.e., gas-to-dust ratio;][]{Bohlin1978} and a \cite{Cardelli1989} extinction law.
This is barely consistent with the highest $\Halpha/\Hbeta$ we measure for AGNs with comparable \NH\ (Fig.~\ref{fig:balmer_dec}, right). For higher \NH, the discrepancy grows substantially and quickly (expected $\Halpha/\Hbeta>500$ by $\logNH=22.5$).
This is consistent with several previous works, which found that the $E(B-V)/\NH$ ratio in AGNs is lower than Galactic by a factor ranging from $\sim$3 and up to $\sim$100 \cite[e.g.][]{Maiolino2001a,Maiolino2001b}, perhaps indicating that the material obscuring the central X-ray source is in part dust-free \cite[e.g.,][]{Burtscher2016}.
Alternatively, {\color{black} the X-ray obscuring material} may be arranged in a compact configuration, which does not (generally) affect the BLR radiation.
The recent study by \cite{JaffarianGaskell2020} further discusses these and other scenarios for the differences between the levels of extinction deduced from Balmer line ratios and from (X-ray) hydrogen column densities.
We'll come back to this issue when discussing intermediate Type AGNs, in Section~\ref{sec:LhaNh}.

\subsection{H$\alpha$ Line Attenuation in Partially-obscured AGNs}
\label{sec:LhaNh}

\begin{figure*}
	\includegraphics[width=0.48\textwidth]{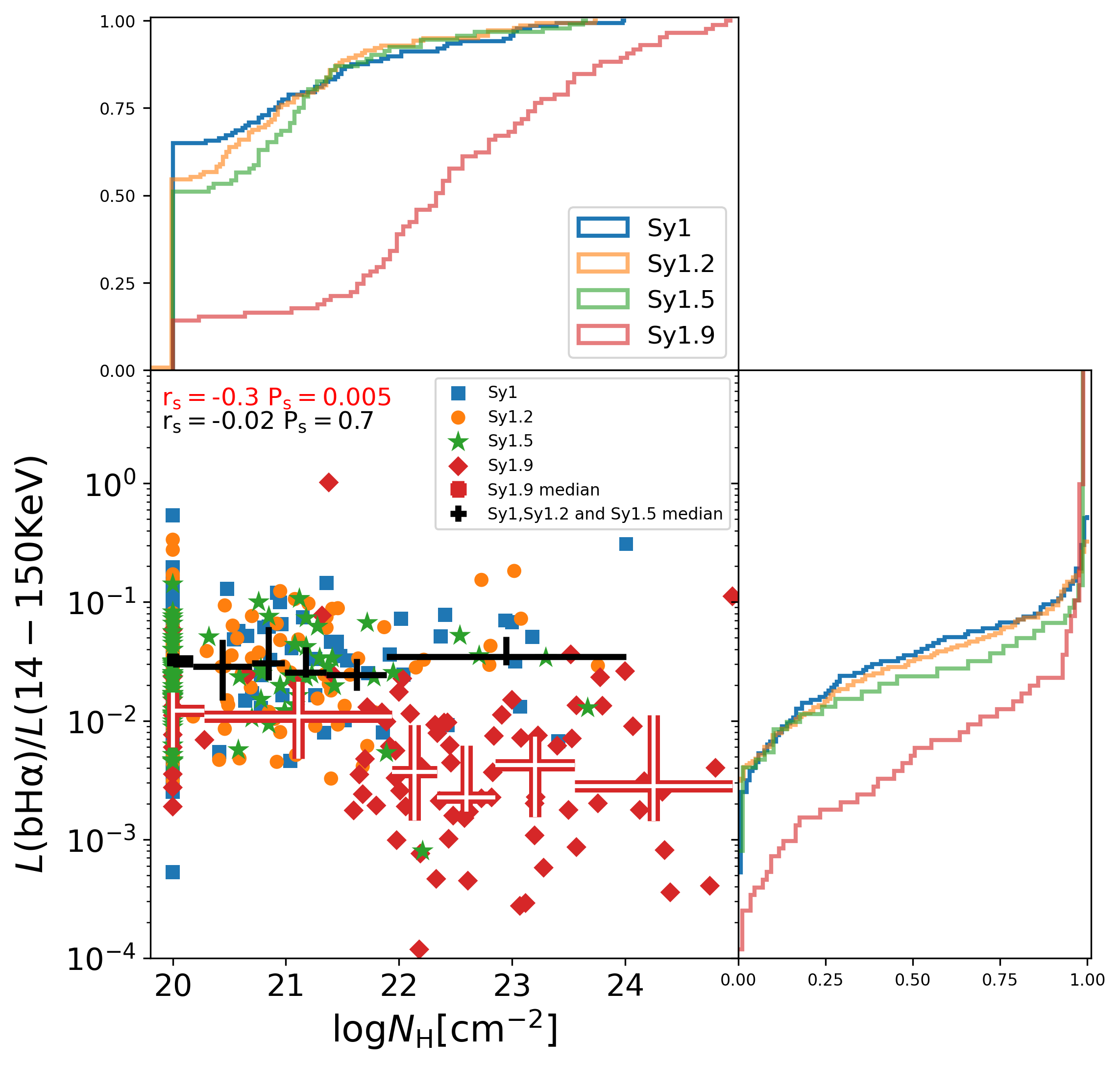}\hfill
    \includegraphics[width=0.48\textwidth]{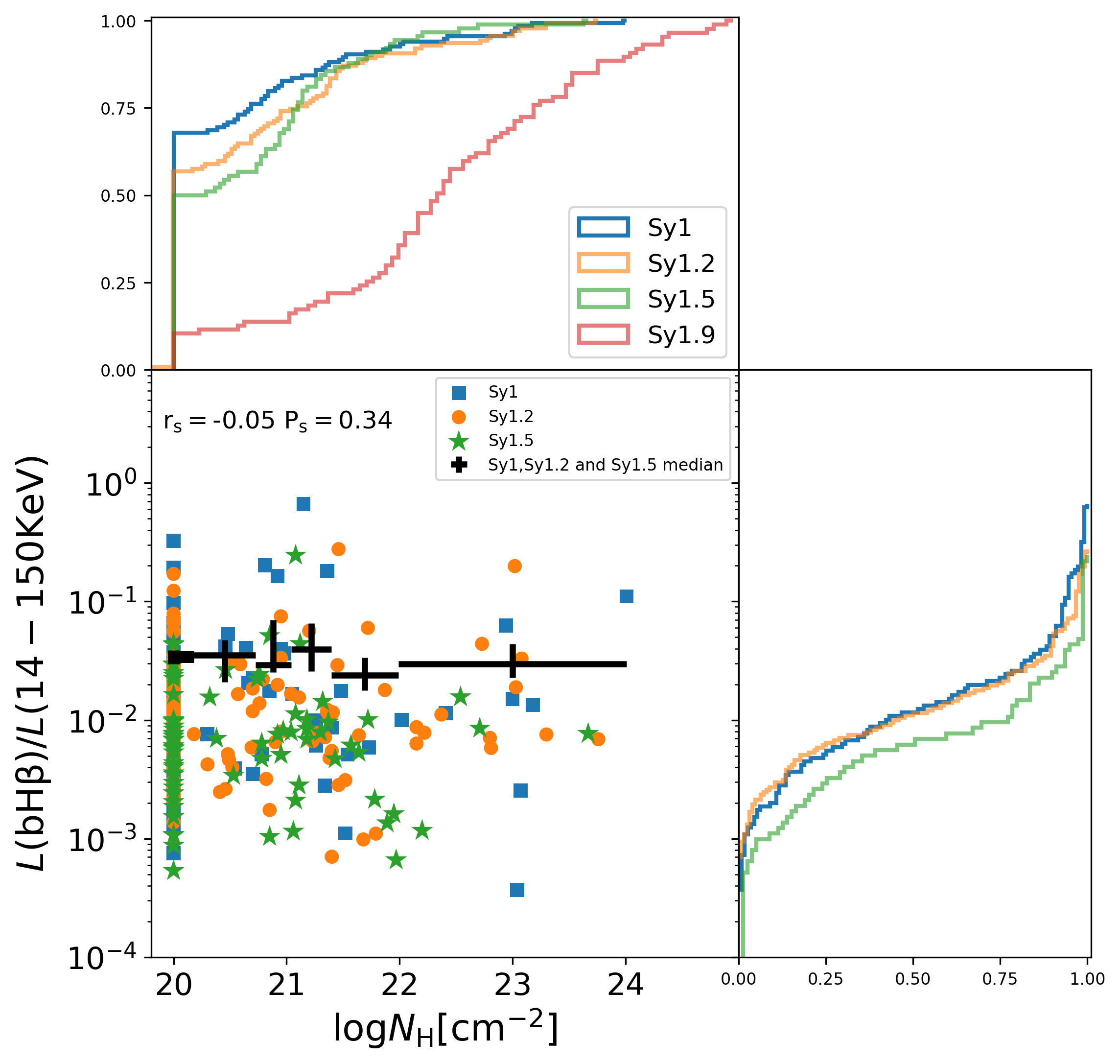}
    \caption{Broad Balmer line strength relative to ultra-hard X-rays for \Lbhax\ (left) and \Lbhbx\ (right), 
    vs. line-of-sight column density, \NH, and the projected distributions of these quantities.
    Broad-line BASS/DR2 AGNs of various AGN sub-types are marked with different colors and symbols (see legend). 
    Large crosses represent the median values of \Lbhax\ and \Lbhbx\ within equally-spaced bins of $\log \NH$, and the corresponding 90\% confidence intervals as determined from bootstrapping. 
    Black thin crosses represent Type 1, 1.2 and 1.5 AGNs, while red thick crosses represent Type 1.9 sources.
    Type 1.9 AGNs typically have \Lbhax\ that are significantly lower than those of Type 1-1.5 AGNs, but only in the $\logNH \gtrsim 22$ regime.
    }
    \label{fig:LX_NH}
\end{figure*}

In Figure~\ref{fig:LX_NH} we show \Lbhax\ (left) and \Lbhbx\ (right) vs.\ \logNH\ for our BASS/DR2 broad-line AGNs, with the respective distributions of these quantities (ancillary panels in each plot), and distinguishing the different AGN sub-classes.
The first thing to notice in Fig.~\ref{fig:LX_NH} is that Sy1.9 sources tend to have higher \NH\ than Sy1-1.5 sources.
This difference is statistically significant, as confirmed by both Kolmogorov-Smirnov (KS) and Wilcoxon rank sum (WRS) tests. 
The $P$-values associated with the null hypotheses, i.e., the probability of having the \logNH\ distribution in Sy1.9s to be drawn from the same \logNHo\ distribution as of Sy1-1.5s, are $\ll 10^{-10}$ (for both tests).

{\color{black} Second, the left panel of Fig.~\ref{fig:LX_NH} shows that the median \Lbhax\ ratio in Sy1-1.5 AGNs  stays roughly constant across the full range in \logNHo\ covered by our sample.}
{\color{black} The same behavior is observed in the right panel where the median \Lbhbx\ is also roughly constant within the full \logNHo\ range.}
For Sy1.9 AGNs, however, the behavior is more complex, and can be split into two different regimes, with sources having column densities either above or below {\color{black} $\logNH = 22$}.
In the {\color{black} $\logNH < 22$} regime, the \Lbhax\ ratios of Sy1.9 sources are broadly consistent with those of Sy1-1.5 sources, with the former being only slightly lower than the latter (red vs.\ black crosses, respectively, in the left panel of Fig. \ref{fig:LX_NH}).
Specifically, for {\color{black} $20 <  \logNH < 22$} AGNs, the median $\Lbhax$ for Sy1.9s is $(11^{+6}_{-4}) \times 10^{-3}$, compared to $ (34^{+4}_{-3}) \times 10^{-3}$ for Sy1-1.5s. 
%
%
In the {\color{black} $\logNH > 22$} regime, the \Lbhax\ ratios of Sy1.9s are significantly lower than those of Sy1-1.5s. 
Specifically, the corresponding median values for sources with {\color{black} $22 < \logNH < 24$}  are $(4_{-1}^{+2}) \times 10^{-3}$ and $ (29_{-7}^{+8}) \times 10^{-3}$ for Sy1.9s and Sy1-1.5s, respectively. 
The \Lbhax\ ratios of Sy1.9 sources with {\color{black} $\logNH > 22$} are thus lower by a factor of $\sim$8.5 times than what is found for the Sy1-1.5 AGN population. 
%
%
This difference is statistically significant, as confirmed by the appropriate KS and WRS tests ($P<10^{-6}$ for both tests).
The more general trend of decreasing \Lbhax\ with increasing \NH\ in Sy1.9 AGNs is only marginally significant, with $P\simeq0.03$ and $P\simeq0.13$ for the Spearman and Pearson correlation tests, respectively.
%

We conclude that for AGNs with relatively weak broad Balmer line emission, that is Type 1.9 AGNs, the (relative) strength of the broad \Halpha\ emission line at fixed ultra-hard X-ray luminosity is linked to the presence of large gas columns along the line of sight, independently determined from X-ray spectral modeling. 
This may suggest that in Type 1.9 AGNs, but not in Type 1-1.5s, the broad \Halpha\ emission is partially absorbed by the same gas that also accounts for the large neutral gas columns.

The association of weak broad \Halpha\ emission with dust obscuration may be challenged by the typical column densities of order $\logNH\simeq23$ in our Sy1.9s: for a standard (Galactic) dust-to-gas ratio \citep{Bohlin1978}, the corresponding optical extinction ($A[\Halpha] \sim 30$ mag) would be expected to completely suppress the optical AGN broad line emission. 
The fact that our Sy1.9s \textit{do} show broad \Halpha\ emission therefore requires either (1) that the Balmer emission is only \textit{partially} obscured, or (2) that the {\color{black} dust-to-gas} ratio of the obscurer is significantly lower than ISM values. 
Partial obscuration of the broad \Halpha\ line could also occur if the line-of-sight to the BLR ``grazes'' the obscuring torus, which completely obscures the line-of-sight to the (X-ray emitting) central engine \cite[see discussion in, e.g.,][]{Goodrich1995,Trippe2010}.

A drastically different interpretation is that the \textit{narrow} \Halpha\ emission in our Sy1.9s is intrinsically strong compared to the broad \Halpha\ emission, as is common in low luminosity AGNs \citep{SternLaor2012b}. Strong narrow line emission may be due to galaxy-scale gas covering a large fraction of sight-lines to the AGN. 
A large abundance of gas in the galaxy may also enhance the typical hydrogen columns along the line of sight to the X-ray source as seen in our Sy1.9s \cite[see also, e.g.,][]{MaiolinoRieke1995,Koss2020_BASS_CO}. 
This latter scenario, however, stands in contrast to some evidence for the high-\NH\ material in (BASS) AGNs to be confined to the nuclear region \cite[e.g.,][]{Ricci2017N}, and in contrast to constraints on galaxy-wide contributions to \NH\ (e.g.,  $\logNH\lesssim22.5$ by \citealt{BuchnerBauer2017}; see also \citealt{RamosAlmeidaRicci2017_rev} for a review).

Since the BASS/DR2 data do not have the spatial information required to thoroughly test this alternative scenario, we next turn our attention to the kinematic information available for our BASS/DR2 AGNs, and particularly for the Sy1.9s, to gain further insight regarding the interplay between (X-ray) obscuration and (suppressed) broad Balmer emission, and the nature of the gas structures at play.

\subsection{Attenuation of the Highest-velocity H$\alpha$ Emission Region}
\label{sec:fwhaNh}

After establishing a link between the detailed attenuation of broad Balmer line emission and X-ray determined line-of-sight column densities, we now use our BASS/DR2 AGN sample to better understand the nature of the relevant obscuring material.

\begin{figure}
	\includegraphics[width=0.475\textwidth]{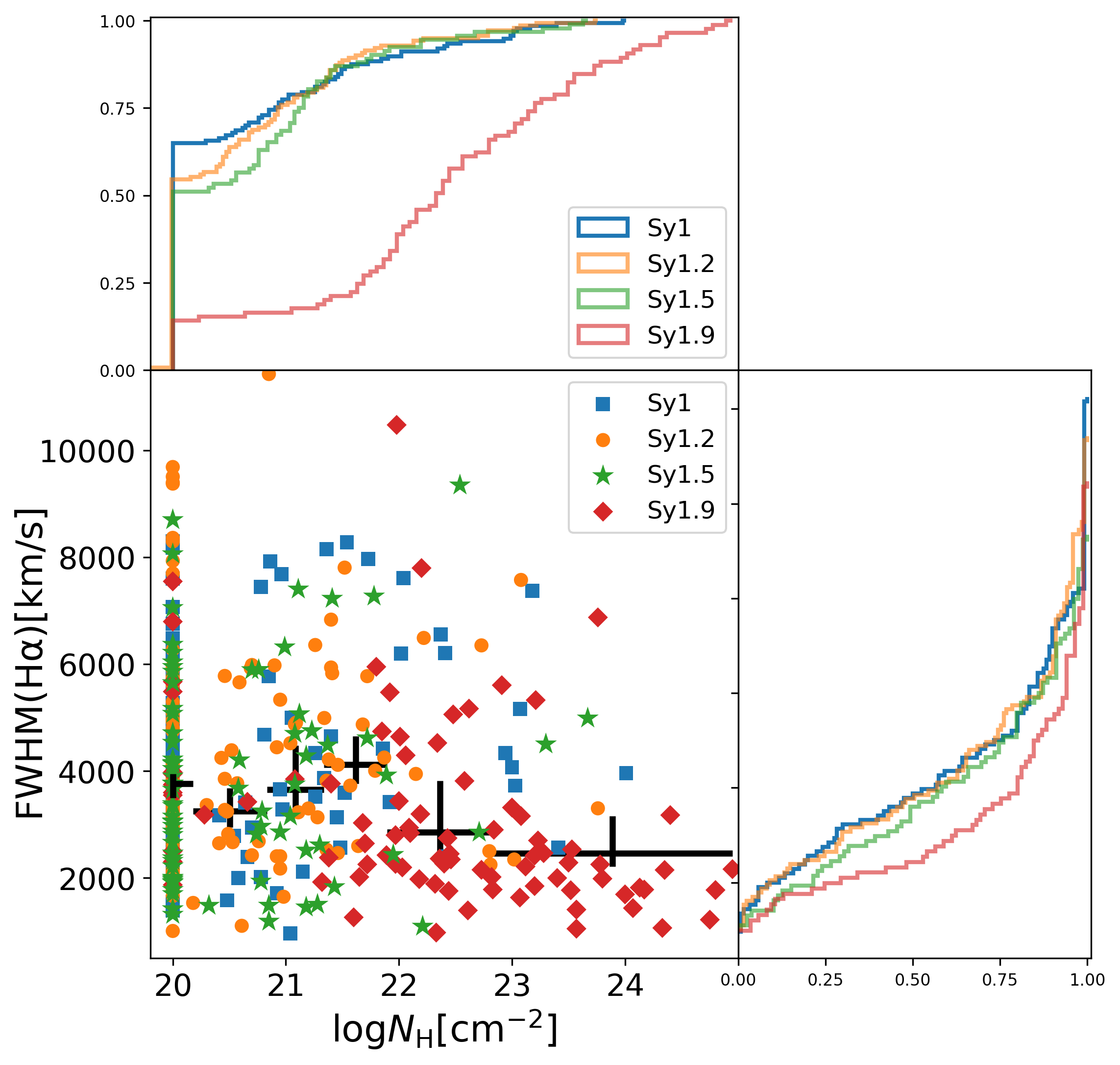}\hfill
    \caption{H$\alpha$ FWHMs vs.\ \NH\ and the projected distribution of these quantities for Sy1 (light-blue), Sy1.2 (orange), Sy1.5 (light-green) and Sy1.9 (red) AGNs. Horizontal error-bars represent the bin edges and vertical error-bars represent the errors in the median \fwhm\ from each bin estimated from bootstrapping. 
    {\color{black} For completeness, Fig.~\ref{fig:FWHB_NH} shows the complementary \fwhb\ vs. \NH\ parameter space.}
    } 
    \label{fig:FWHA_NH}
\end{figure}

In Figure~\ref{fig:FWHA_NH} we show \fwha\ 
vs. \logNHo, as well as the respective projected cumulative distributions for these quantities, for our sample of broad-line BASS/DR2 AGNs. 
{\color{black} For the sake of completeness, we also show a similar figure for \fwhb\ in Fig.~\ref{fig:FWHB_NH} (in Appendix~\ref{app:FWHB_NH}).}
A simple visual inspection of Fig.~\ref{fig:FWHA_NH} suggests that sources with no detected broad \Hbeta\ emission (i.e., Sy1.9 sources) are clustered towards higher column densities ($\logNH\gtrsim22$) and narrower \Halpha\ ($\fwha\lesssim3000\,\kms$), compared with the \logNHo\ and  \fwha\ distribution of sources {\it with} detected broad \Hbeta\ (i.e., Sy1-1.5s). 
Indeed, formal KS and WRS statistical tests indicate that the \fwha\ distribution in Sy1.9 is significantly different from that of Sy1-1.5s 
($P\lesssim10^{-5}$ for the null hypotheses of both tests).
The broad \Halpha\ emission lines in Sy1.9s are thus generally narrower than in Sy1-1.5s.
More specifically, most Sy1.9s with $22 < \logNH < 24$ have $\fwha\lesssim 2500\,\kms$, and the median value for such sources is $2452^{+494}_{-197}\,\kms$, compared to a median \fwha\ in Sy1-1.5s of $4337.0^{+1159}_{-610}\,\kms$ (across the entire  \logNHo\ range). 
In contrast, the median \fwha\ in Sy1.9s with $\logNH < 22$ ($3598^{+358}_{-316}\,\kms$) is consistent with that of Sy1-1.5s ($3677^{+142}_{-157}\,\kms$).

\begin{figure}
	\includegraphics[trim={2cm 0.5cm 21.2cm 2cm}, clip,width=0.475\textwidth]{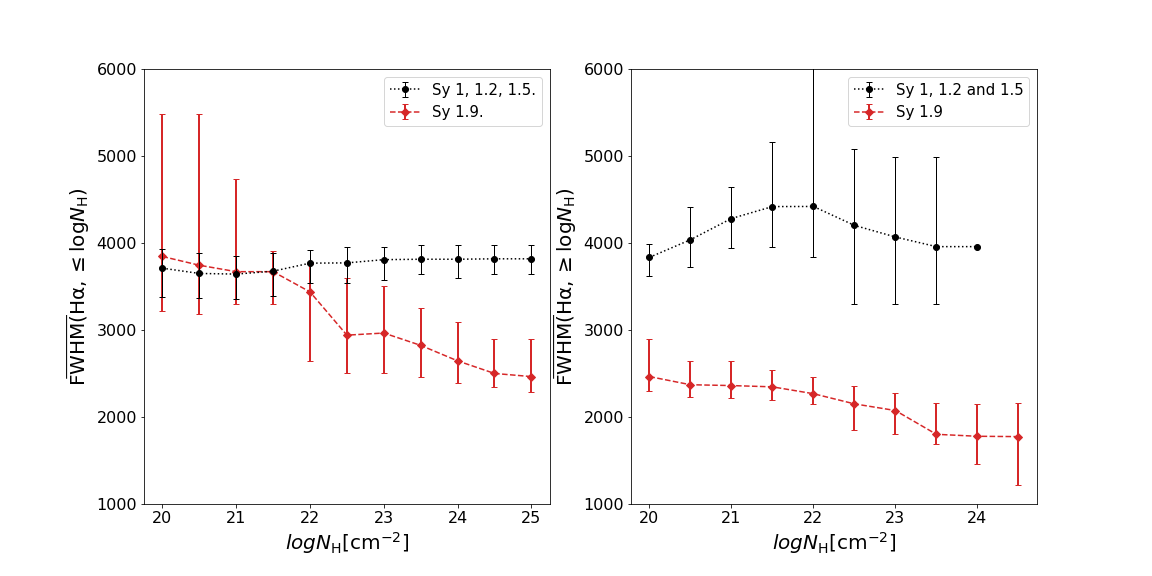}
    \caption{Median  \fwha\ for objects with \NH\ smaller than the \NH\ in the \logNHo\ (horizontal) axis. 
     The error-bars are obtained from bootstrapping and correspond to a confidence level of 90\%.
     Type 1.9 AGNs, which tend to have higher column densities ($\logNH>22$) typically also have narrower \Halpha\ broad emission lines.
    }
    \label{fig:FWLX_NH_MED}
\end{figure}

To further illustrate this point, in Figure~\ref{fig:FWLX_NH_MED} we show the median \fwha\ of Sy1-1.5s and Sy1.9s which have \logNHo\ smaller than (or equal to) the corresponding value on the \logNHo\ (horizontal) axis. 
Evidently, for $\logNH \lesssim 21.5$ the median values of \fwha\ in Sy1-1.5s and in Sy1.9s are in good agreement. However, when $\logNH \gtrsim 21.5$, Sy1.9s start to show narrower profiles than Sy1-1.5s, with a clear break point around $\logNH \gtrsim 23$ where the difference becomes more prominent and exceeds the 90\% confidence level (that is, exceeds the corresponding error bars).


\begin{figure*}
    \includegraphics[width=0.5\textwidth]{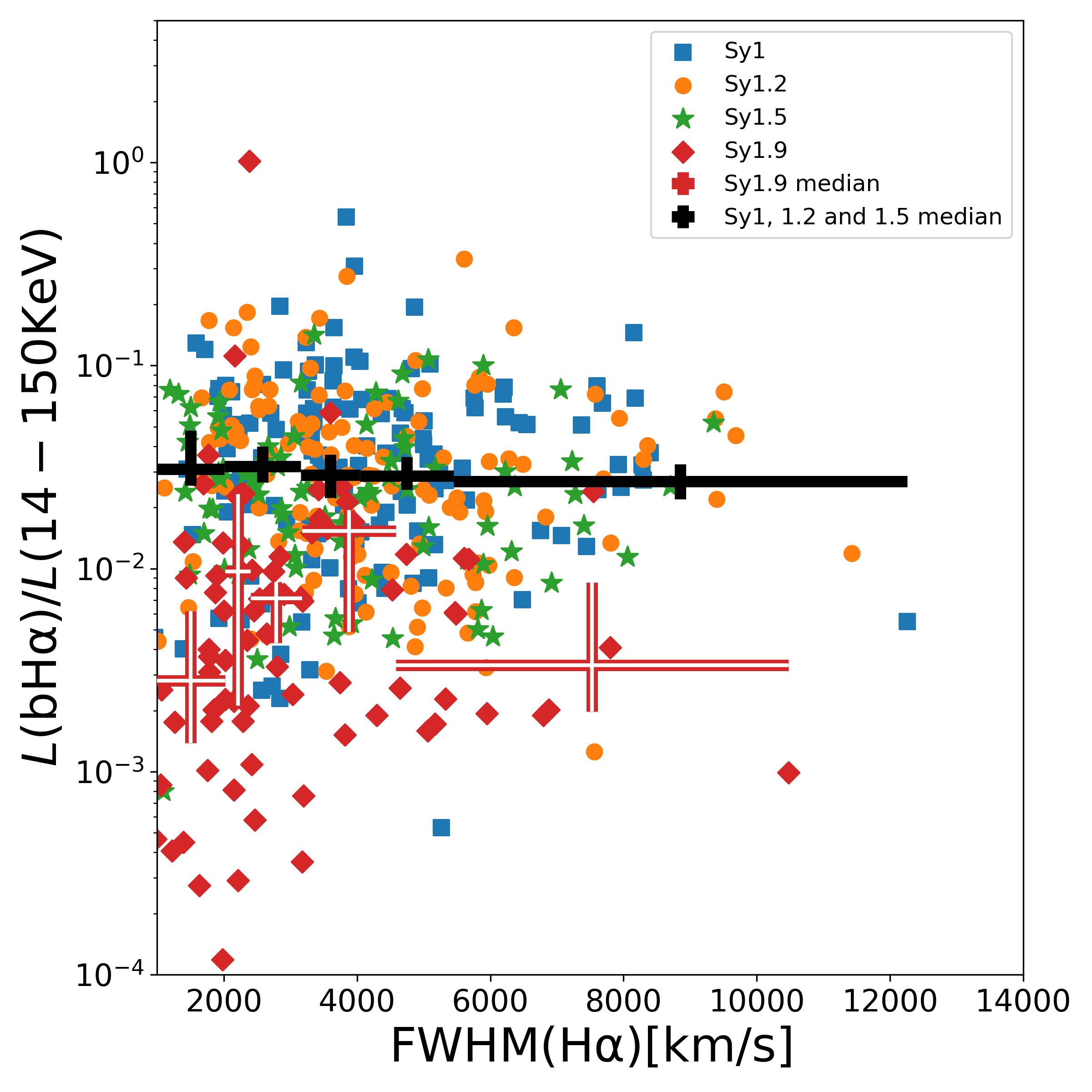}
    \hfill
	\includegraphics[width=0.5\textwidth]{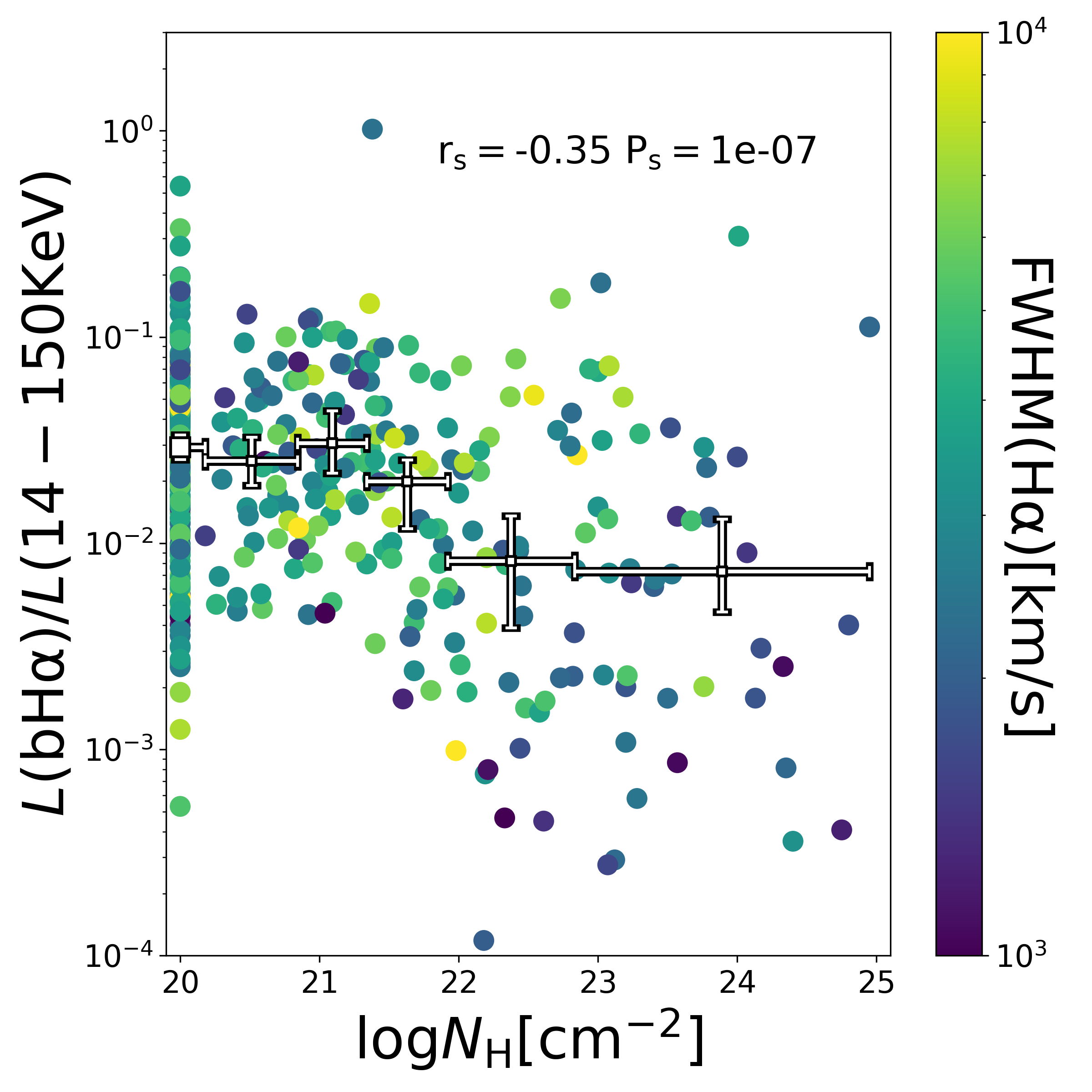}
    \caption{{\it Left:} $L\left({\rm H\alpha}\right)/L\left({\rm 14-150 KeV}\right)$ vs FWHM$\left({\rm H}\alpha\right)$. 
    {\it Right:} $L\left({\rm H\alpha}\right)/L\left({\rm 14-150 KeV}\right)$ vs  $\log N_{\rm H}$ color-coded by FWHM$\left({\rm H}\alpha\right)$. The vertical error-bars squares  represent the median values of $L\left({\rm H\alpha}\right)/L\left({\rm 14-150 KeV}\right)$ after binning in  FWHM$\left({\rm H}\alpha\right)$ (left panel) and  $\log N_{\rm H}$ (right panel) in equally spaced quantiles.  Horizontal error-bars  represent the bin edges and vertical error bars the errors in the median $L\left({\rm H\alpha}\right)/L\left({\rm 14-150 KeV}\right)$ estimated from bootstrapping.}
    \label{fig:LHAX_FWHM_NH}
\end{figure*}

In order to further characterize the apparent high velocity suppression in the broad \Halpha\ profiles, in the left panel of Figure~\ref{fig:LHAX_FWHM_NH} we show \Lbhax\ vs.\ \fwha\ for Sy1-1.5 and Sy1.9 sources, with large crosses representing the median values within \fwha\ bins (and corresponding error-bars; see figure caption). 
Figure~\ref{fig:LHAX_FWHM_NH} (left) shows that, in general, Sy1.9s tend to have systematically lower \Lbhax\ ratios across the full range of \fwha, compared to Sy1-1.5 sources. 
Moreover, the Sy1.9s with the narrowest \Halpha\ profiles ($\fwha\lesssim 3000\,\kms$) show yet weaker broad \Halpha\ (in terms of \Lbhax) than their broader-profile counterparts (i.e., Sy1.9s with $3000 \lesssim \fwha / \kms \lesssim 5000$). 

{\color{black} 
The right panel of Figure~\ref{fig:LHAX_FWHM_NH} shows \Lbhax\ vs.\ \logNHo\ for all our broad-line AGNs, irrespective of their sub-class (c.f.\ Fig.~\ref{fig:LX_NH}, left), with each AGN color-coded by its \fwha. It is again evident that the \fwha\ of heavily obscured AGNs, mostly dominated by Type 1.9 sources, show narrow and weak broad \Halpha\ emission lines.
We note here that the general trend of decreasing \Lbhax\ with increasing \logNHo, among all AGNs in our sample, is highly significant ($P\simeq10^{-7}$, as indicated).
} 

With the insights gained from Figures \ref{fig:LX_NH}, \ref{fig:FWHA_NH}, and \ref{fig:LHAX_FWHM_NH}, we infer that the heavily obscured Type 1.9 AGNs ($\logNH\gtrsim22.5$) generally show narrower and weaker \Halpha\ broad emission line profiles (i.e., lower \Lbhax), compared with (lower-\NH) Type 1-1.5 AGNs.

Combining these findings with those presented in Section \ref{sec:LhaNh}, we conclude that our BASS/DR2 sample shows evidence that the attenuation of the broad \Halpha\ line emission in Type 1.9 AGNs predominantly affects the highest-velocity line emitting gas. Thus, the obscuring material (which is also related to the higher column densities) must be, at least partially, located on scales comparable with the innermost parts of the BLR.

In two parallel BASS studies, NIR spectroscopy is used to model (broad) Paschen emission lines \citep{denBrok_DR2_NIR,Ricci_DR2_NIR_Mbh}. 
One of the results of the \cite{denBrok_DR2_NIR}  study is that the FWHM ratio between NIR and \Halpha\ lines in Sy1.9s increases monotonically  (from $\sim$1.2 to 2) with increasing line-of-sight obscuration (from $\logNH=21$ to $\logNH=25$). 
In principle, this may further support the scenario in which the highest-velocity \Halpha\ emitting regions tend to be suppressed by obscuration.
However, this finding is based on a limited number of sources ($\sim$10).
Moreover, the \cite{Ricci_DR2_NIR_Mbh} study essentially finds \textit{no} statistically significant trend between the FWHM ratio and \NH, at least up to $\logNH\simeq23$.

%
%

\begin{figure*}
	\includegraphics[width=0.475\textwidth]{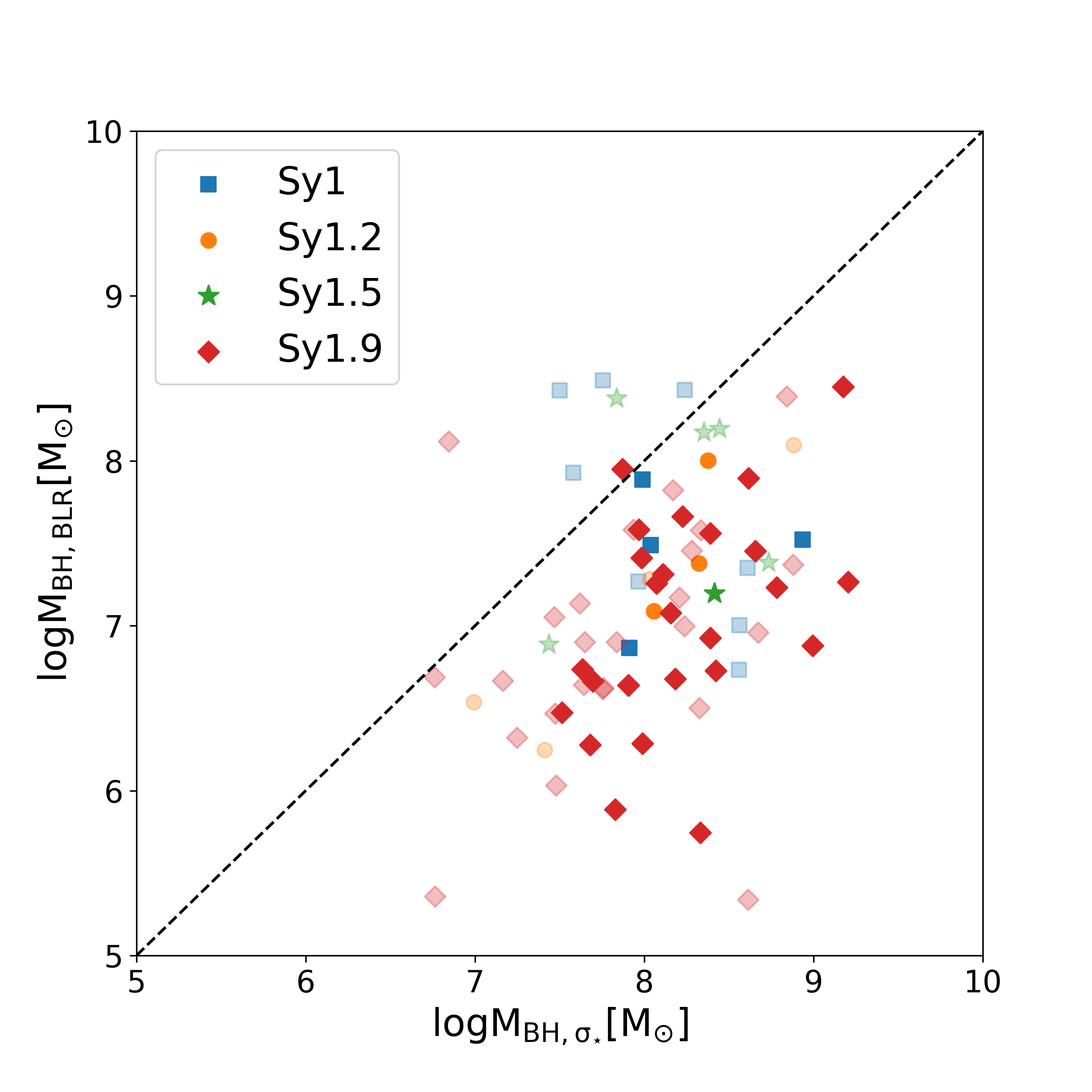}\hfill
	\includegraphics[width=0.475\textwidth]{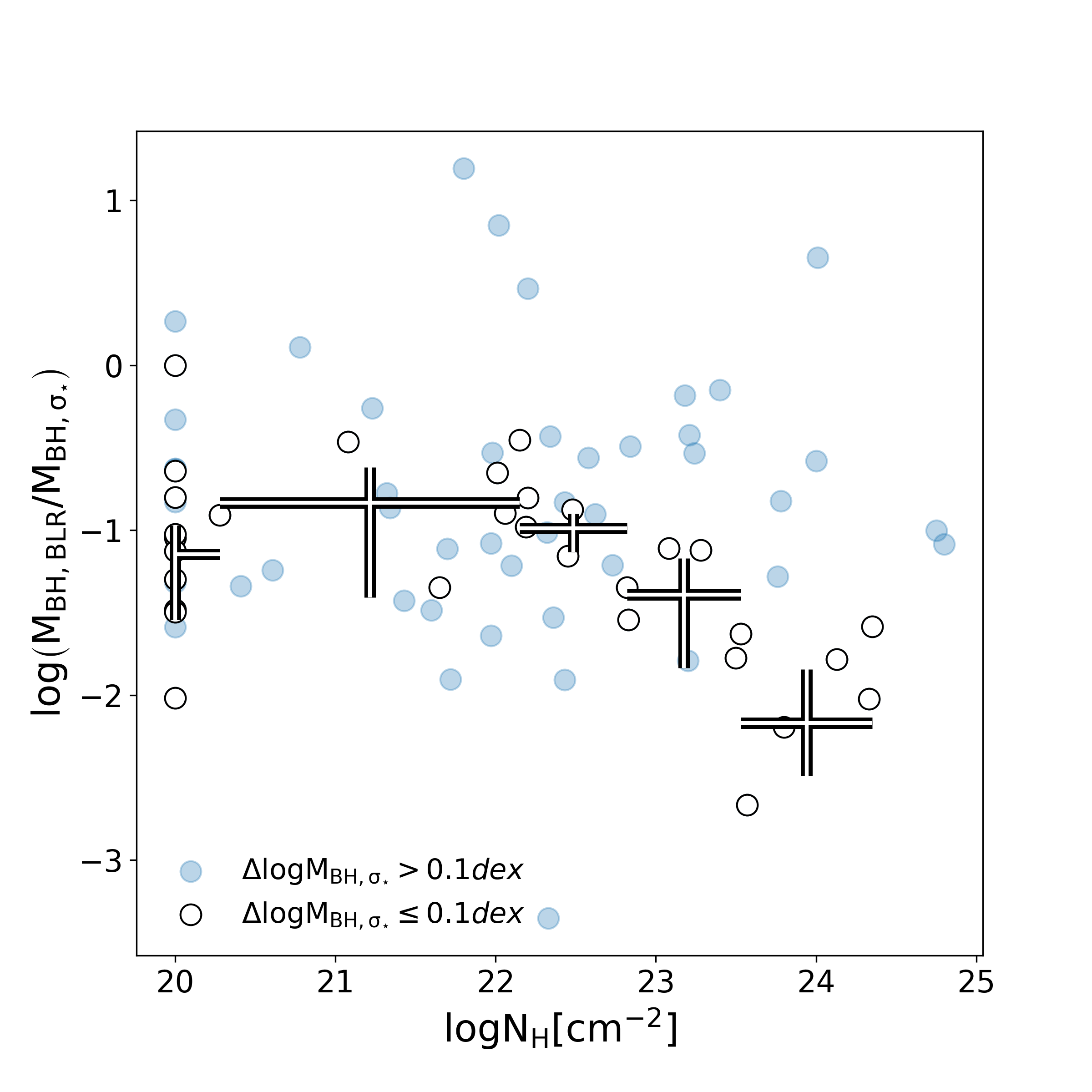}
    \caption{{\it Left:} Broad \Halpha\ based versus $\sigma_{\star}$ based \Mbh\ estimates.     
    {\it Right:} $\log\left({ M}_{\rm BH, BLR}/{M}_{\rm BH, \sigma_{\star}}\right)$ vs \logNHo. 
    In both panels, light-shaded symbols mark objects for which the measurement-related (fitting-related) uncertainties on \Mbh\ estimates from \sigmas, $\Delta \log{M}_{\rm BH, \sigma_{\star}}$, are larger than 0.1 dex. }
    \label{fig:MBHcomparison}
\end{figure*}

\subsection{Comparing Broad-line Based and Stellar-velocity Based $M_{\rm BH}$ Estimates in AGNs}
\label{sec:mbh_comp}

In this Section we provide a preliminary analysis of the differences that we find between \Mbh\ estimates derived from broad \Halpha\ emission lines ($M_{\rm BH, BLR}$, from this paper) and those derived from the stellar velocity dispersion (\sigs) measured in the AGN hosts ($M_{\rm BH, \sigs}$). 

The \sigs\ measurements are described in detail in a dedicated BASS/DR2 paper (\citealt{Koss_DR2_sigs}; see also \citealt{Caglar_DR2_Msigma}). 
Here we briefly note that {\color{black} these \sigs\ measurements are based on high-quality spectroscopy and analysis of the spectral regions that include the \Cahk, \Mgb, and/or Calcium triplet (near 8500 \AA) absorption features. 
The corresponding $M_{\rm BH, \sigs}$ estimates are then derived through the relation of \cite{KormendyHo2013}.} 
In principle, aperture size effects may be an important factor in \sigs\ estimates, particularly for surveys that cover a wide redshift range. 
In practice, however, most of our spectra were taken with slits of $\sim$1.5\arcsec width, corresponding to $\sim$0.5-3.6 kpc scales for BASS AGNs at $z\simeq0.015-0.14$, which encompasses 80\% of our sources.
Moreover, large galaxy samples show a rather limited diversity of \sigs\ radial profiles \citep[$\lesssim$15\% variation][]{Ziegler1997,Cappellari2006,FalconBarroso2017}.
We therefore expect only about 15\% systematic uncertainty in our \sigs\ estimates.

In the left panel of Figure~\ref{fig:MBHcomparison} we directly compare the two sets of \Mbh\ estimates -- from broad \Halpha\ emission ($M_{\rm BH, BLR}$) and from \sigs\ ($M_{\rm BH, \sigs}$), for the \NbothM\ BASS/DR2 AGNs for which both types of measurements are available. 
$M_{\rm BH, \sigs}$ estimates are generally larger than $M_{\rm BH, BLR}$, with median deviations of $\sim$0.69 and $\sim$0.89 dex for Sy1-1.5 and Sy1.9 sources, respectively. 
This result is in agreement with the recent studies of \cite{Caglar2020} on a sample of 19 local X-ray selected AGNs from the LLAMA project \citep{Davies2015}, where they find median offsets of 0.60 and 1.0 dex for Sy1s and Sy1.9s, respectively.\footnote{The {\color{black} virial} factor $f$ of $M_{\rm BH, BLR}$ in Fig.~\ref{fig:MBHcomparison} has been rescaled to 1.14 to better match the virial factor used in \cite{KormendyHo2013}, which is in turn used to estimate $M_{\rm BH, \sigs}$ in BASS/DR2 \citep{Koss_DR2_sigs}.} 

In the right panel of Figure~\ref{fig:MBHcomparison}, we present the differences between the two types of \Mbh\ estimates, in terms of $\Delta \log \Mbh \equiv \log (M_{\rm BH, BLR} / M_{\rm BH, \sigs})$, vs.\ line-of-sight column densities, $\log \NH$. 
When considering all available data points, there is a large scatter and no clear correlation between the two quantities. 
However, given the difficulties to measure \sigs, especially in systems where the optical continuum is AGN-dominated, we also consider a restricted subset of measurements, for which the uncertainties on $M_{\rm BH, \sigs}$ measurements are below 0.1 dex (white filled circles).
For this subset of higher-quality measurements, we can see that $\Delta \log \Mbh$ is roughly $-0.74$ dex for unobscured and mildly obscured AGNs, i.e., $\logNH \leq 22.5$. 
This is in agreement with the findings of previous studies, such as \citet{Woo2013,Woo2015} and \cite{Shankar2016}, and more recently by \citet{Shankar2019} and \citet{Caglar2020}. 
These works explored several scenarios to explain this offset, which we discuss below. 
For higher column densities, above $\logNH \approx 22$ -- that is, the regime dominated by Sy1.9s and where dust is expected to more strongly affect \Lbha\ measurements -- $\Delta \log \Mbh$ further decreases, strongly and monotonically, from about $-0.74$ dex to $-1.94$ dex at $\logNH \simeq 24$. 
A formal fit of our robustly-measured AGNs with $\logNH \geq 22.5$, derived using the \texttt{emcee} MCMC sampler yields the best-fit relation
\begin{multline}\label{eq:deltaM_vs_NH_highNH}
\Delta \log \Mbh =\\ (-0.64^{+0.26}_{-0.29}) 
    \times \log(N_{\rm H}/10^{22}\,{\rm cm}^{-2}) 
    - (0.52^{+0.20}_{-0.23}) \, , 
\end{multline}
where the quoted uncertainties represent 95\% confidence intervals. 
A fit using the BCES(Y$|$X) method provides a highly consistent relation, with slope and intercept of $-0.59 \pm 0.09$ and $0.62 \pm 0.11$, respectively.
The reason for this difference can be directly attributed to the fact that Sy1.9 sources show systematically lower \Lbhax\ and narrower \fwha\ (as shown in detail in the preceding sections), which contributes to lower $M_{\rm BH, BLR}$ (see Eq.~\ref{eq:mbh_general} and Table~\ref{tab:mass}).


One possible explanation for the discrepancy between broad-line-based and host-based determinations of \Mbh\ in nearby AGNs, as discussed in \citet[][]{Shankar2016,Shankar2019}, is that the \Mbh-\sigs\ relation determined for {\it inactive} galaxies is biased against low mass BHs because of the difficulties in resolving the sphere of influence and subsequently determine the black hole mass in these systems. 
According to these analyses, this bias artificially flattens the power-law index and enhances the intercept of the observed $\Mbh-\sigs$ relation of inactive galaxies. These, in turn, may amount to a discrepancy of about 0.7 dex with respect to the (assumed) intrinsic $\Mbh-\sigs$ relation -- in broad agreement with what is seen in our analysis of the BASS/DR2 sample, as well as other AGN samples. 

Two additional possible explanations are related to selection biases against low and high luminosities in the sample of reverberation-mapped, broad-line AGNs that is used to calibrate BLR-based mass prescriptions, as discussed in \cite{Woo2013}. 
On one hand, this RM sample can be slightly biased against low luminosity AGNs and therefore against low mass SMBHs because of their weak broad emission lines. 
On the other hand, a more important bias in such a sample is against luminous AGNs that are expected to preferentially harbor high mass SMBHs. 
This is due to a variability bias caused by the anti-correlation between the amplitude of variability and AGN luminosity \cite[e.g.,][and references therein]{Caplar2017_PTF}, that makes it difficult to measure the reverberation time lags in the most luminous systems. 
Another issue with highly luminous systems highlighted by \citet{Woo2013} is the great difficulty in measuring \sigs\ when the optical spectrum is dominated by a prominent, accretion disk powered component, which dilutes the weak stellar absorption features {\color{black} (see, e.g., \citealt{Grier2013})}. 
The study by \citet{Woo2013} explicitly showed that addressing these limitations of the RM sample can indeed account for the observed discrepancies seen between BLR-based and host-based determinations of \Mbh.

A final possibility is that discrepancies between broad-line-based and \sigs-based \Mbh\ estimates are caused by an overall different phase of evolution of the inactive and active galaxies populations. 
In such a scenario, the SMBHs of those galaxies observed to be active are still growing, and have yet to reach their ``final'' location in the $\Mbh-\sigs$ plane. 
While growing, active systems may indeed be located ``below'' the BH-host relations of inactive galaxies, and will eventually reach them, as expected from some co-evolutionary models \cite[e.g.,][]{SilkRees1998,King2003} and simulations \cite[e.g.,][]{AnglesAlcazar2017_FIRE_BHs,Bower2017,Lapiner2021}.
%
We note that the (late) evolution of active galaxies in the $\Mbh-\sigs$ plane is far from being well-understood, and radically different scenarios have been explored in numerous studies that address the (redshift resolved) AGN and galaxy populations 
\cite[e.g.,][and references therein]{Caplar2018}.

Unfortunately, the BASS/DR2 sample cannot be used to directly address these previously published scenarios as the vast majority (\NbothMinter\ out of \NbothM, or 66\%) of the objects in our sample with both broad-line-based and \sigs-based \Mbh\ estimates are Sy1.9 sources, which exhibit much larger mass discrepancies (Fig.~\ref{fig:MBHcomparison}). 
Taken at face value, these large discrepancies in dust-obscured Sy1.9 mass estimates (of up to 2 dex) may hint at the possibility that dust obscuration and/or circumnuclear (dusty) gas may play a role in where a given AGN appears in the $\Mbh-\sigs$ plane. 
However, our analysis has demonstrated that it is much more likely that the seemingly low broad-line-based \Mbh\ estimates of Sy1.9s are due to the diminished emission of the (high-velocity) \Halpha\ line.

In order to more directly address the issue of \Mbh\ discrepancies, the BASS team is pursuing two complementary directions. 
\cite{Caglar_DR2_Msigma} focuses on a highly-complete sample of Sy1 sources with both broad-line-based and \sigs-based estimates of \Mbh, and little sign of obscuration ($\logNH\lesssim22$). 
As mentioned above, \cite{Ricci_DR2_NIR_Mbh} uses NIR broad-line based \Mbh\ estimates in Sy1.9s using, e.g., broad Pa$\alpha$ and Pa$\beta$\ lines, which are far less affected by dust (compared to \Halpha). 

The findings presented here have important implications for determination of \Mbh\ in individual AGNs, and of the distributions of \Mbh\ (i.e., the BHMF) in AGN samples that are based solely on the identification of broad \Halpha\ emission. 
In such surveys, some portion of Sy1.9 sources may not be robustly identified (and excluded), while some portion of the ones that are identified will have \Mbh\ measurements that are underestimated by as much as 2 dex. 
Conversely, this would lead to \lledd\ being overestimated by up to 2 dex.
To remedy this when using large samples, one may consider focusing on those sources which have a robust identification of broad \Hbeta\ emission, or in which broad-band (X-ray) spectral analysis suggests limited dust obscuration ($\logNH\lesssim22$).

Another practical remedy would be to derive empirical corrections for the key observables, and the \Mbh\ estimates, of Type 1.9 sources. 
We calibrate such corrections in the next section.

\subsection{Correcting Single epoch $\Mbh\left({\rm b\Halpha}\right)$ Estimates in Type 1.9 AGNs}
\label{sec:correcting_mbh}

Our analysis shows that Type 1.9 AGNs exhibit suppression of the broad \Halpha\ line emission, particularly the highest-velocity emission, likely caused by dust obscuration. 
These effects become more prominent with increasing $N_{\rm H}$.
Given that the determination of \Mbh\ from BLR properties depends (almost solely) on \Lbha\ and \fwha\ measurements, these effects may have a direct impact on the determination of \Mbh\ in AGN samples, introducing a bias of underestimated \Mbh\ in (partially) obscured AGNs.

How can one overcome this tendency to underestimate \Mbh\ in Sy1.9 sources?
Given that our BASS/DR2 AGNs sample has only \NbothMinter\ Sy1.9 sources with both types of \Mbh\ estimates,
we prefer to provide only simple, median corrections -- that is, corrections that will bring the median quantities to agreement --  which can be applied to Sy1.9s in various regimes of key observables. 
Below we provide such corrections to \Lbha\ and \fwha\ in Sy1.9 sources, using \Lx\ (whenever it is available). 
To this end, we divide the $\Lbhax- \fwha$ parameter space into three regimes. 
We then simply identify the multiplicative corrections in \Lbha\ and \fwha\ that bring the median values of these quantities in Sy1.9s to agree with the medians of the Sy1-1.5s.
The uncertainties on these corrections were derived through a bootstrapping procedure, and represent the central 68th percentiles (i.e., $1\sigma$ equivalent). 
We also report the corresponding corrections to $\log\Mbh$, which are derived by combining the corrections in \Lbha\ and \fwha, through our \Mbh\ prescription.

The corrections for various ranges in \Lbhax\ and \fwha\ are:
\begin{enumerate}
    \item {\bf {\boldmath $\fwha_{\rm obs}{\lesssim}2400\,\kms$} and {\boldmath $\Lbhax{<}10^{-2}$}:} 
    
    \begin{itemize}
        \item $\Lbha_{\rm corr}=\left(21.6^{+20.2}_{-9.8} \right) \times \Lbha_{\rm obs}$
        \item $\fwha_{\rm corr}=\left(2.05^{+0.66}_{-0.32}\right) \times \fwha_{\rm obs}$
        \item $\Delta \log \Mbh \simeq 1.28^{+0.41}_{-0.30}$ dex
    \end{itemize}
    
    \item {\bf {\boldmath $\fwha_{\rm obs}{\gtrsim}2400\,\kms$} and {\boldmath $\Lbhax{<}10^{-2}$}:} 
    \begin{itemize}
        \item $\Lbha_{\rm corr}=\left(8.4^{+3.9}_{-2.5}\right) \times \Lbha_{\rm obs}$
        \item $\fwha_{\rm corr}=\left(0.93^{+0.32}_{-0.22} \right) \times \fwha_{\rm obs}$\\ (consistent with no correction).
        \item $\Delta \log \Mbh \simeq 0.53 ^{+0.09}_{-0.09}$ dex
    \end{itemize}
    
    \item {\bf {\boldmath $\Lbhax{>}10^{-2}$}: }
    \begin{itemize}
        \item 
        $\Lbha_{\rm corr}=\left(1.4^{+0.3}_{-0.4}\right) \times \Lbha_{\rm obs}$. \\
        (consistent with no correction).
        \item $\fwha_{\rm corr}=\left(0.99^{+0.12}_{-0.34}\right) \times \fwha_{\rm obs}$\\ (no correction needed).
        \item No correction needed for $\Mbh$.
    \end{itemize}
\end{enumerate}

In practice, most AGN surveys lack measurements of \Luhard, which would render the above corrections impractical. 
First, we note that the much more common, lower-energy measurements of \Lhard\ may be used as a proxy for \Luhard. Specifically, for a photon index of $\Gamma_{\rm X}=1.8$, the luminosities scale as $\Lhard = 0.42 \times \Luhard$.
Second, we have also derived an additional set of corrections, where the infrared (IR) emission serves as a proxy of the (ultra-hard) X-rays, motivated by many previous studies of the link between these spectral regimes in AGNs \cite[e.g.,][and references therein]{Lutz2004_IR_Xray,Fiore2009_COSMOS_obscured,Gandhi2009_IR_Xray,Asmus2015,Stern2015,Lansbury2017,Ichikawa2017}.
Specifically for our BASS/DR2 sample, we used the IR measurements described in \cite{Ichikawa2019}, and find that the flux at (rest-frame) 12$\mu$m shows the tightest correlation with ultra-hard X-ray emission ($r_{\rm s}=0.55$, $P_{\rm s} \ll 10^{-10}$), again consistent with previous studies \citep{Asmus2015,Ichikawa2017}. 
We also confirmed that \Lbhair\ preserves the correlation with \logNHo\ with a similar significance (see Fig. \ref{fig:LHAIR_FWHM_NH} in Appendix~\ref{app:LHAIR_FWHM_NH}).  
Below we provide median corrections to {\color{black} \Lbha, \fwha\ and \Mbh} for Sy1.9 sources whenever \Lir\ is available. 
For this, we have repeated our analysis while dividing the Sy1.9s in our sample into three regimes in \Lbhair\ and \fwha.
The corresponding median corrections are:

\begin{enumerate}
    \item {\bf {\boldmath $\fwha_{\rm obs}{\lesssim} 2400\,\kms$} and {\boldmath $\Lbhair{<}7{\times}10^{-3}$}:} 
    \begin{itemize}
        \item $\Lbha_{\rm corr}=\left(17.0^{+16.9}_{-7.6}\right) \times \Lbha_{\rm obs}$
        \item $\fwha_{\rm corr}=\left(1.92^{+0.41}_{-0.22} \right) \times \fwha_{\rm obs}$
        \item $\Delta \log \Mbh \sim 1.40^{+0.23}_{-0.37}$ dex
    \end{itemize}
    
    \item {\bf {\boldmath $\fwha_{\rm obs}{\gtrsim}2400\,\kms$} and {\boldmath $\Lbhair{<}10^{-2}$}:} 
    \begin{itemize}
        \item $\Lbha_{\rm corr}=\left(16.0^{+5.4}_{-5.3}\right) \times \Lbha_{\rm obs}$
        \item $\fwha_{\rm corr}=\left(0.88^{+0.42}_{-0.18} \right) \times \fwha_{\rm obs}$\\ (consistent with no correction).
        \item $\Delta \log \Mbh \sim 0.68 ^{+0.09}_{-0.10}$ dex
    \end{itemize}
    
    \item {\bf {\boldmath $\Lbhair{>}10^{-2}$}: }
    \begin{itemize}
        \item $\Lbha_{\rm corr}=\left(2.4^{+1.1}_{-0.8}\right) \times \Lbha_{\rm obs}$ 
        \item $\fwha_{\rm corr}=\left(1.03^{+0.28}_{-0.25}\right) \times \fwha_{\rm obs}$\\
        (no correction needed).
        \item $\Delta \log \Mbh \sim 0.21 ^{+0.10}_{-0.09}$ dex
    \end{itemize}
\end{enumerate}


We finally note that, 
as part of our search for ways to improve \Mbh\ estimates in (Sy1.9) BASS AGNs, 
we have also checked the possibility that \fwha\ is correlated with $L(\oiii)/L({\rm n}\Hbeta)$, as found by \cite{BaronMenard2019} in their (spectral stacking) analysis of the SDSS/DR7 quasar sample. 
This correlation is proposed as a promising method to provide \Mbh\ estimates for narrow-line AGNs and -- in the context of the present study -- may thus be used to improve mass estimates in Sy1.9 sources. 
We do find that \fwha\ and $L(\oiii)/L({\rm n}\Hbeta)$ are correlated in our BASS/DR2 sample, with the Pearson and Spearman correlating tests resulted in $P=0.006$ and ${\approx}10^{-4}$, respectively. 
However, these correlations are weak ($r_{\rm s}=0.14$ and $0.19$, respectively) and the scatter is huge, which prevents us from using the correlation to improve our \Mbh\ estimates.
We stress that we are not evaluating the correlation on {\it stacked} data, as was done in \cite{BaronMenard2019}, but rather on individual spectra in which measuring $L(\oiii)/L({\rm n}\Hbeta)$ is much more challenging. 
Proper stacking analysis is beyond the scope of the present study. 


\section{Summary and conclusions}

In this paper we presented broad emission line measurements for the 2nd data release of the BAT AGN Spectroscopic Survey (BASS/DR2), which consists of \NblrGood\ AGNs selected in the ultra-hard X-rays and for which high-quality fits of the \Halpha, \Hbeta, \mgii, and/or \civ\ emission lines are now made available. 
These detailed spectral measurements are used to also determine the masses (\Mbh) and Eddington ratios (\lledd) of the SMBHs that power these AGNs.
The key features of this new catalog, compared to BASS/DR1, are:
\begin{enumerate}
    
    \item We provide broad line measurements and derived BH masses for \Nfullgood\ AGNs, of which \NblrGood\ are drawn from the 70-month Swift/BAT catalog (i.e., almost 60\% of the 70-month sources, that constitute the main DR2 sample). 
    In addition, we provide measurements for \Nbonusgood\ AGNs detected in deeper BAT data
    
    \item At its core, lower-redshift focus, our BASS/DR2 catalog has \Nhgoodnb\ unbeamed, $z<0.7$ AGNs drawn from the 70-months Swift/BAT catalog, with reliable determinations of \Mbh\ from the broad \Halpha\ and/or \Hbeta\ emission lines. 
    
    \item We provide improved spectral measurements and BH determinations for $>$200 BASS AGNs, for which the BASS/DR2 efforts resulted in higher-quality data and/or analysis.
    
    \item The larger fraction of sources with a wide spectral coverage allows for a more complete identification of sub-classes using optical line ratios. 
    
    
    \item BH masses are estimated using a more consistent set of prescriptions, particularly the virial factor ($f=1$).

\end{enumerate}
The BASS/DR2 broad emission line catalog is released as part of this paper (in machine-readable form) and is available on the BASS website.\footnote{\url{http://www.bass-survey.com}.}


In the second part of the paper, we used the unprecedentedly large compilation of BASS/DR2 multi-wavelength data, to investigate the properties of ``partially obscured'' broad-line systems--so-called Type 1.9 AGNs (or Sy1.9s), which show broad \Halpha\ emission lines but no bluer broad (Balmer) lines.
We compared these Type 1.9 sources to those AGNs with both broad \Halpha\ and \Hbeta\ emission lines, i.e. Type 1-1.5 sources.
Our main findings regarding partially obscured, Type 1.9 AGNs can be summarized as follows:
\begin{enumerate}
    
    \item Type 1.9 AGNs tend to exhibit high column densities, typically $\logNH \gtrsim 22$, compared to Type 1-1.5 AGNs which typically have $\logNH \lesssim 22$.
    
    \item The strength of the broad \Halpha\ emission line (relative to the X-ray continuum) decreases with increasing $N_{\rm H}$, and is particularly suppressed in Type 1.9 AGNs. 
    This suggests that the broad line emission is affected by dust.
    
    \item The broad \Halpha\ suppression particularly affects the highest-velocity parts of the line profile, that is the inner-most parts of the \Halpha\ emitting region in the BLR.
    
    \item These effects result in a significant underestimation of BLR-based \Mbh\ determinations in Type 1.9 AGNs, with a discrepancy of 0.8 dex at $\logNH \simeq 22.5$ and up to 2 dex at $\logNH \simeq 24$.
    
    \item To remedy the potential \Mbh\ discrepancies, we provide simple, empirical corrections for \Lbha\ and \fwha, applicable to Type 1.9 AGNs with either (ultra-hard) X-ray or near-IR measurements.

\end{enumerate}
%
%
As an alternative to our corrections, if near-IR spectroscopy is available, then one should consider using \Mbh\ prescriptions that are based on broad Paschen emission lines \cite[e.g.,][]{RicciFed2017_Mvir,Kim2018}, as this spectral regime is less affected by dust. 

Our work provides the community with a large, highly-complete compilation of reliable determinations of \Mbh\ (and \lledd), while also highlighting some of the challenges associated with partially-obscured sources, and with AGN surveys where broad Balmer emission lines are used for \Mbh\ determinations.
As such, we hope our catalog and analysis can be useful for detailed investigations of individual AGN and/or of SMBH demographics in the local Universe, particularly when combined with the rich compilation of multi-wavelength measurements available through BASS.
Several complementary works, published as part of the BASS/DR2 effort, indeed pursue such investigations.


\section*{}
\label{ack}

{\color{black} We thank the anonymous referee for their constructive comments, which helped us improve the paper.}
We also thank Lea Marcotulli for her assistance with identifying beamed AGN candidates, and Jonathan Stern for his insightful comments.

B.T. acknowledges support from the Israel Science Foundation (grant number 1849/19) and from the European Research Council (ERC) under the European Union's Horizon 2020 research and innovation program (grant agreement number 950533).
M.K. acknowledges support from NASA through ADAP award NNH16CT03C.
K.O. acknowledges support from the National Research Foundation of Korea (NRF-2020R1C1C1005462).
C.R. acknowledges support from the Fondecyt Iniciacion grant 11190831. 
We also acknowledge support from 
ANID grants CATA-Basal AFB-170002 (F.R., F.E.B., E.T.) and FB210003 (C.R., F.E.B., E.T., R.J.A.); 
FONDECYT Regular 1190818 (E.T., F.E.B.), 1191124 (R.J.A.), and 1200495 (F.E.B., E.T.);
FONDECYT Postdoctorado 3180506 (F.R.) and 3210157 (A.R.);
Anillo ACT172033 (E.T.); and
Millennium Science Initiative ICN12\_009 (MAS; F.E.B.) and Millenium Nucleus NCN19\_058 (TITANs; E.T.).
D.A. acknowledges funding through the European Union's Horizon 2020 and Innovation programme under the Marie Sklodowska-Curie grant agreement no. 793499 (DUSTDEVILS). 
The work of K.I.\ is supported by the Japan Society for the Promotion of Science (JSPS) KAKENHI (18K13584, 20H01939). 

This work relies on data collected with a large variety of facilities and analyzed using several tools.
We acknowledge the work that the Swift BAT team has done to make this project possible, and the teams of the various observatories that obtained the data used in this paper.  
Specifically, this work is based on observations collected at the European
Organisation for Astronomical Research in the Southern
Hemisphere under 29 ESO programmes: 60.A-9024(A), 60.A-9421(A), 062.H-0612(A), 086.B-0135(A),
089.B-0951(A), 089.B-0951(B),
090.A-0830(A), 090.D-0828(A),
091.B-0900(B), 091.C-0934(B),
092.B-0083(A), 093.A-0766(A),
095.B-0059(A), 098.A-0062,
098.A-0635(B), 099.A-0403(A),
099.A-0403(B), 099.A-0442(A),
099.B-0785(A), 0101.A-0765(A),
0101.B-0456(B), 0101.B-0739(A), 0102.A-0433(A),
0103.A-0521(A), 0103.B-0566(A), 0104.A-0353(A), 0104.B-0959(A), 0106.A-0521(A), 385.B-1035(A), and 2100.B-5018(B). 
%
%
%

%
BASS/DR2 also relies on observations from seven CNTAC programs:  CN2016A-80, CN2018A-104, CN2018B-83, CN2019A-70, CN2019B-77, CN2020A-90, and CN2020B-48 (PI C. Ricci); and from NOIRLab program 2012A-0463 (PI M. Trippe).
%
%
Funding for SDSS-III has been provided by the Alfred P. Sloan Foundation, the Participating Institutions, the National Science Foundation, and the U.S. Department of Energy Office of Science. 
The SDSS-III web site is http://www.sdss3.org/.
The authors wish to recognize and acknowledge the very significant cultural role and reverence that the summit of Mauna Kea has always had within the indigenous Hawaiian community.  
We are most fortunate to have the opportunity to conduct observations from this mountain.

This research made use of the NASA/IPAC Extragalactic Database (NED), which is operated by the Jet Propulsion Laboratory, California Institute of Technology, under contract with the National Aeronautics and Space Administration and the SIMBAD database, operated at CDS, Strasbourg, France \citep{Wenger:2000:9}. 


%

\vspace{5mm}
\facilities{Keck:I (LRIS), Magellan:Clay, Hale (Doublespec), NuSTAR, Swift (XRT and BAT), VLT:Kueyen (X-Shooter), VLT:Antu (FORS2), SOAR (Goodman)}



\software{{\tt AstroPy} \citep{Astropy2013,Astropy2018},
{\tt Matplotlib} \citep{Matplotlib2007}, 
{\tt NumPy} \citep{NumPy20_Nature}}




\clearpage
\newpage

\appendix

\section{Fit quality examples}
\label{app:fitq}

\begin{figure*}
\includegraphics[width=0.3\textwidth]{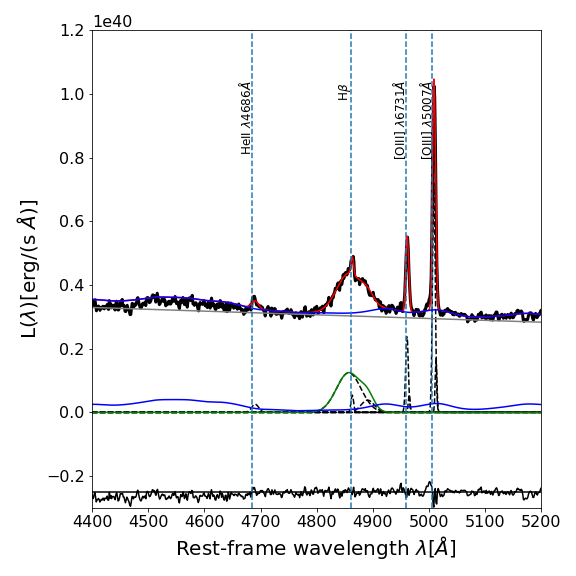}
\includegraphics[width=0.3\textwidth]{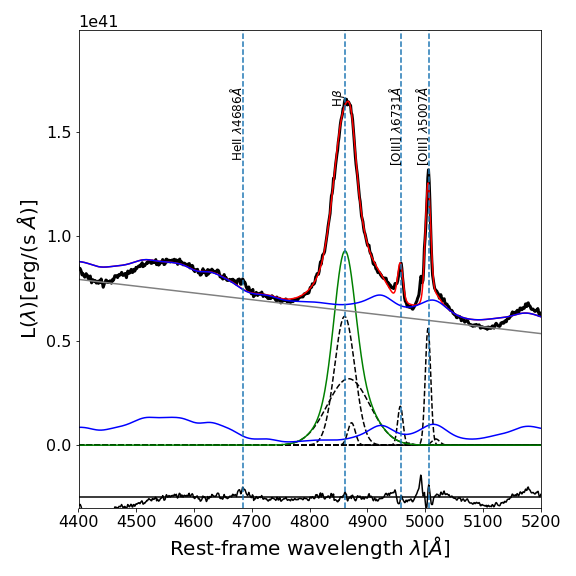}
\includegraphics[width=0.3\textwidth]{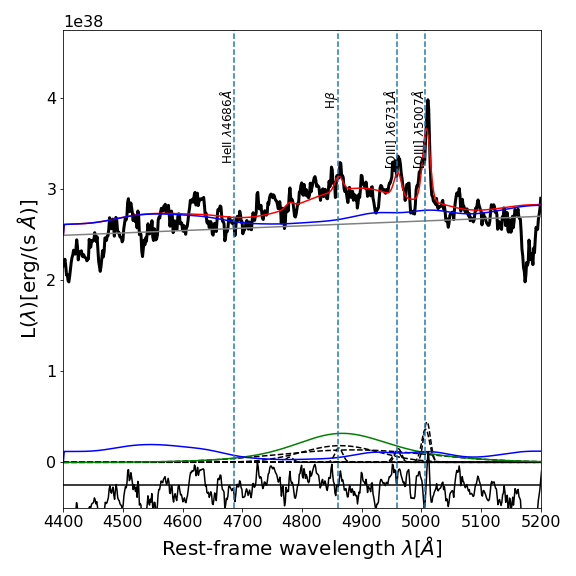}\\
\includegraphics[width=0.3\textwidth]{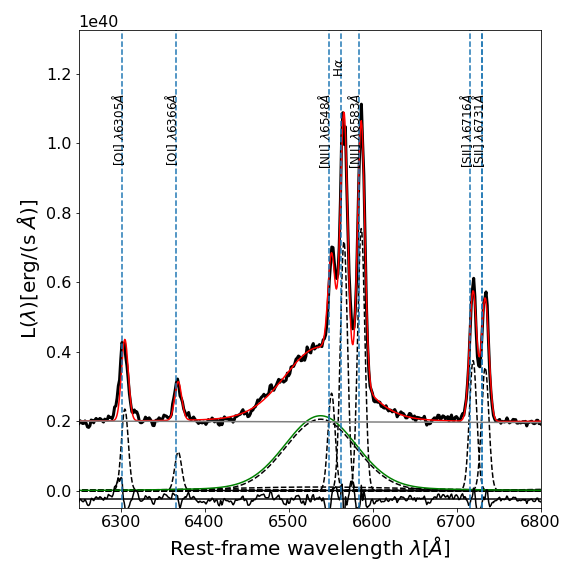}
\includegraphics[width=0.3\textwidth]{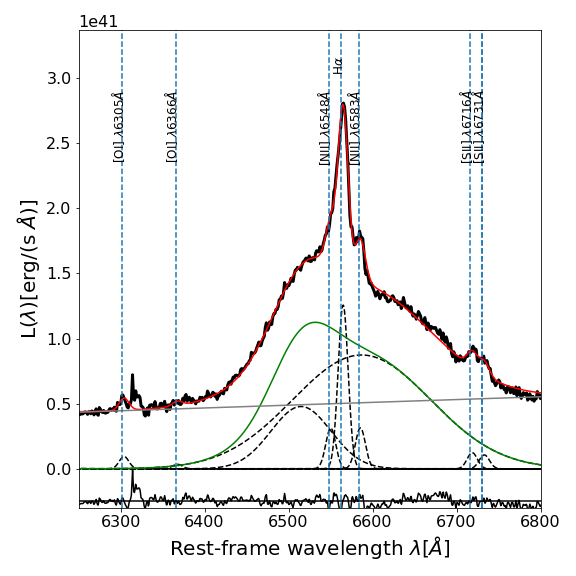}
\includegraphics[width=0.3\textwidth]{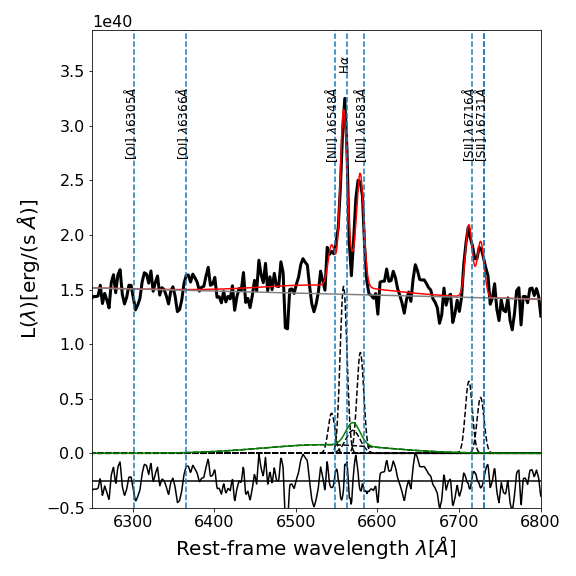}
\caption{Spectral fit quality flag (\FQ) examples, in order of descending quality for the \Hbeta\ (top) and \Halpha\ (bottom) spectral complexes: $\FQ=1$ (left), $\FQ=2$ (center) and $\FQ= 3$ (right). 
In each case, the observed spectrum (solid black line) should be compared with the total spectral model (red).
We also show the broad Balmer emission lines model (green), the blended iron emission (blue), and the narrow emission lines (dashed black).
The black solid line at the bottom of each panel represents the fit residuals.}
\label{fig:fitq}
\end{figure*}

In Figure \ref{fig:fitq} we show several examples of \Halpha\ and \Hbeta\ fits of different fit-quality classes  ($\FQ=1, 2$, and $3$), as assigned during our visual inspection of the spectra and best-fit models. 
We recall that only sources with $\FQ<3$ provide acceptable spectral measurements, while those with $\FQ=3$ should be discarded from any analysis. 
For the most cautious analyses we further recommend to focus on $\FQ\leq2$ (i.e., omitting objects with $\FQ=2.5$, as we did in the present study).

\section{Comparing broad line measurements in BASS DR2 and DR1}
\label{sec:DR2vsDR1}

Here we compare the line width and \Mbh\ measurements from the new BASS/DR2 catalog presented here to those of our our previous release, DR1. 
As mentioned in section \ref{sec:modeling}, compared to DR1, DR2 includes not only new optical spectra but also a more homogeneous spectral modeling procedure to derive broad line properties and black hole masses. 

\subsection{FWHM Comparison}
In the top-left panel of Fig.~\ref{fig:MbhHADR12} we compare the \fwha\ obtained from BASS/DR2 catalog to those measured in DR1, for sources which were included in both catalogs. 
The DR2 measurements are slightly narrower than the DR1 ones, with a median offset of about 7\% (see diagonal lines in Fig.~\ref{fig:MbhHADR12}).
Similarly, in the top-right panel of Fig.~\ref{fig:MbhHADR12} we compare the \fwhb\ measurements in the DR1 and DR2 catalogs. 
In this case, the two sets of measurements are in very good agreement up to 8000 \kms, with a median offset from the 1:1 relation only 2\%. 
The reason for the good agreement between DR1 and DR2 \fwhb\ measurements is that in both cases we followed a very similar fitting procedure. 
On the other hand, the slightly larger offset in the \fwha\ measurements is very likely caused by the differences in the fitting procedures - in DR1 the \Halpha\ spectral complex was modeled with rather simplistic, ad-hoc procedures, 
while for DR2 we adopt the more elaborate and AGN-tailored procedures of \cite{MejiaRestrepo2016a}.

\subsection{Black hole Mass Comparison}

\begin{figure*}
\includegraphics[width=0.5\textwidth]{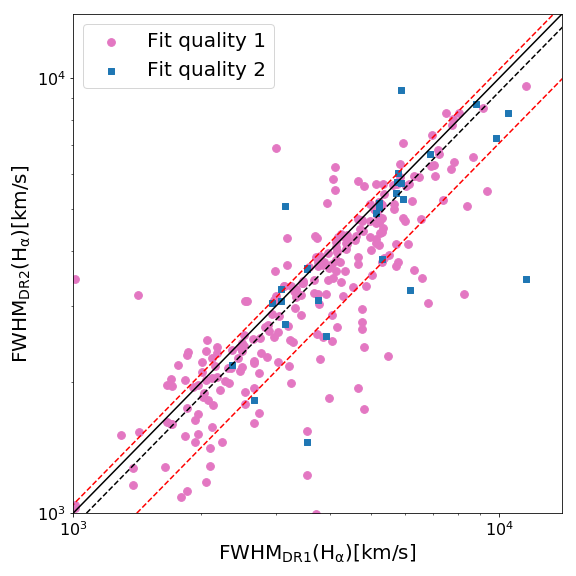}
\includegraphics[width=0.5\textwidth]{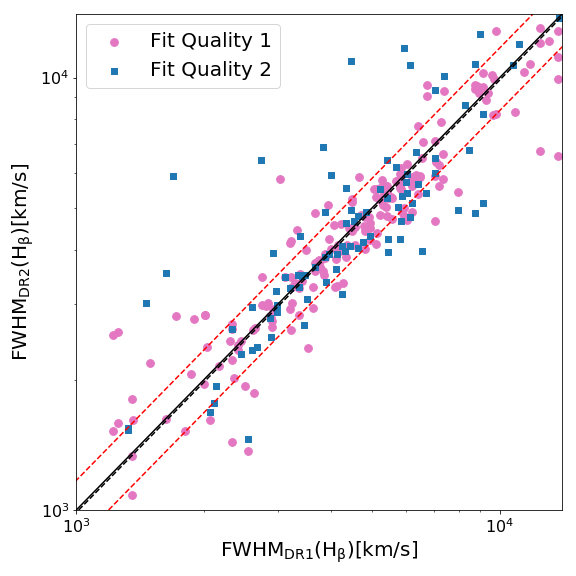} \\
\includegraphics[width=0.5\textwidth]{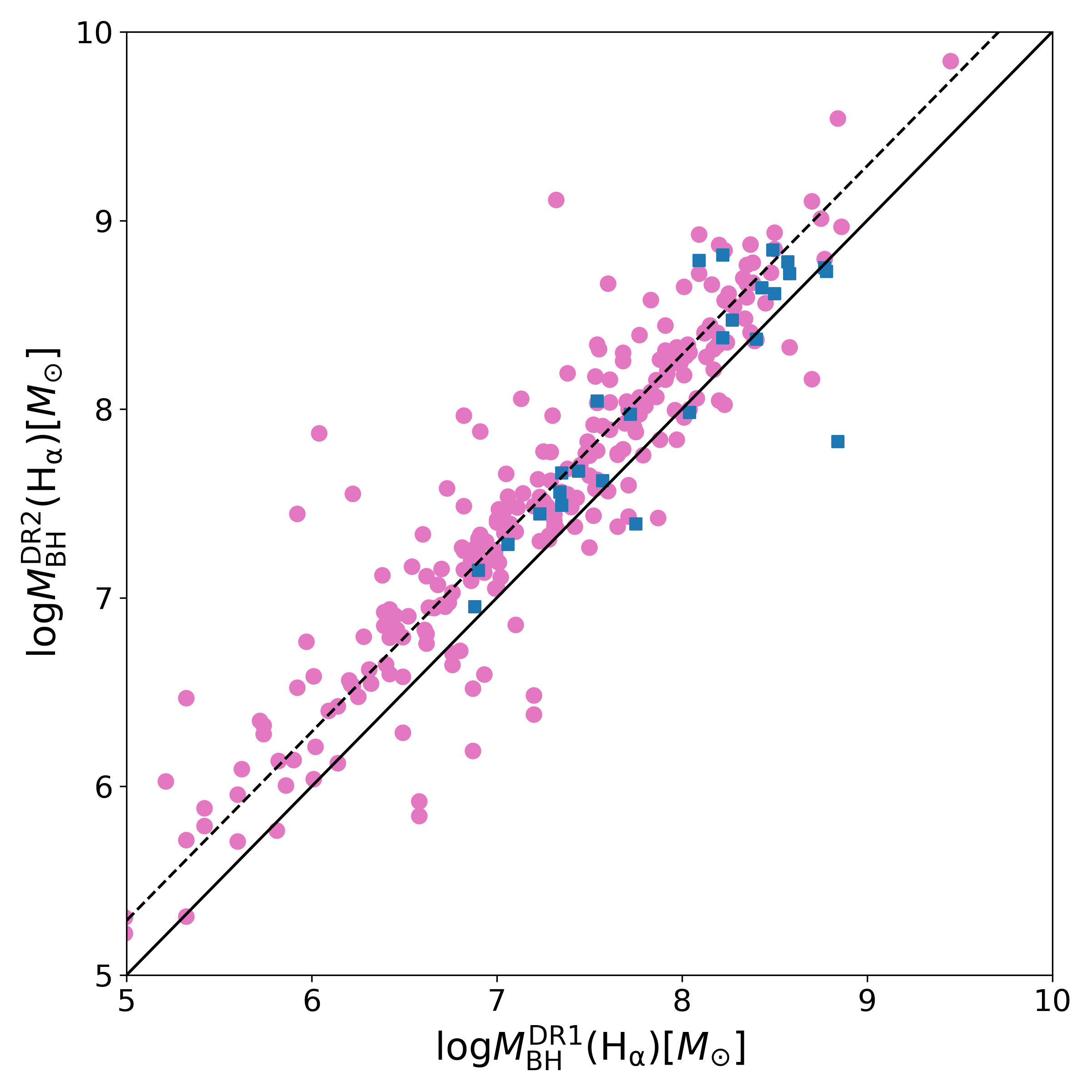}
\includegraphics[width=0.5\textwidth]{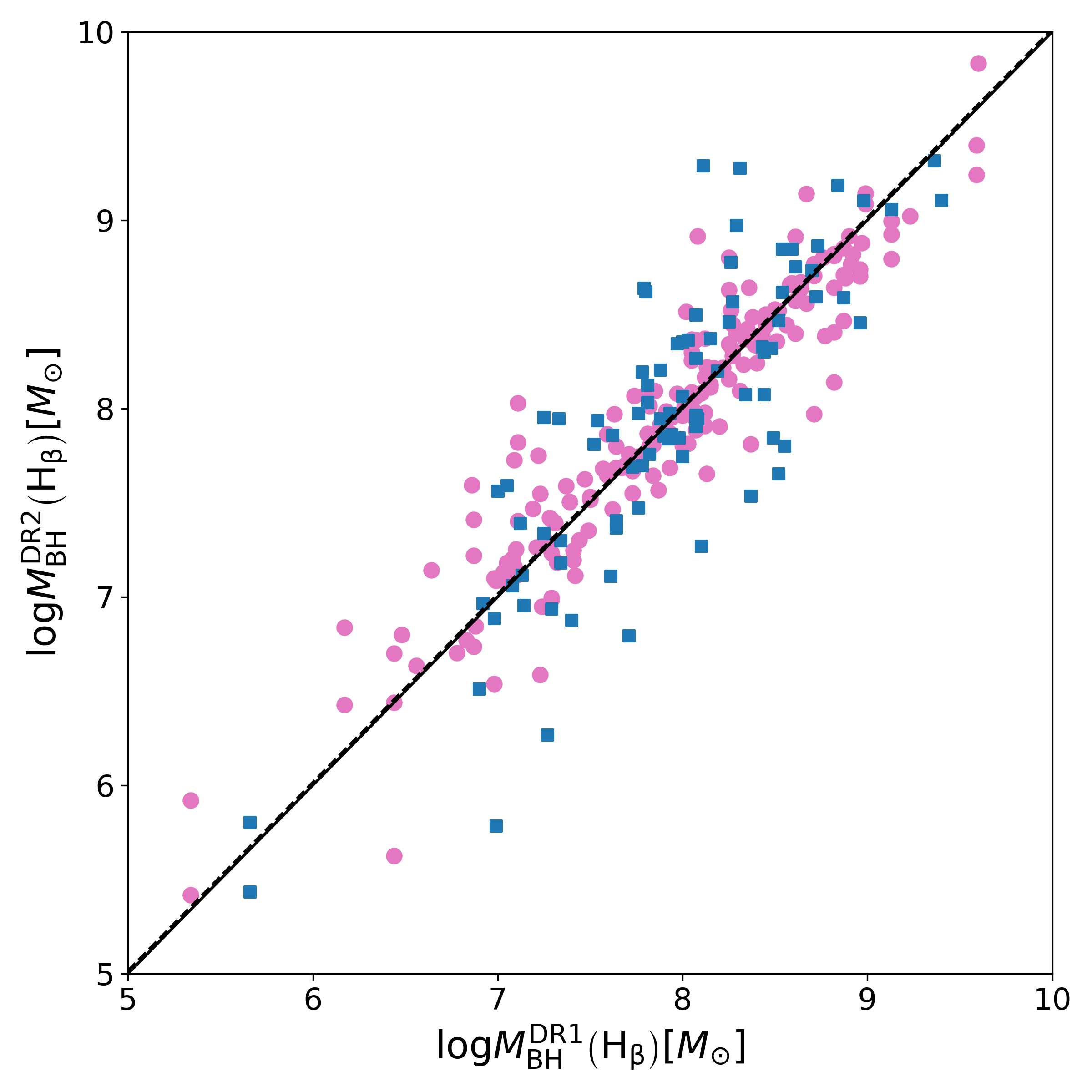}
\caption{Comparison DR1 and DR2 measurements: line widths (\fwhm; top) and black hole masses (\Mbh; bottom), for both H$\alpha$ (left) and H$\beta$ (right). 
In all panels, black solid lines represent the 1:1 relation and black dashed lines trace the median offsets between DR1 and DR2 measurements.}
\label{fig:MbhHADR12}
\end{figure*}

In the bottom panels of Fig.~\ref{fig:MbhHADR12} we compare the \Halpha-based (bottom-left) and \Hbeta-based (bottom-right) BH mass estimates obtained in DR2 to those obtained in DR1. 
The \Hbeta-based \Mbh\ estimates from both DRs are in very good agreement, with a negligible offset (median of -0.02 dex). 
However, when it comes to \Halpha\ there is a clear disagreement of 0.23 dex between DR1 and DR2 \Mbh\ measurements, in the sense that DR2 measurements are systematically larger than DR1 ones. 
One of the main reasons for this discrepancy is the usage of different virial factors in DR1 and DR2: while for DR1 we used $f_{\fwha}=0.75$ and $f_{\fwhb}=1$, in DR2 we instead use $f_{\fwha}=f_{\fwhb}=1$. The reason for this choice is to keep consistency between the masses derived through the two emission lines, and to more recent calibrations that are based on the comparison of virial (SE) and \sigs-based \Mbh\ estimates \cite[e.g.,][]{Woo2015}. 
This update of the virial factor accounts for 0.13 dex on the total offset. 
The  remaining $\sim$0.1 dex is explained by the usage in DR2 of an alternative $\RBLR-\Lbha$ calibration, which includes more RM measurements towards the low luminosity end \citep{GreeneHo2005}, together with slight differences between the DR1 and DR2 \fwha\ measurements.

\clearpage

\section{Sources with double-peaked broad emission lines}
\label{app:double_peaked}

Table~\ref{tab:double_peaked} lists the BASS DR2 AGNs which we've identified to have double-peaked \Halpha\ and/or \Hbeta\ broad emission lines.

\begin{deluxetable}{rlccc}
\label{tab:double_peaked}
\tablecaption{AGNs with double-peaked broad Balmer emission lines.}
\tablewidth{1.0\textwidth}
\tablehead{
\colhead{BAT ID} & \colhead{AGN Name\tablenotemark{a}} & \nocolhead{} &  \multicolumn{2}{c}{Double-peaked...\tablenotemark{b}}\\
\nocolhead{} & \nocolhead{} & \nocolhead{} & \colhead{\Halpha} & \colhead{\Hbeta}
}
\startdata
  45	&	SWIFT J$0101.5-0308	$	&	&	0	&	1	\\
  48	&	SWIFT J$0103.8-6437	$	&	&	1	&	0	\\
 107	&	SWIFT J$0207.0+2931	$	&	&	1	&	0	\\
 135	&	SWIFT J$0235.3-2934	$	&	&	1	&	0	\\
 136	&	SWIFT J$0238.2-5213	$	&	&	1	&	0	\\
 147	&	SWIFT J$0244.8+6227	$	&	&	0	&	1	\\
 170	&	SWIFT J$0312.9+4121	$	&	&	1	&	0	\\
 172	&	SWIFT J$0318.7+6828	$	&	&	1	&	0	\\
 207	&	SWIFT J$0405.3-3707	$	&	&	1	&	1	\\
 270	&	SWIFT J$0519.5-4545	$	&	&	1	&	0	\\
 371	&	SWIFT J$0726.5+3659	$	&	&	1	&	0	\\
 372	&	SWIFT J$0727.4-2408	$	&	&	1	&	0	\\
 403	&	SWIFT J$0803.4+0842	$	&	&	1	&	0	\\
 413	&	SWIFT J$0818.1+0120	$	&	&	1	&	0	\\
 414	&	SWIFT J$0819.2-2259	$	&	&	0	&	1	\\
 454	&	SWIFT J$0923.6-2136	$	&	&	1	&	0	\\
 525	&	SWIFT J$1103.4+3731	$	&	&	1	&	1	\\
 607	&	SWIFT J$1217.3+0714	$	&	&	1	&	1	\\
 618	&	SWIFT J$1227.8-4856	$	&	&	0	&	1	\\
 686	&	SWIFT J$1341.9+3537	$	&	&	0	&	1	\\
 690	&	SWIFT J$1347.1+7325	$	&	&	1	&	0	\\
 713	&	SWIFT J$1416.9-1158	$	&	&	1	&	1	\\
 715	&	SWIFT J$1417.7+6143	$	&	&	1	&	0	\\
 776	&	SWIFT J$1542.0-1410	$	&	&	1	&	0	\\
 800	&	SWIFT J$1617.8+3223	$	&	&	1	&	1	\\
 876	&	SWIFT J$1719.7+4900	$	&	&	1	&	1	\\
 882	&	SWIFT J$1723.2+3418	$	&	&	1	&	1	\\
 907	&	SWIFT J$1742.2+1833	$	&	&	1	&	1	\\
 923	&	SWIFT J$1747.7-2253	$	&	&	1	&	1	\\
1079	&	SWIFT J$2030.2-7532	$	&	&	1	&	1	\\
1151	&	SWIFT J$2223.9-0207	$	&	&	1	&	1	\\
1179	&	SWIFT J$2301.4-5916	$	&	&	1	&	1	\\
1196	&	SWIFT J$2327.4+1525	$	&	&	1	&	0	\\
1200	&	SWIFT J$2333.9-2342	$	&	&	1	&	0	\\
%
%
\enddata
\tablenotetext{a}{Unique AGN names from the Swift/BAT 70-month catalog.}
\tablenotetext{b}{Binary flags indicating which of the two key broad Balmer lines has a double-peaked profile.}
\end{deluxetable}

\clearpage

\section{Broad Line Measurements Catalog}
\label{app:blr_catalog}

Tables \ref{tab:HA}, \ref{tab:HB}, \ref{tab:MG}, and \ref{tab:CIV} describe the contents of our measurement catalogs, for the spectral regions including the broad \Halpha, \Hbeta, \MgII, and \CIV\ emission lines (respectively).

\startlongtable
\begin{deluxetable}{rll}
\tabletypesize{\small}
\label{tab:HA}
\tablecaption{Column description for the H$\alpha$ measurements table.}
\tablewidth{0pt}
\tablehead{
\colhead{Column} & \colhead{Name} & \colhead{Description}}
\startdata
 1 & BAT\_ID                            & BASS identifier \\
 2 & Subsample                          & The source of the optical spectrum (facility, instrument, etc.).\\
 3 & \FQ(\Halpha)                       & Quality flag for the fit: 1 for a good fit and 2 for an acceptable one \\
 4 & FWHM(b\Halpha) [\kms]              & FWHM of the broad H$\alpha$ emission line. \\
 5 & $\Delta^-$ FWHM(b\Halpha) [\kms]   & Lower error on FWHM(b\Halpha)\\
 6 & $\Delta^+$ FWHM(b\Halpha) [\kms]   & Upper error on FWHM(b\Halpha)\\
 7 & $\log \Lbha$ [\ergs]               & Broad H$\alpha$ line luminosity \\
 8 & $\log \Fbha$ [\ergscm]             & Broad H$\alpha$ line integrated flux \\
 9 & $\Delta^-\log \Lbha$ [\ergs]       & Lower error on $\log\Lbha$\\
10 & $\Delta^+\log \Lbha$ [\ergs]       & Upper error on $\log\Lbha$\\
11 & $\log \Lsix$ [\ergs]               & Monochromatic luminosity at rest-frame 6200 \AA, \lamLlam(6200 \AA) \\
12 & $\log \Fsix$ [\ergscm]             & Monochromatic flux at rest-frame 6200 \AA, \lamFlam(6200 \AA) \\
13 & $\Delta^-\log \Lsix$ [\ergs]       & Lower error on $\log\Lsix$\\
14 & $\Delta^+\log \Lsix$ [\ergs]       & Upper error on $\log\Lsix$\\
15 & EW(b\Halpha) [\AA]                 & Rest-frame equivalent width of the broad \Halpha\ line \\
16 & $\Delta^-$EW(b\Halpha) [\AA]       & Lower error on EW(b\Halpha)\\
17 & $\Delta^+$EW(b\Halpha) [\AA]       & Upper error on EW(b\Halpha)\\
18 & $\log L_{\rm peak}$(b\Halpha) [\ergsA]             & Monochromatic luminosity of the broad \Halpha\ line at peak \\
19 & $\log F_{\rm peak}$(b\Halpha) [\ergcmsA]           & Monochromatic flux of the broad \Halpha\ line at peak \\
20 & $\Delta^- \log L_{\rm peak}$(b\Halpha) [\ergsA]   & Lower error on $\log L_{\rm peak}$(b\Halpha)\\
21 & $\Delta^+ \log L_{\rm peak}$(b\Halpha) [\ergsA]   & Upper error on $\log L_{\rm peak}$(b\Halpha)\\
22 & $\Delta v_{\rm peak}$(b\Halpha) [\kms]            & Velocity shift of the broad \Halpha\ line peak \\
23 & $\Delta^- \Delta v_{\rm peak}$(b\Halpha) [\kms]   & Lower error on $\Delta v_{\rm peak}$(b\Halpha)\\
24 & $\Delta^+ \Delta v_{\rm peak}$(b\Halpha) [\kms]   & Upper error on $\Delta v_{\rm peak}$(b\Halpha)\\
25 & $\Delta v_{\rm cent}$(b\Halpha) [\kms] & Velocity shift of the broad \Halpha\ line centroid \\
26 & $\Delta^- \Delta v_{\rm cent}$(b\Halpha) [\kms]   & Lower error on $\Delta v_{\rm cent}$(b\Halpha)\\
27 & $\Delta^+ \Delta v_{\rm cent}$(b\Halpha) [\kms]   & Upper error on $\Delta v_{\rm cent}$(b\Halpha)\\
28 & $\log\Mbh$(b\Halpha) [\Msun, GH05] & Broad \Halpha-based BH mass estimate, following \citet{GreeneHo2005}\\
29 & $\Delta^-\log\Mbh$(b\Halpha) [\Msol] &  Lower error on $\log\Mbh$(b\Halpha)\\
30 & $\Delta^+\log\Mbh$(b\Halpha) [\Msol] &  Lower error on $\log\Mbh$(b\Halpha)\\
%
%
%
31 & $\log L$(\sii) [\ergs]             & Luminosity of the narrow \SII\ emission line \\
32 & $\log F$(\sii) [\ergscm]             & Integrated flux of the narrow \SII\ emission line \\
33 & $\Delta \log L$(\sii) [\ergs]      & Error on $\log L$(\sii) \\
34 & $\Delta v$(\sii) [\kms]            & Velocity shift of the \SII\ line peak/centroid \\
35 & $\Delta \Delta v$(\sii) [\kms]     & Error on $\Delta v$(\sii) \\
36 & FWHM(\sii) [\kms]                  & FWHM of the narrow \SII\ emission line \\
37 & $\Delta$FWHM(\sii) [\kms]          & Error on FWHM(\sii) \\
38 & $\log L$(n\Halpha) [\ergs]         & Luminosity of the narrow \Halpha\ line \\
39 & $\log F$(n\Halpha) [\ergscm]         & Integrated flux of the narrow \Halpha\ line \\
40 & $\Delta \log L$(n\Halpha) [\ergs]  & Error on $\log L$(n\Halpha) \\
41 & $\Delta v$(n\Halpha) [\kms]        & Velocity shift of the narrow \Halpha\ line peak/centroid \\
42 & $\Delta \Delta v$(n\Halpha) [\kms] & Error on $\Delta v$(n\Halpha) \\
43 & FWHM(n\Halpha) [\kms]              & FWHM of the narrow \Halpha\ emission line \\
44 & $\Delta$FWHM(n\Halpha) [\kms]      & Error on FWHM(n\Halpha) \\
45 & $\log L$(\nii) [\ergs]             & Luminosity of the narrow \NII\ line \\
46 & $\log F$(\nii) [\ergscm]             & Integrated flux of the narrow \NII\ line \\
47 & $\Delta \log L$(\nii) [\ergs]      & Error on $\log L$(\nii) \\
48 & $\Delta v$(n\Halpha) [\kms]        & Velocity shift of the narrow \NII\ line peak/centroid \\
49 & $\Delta \Delta v$(\nii) [\kms]     & Error on $\Delta v$(\nii) \\
50 & FWHM(\nii) [\kms]                  & FWHM of the narrow \NII\ emission line \\
51 & $\Delta$FWHM(\nii) [\kms]          & Error on FWHM(\nii) \\
52 & $z$(DR2)                    & The source redshift, as reported in the BASS/DR2 catalog \citep{Koss_DR2_catalog}  \\
53 & $z_{\rm corr}$(DR2, \sii)   & Updated redshift, based on NLR measurements of the \sii\ emission lines from this paper \\
54 & DR2Type & DR2 Seyfert type according to \cite{Winkler1992} classification (see Section \ref{subsec:types} for details).\\ 
\enddata
\tablecomments{All errors are $1\sigma$ equivalent, and were obtained obtained through our spectral bootstrapping  procedure. When both lower and upper errors are reported, these correspond to the 16th and 84th percentiles of the corresponding distribution. When a single error is reported, it corresponds to the standard deviation.}
\end{deluxetable}

\section*{~}

\clearpage\clearpage\clearpage
\newpage
\newpage
\newpage

\startlongtable
\begin{deluxetable}{rll}
\tabletypesize{\scriptsize}
\label{tab:HB}
\tablecaption{Column description for the H$\beta$ measurements table.}
\tablewidth{0pt}
\tablehead{
\colhead{Column} & \colhead{Name} & \colhead{Description}}
\startdata
 1 & BAT\_ID                            & BASS identifier \\
 2 & Subsample                          & The source of the optical spectrum (facility, instrument, etc.)\\
 3 & \FQ(\Hbeta)                        & Quality flag for the fit: 1 for a good fit and 2 for an acceptable one \\
 4 & FWHM(b\Hbeta) [\kms]               & FWHM of the broad \Hbeta\ emission line \\
 5 & $\Delta^-$ FWHM(b\Hbeta) [\kms]    & Lower error on FWHM(b\Hbeta)\\
 6 & $\Delta^+$ FWHM(b\Hbeta) [\kms]    & Upper error on FWHM(b\Hbeta)\\
 7 & $\log \Lbhb$ [\ergs]               & Broad \Hbeta\ line luminosity \\
 8 & $\log \Fbhb$ [\ergscm] & Broad \Hbeta\ line integrated flux  \\
 9 & $\Delta^-\log \Lbhb$ [\ergs]       & Lower error on $\log\Lbhb$\\
 10 & $\Delta^+\log \Lbhb$ [\ergs]       & Upper error on $\log\Lbhb$\\
11 & $\log \Lop$ [\ergs]                & Monochromatic luminosity at rest-frame 5100 \AA, \lamLlam(5100 \AA) \\
12 & $\log \Fop$ [\ergscm]                & Monochromatic Flux at rest-frame 5100 \AA, \lamFlam(5100 \AA) \\
13 & $\Delta^-\log \Lop$ [\ergs]        & Lower error on $\log\Lop$\\
14 & $\Delta^+\log \Lop$ [\ergs]        & Upper error on $\log\Lop$\\
15 & EW(b\Hbeta) [\AA]                  & Rest-frame equivalent width of the broad \Hbeta\ line \\
16 & $\Delta^-$EW(b\Hbeta) [\AA]        & Lower error on EW(b\Hbeta)\\
17 & $\Delta^+$EW(b\Hbeta) [\AA]        & Upper error on EW(b\Hbeta)\\
18 & $\log L_{\rm peak}$(b\Hbeta) [\ergsA]              & Monochromatic luminosity of the broad \Hbeta\ line at peak \\
19 & $\log F_{\rm peak}$(b\Hbeta) [\ergcmsA]              & Monochromatic flux of the broad \Hbeta\ line at peak \\
20 & $\Delta^- \log L_{\rm peak}$(b\Hbeta) [\ergsA]     & Lower error on $\log L_{\rm peak}$(b\Hbeta)\\
21 & $\Delta^+ \log L_{\rm peak}$(b\Hbeta) [\ergsA]     & Upper error on $\log L_{\rm peak}$(b\Hbeta)\\
22 & $\Delta v_{\rm peak}$(b\Hbeta) [\kms]              & Velocity shift of the broad \Hbeta\ line peak \\
23 & $\Delta^- \Delta v_{\rm peak}$(b\Hbeta) [\kms]     & Lower error on $\Delta v_{\rm peak}$(b\Hbeta)\\
24 & $\Delta^+ \Delta v_{\rm peak}$(b\Hbeta) [\kms]     & Upper error on $\Delta v_{\rm peak}$(b\Hbeta)\\
25 & $\Delta v_{\rm cent}$(b\Hbeta) [\kms]              & Velocity shift of the broad \Hbeta\ line centroid \\
26 & $\Delta^- \Delta v_{\rm cent}$(b\Hbeta) [\kms]     & Lower error on $\Delta v_{\rm cent}$(b\Hbeta)\\
27 & $\Delta^+ \Delta v_{\rm cent}$(b\Hbeta) [\kms]     & Upper error on $\Delta v_{\rm cent}$(b\Hbeta)\\
28 & $\log\Mbh$(b\Hbeta) [\Msun, TN12]   & Broad \Hbeta-based BH mass estimate, following \citet{TrakhtenbrotNetzer2012} \\
29 & $\Delta^-\log\Mbh$(b\Hbeta) [\Msol] &  Lower error on $\log\Mbh$(b\Hbeta)\\
30 & $\Delta^+\log\Mbh$(b\Hbeta) [\Msol] &  Lower error on $\log\Mbh$(b\Hbeta)\\
%
%
%
31 & $\log L$(\oiii) [\ergs]             & Luminosity of the narrow \OIII\ emission line \\
32 & $\log F$(\oiii) [\ergscm]             & Integrated flux of the narrow \OIII\ emission line \\
33 & $\Delta \log L$(\oiii) [\ergs]      & Error on $\log L$(\oiii) \\
34 & $\Delta v$(\oiii) [\kms]            & Velocity shift of the \OIII\ line peak/centroid \\
35 & $\Delta \Delta v$(\oiii) [\kms]     & Error on $\Delta v$(\oiii) \\
36 & FWHM(\oiii) [\kms]                  & FWHM of the narrow \OIII\ emission line \\
37 & $\Delta$FWHM(\oiii) [\kms]          & Error on FWHM(\oiii) \\
38 & $\log L$(n\Hbeta) [\ergs]         & Luminosity of the narrow \Hbeta\ line \\
39 & $\log F$(n\Hbeta) [\ergscm]         & Integrated flux of the narrow \Hbeta\ line \\
40 & $\Delta \log L$(n\Hbeta) [\ergs]  & Error on $\log L$(n\Hbeta) \\
41 & $\Delta v$(n\Hbeta) [\kms]        & Velocity shift of the narrow \Hbeta\ line peak/centroid \\
42 & $\Delta \Delta v$(n\Hbeta) [\kms] & Error on $\Delta v$(n\Hbeta) \\
43 & FWHM(n\Hbeta) [\kms]              & FWHM of the narrow \Hbeta\ emission line \\
44 & $\Delta$FWHM(n\Hbeta) [\kms]      & Error on FWHM(n\Hbeta) \\
45 & $z$(DR2)                    & The source redshift, as reported in the BASS/DR2 catalog \citep{Koss_DR2_catalog} \\
46 & $z_{\rm corr}$(DR2, \oiii)   & Updated redshift, based on NLR measurements of the \oiii\ emission lines from this paper \\
47 & DR2Type & DR2 Seyfert type according to \cite{Winkler1992} classification (see subsection \ref{subsec:types} for details)
\enddata
\tablecomments{All errors are $1\sigma$ equivalent, and were obtained obtained through our spectral bootstrapping  procedure. When both lower and upper errors are reported, these correspond to the 16th and 84th percentiles of the corresponding distribution. When a single error is reported, it corresponds to the standard deviation.}
\end{deluxetable}

\clearpage
\newpage

\startlongtable
\begin{deluxetable}{rll}
\tabletypesize{\small}
\label{tab:MG}
\tablecaption{Column description for the \MgII\ measurements table.}
\tablewidth{0pt}
\tablehead{
\colhead{Column} & \colhead{Name} & \colhead{Description}}
\startdata
 1 & BAT\_ID                            & BASS identifier \\
 2 & Subsample                          & The source of the optical spectrum (facility, instrument, etc.)\\
 3 & \FQ(\mgii)                         & Quality flag for the fit: 1 for a good fit and 2 for an acceptable one \\
 4 & FWHM(\mgii) [\kms]                 & FWHM of the broad \mgii\ emission line \\
 5 & $\Delta^-$ FWHM(\mgii) [\kms]    & Lower error on FWHM(\mgii)\\
 6 & $\Delta^+$ FWHM(\mgii) [\kms]    & Upper error on FWHM(\mgii)\\
 7 & $\log L$(\mgii) [\ergs]               & Broad \mgii\ line luminosity \\
 8 & $\log F$(\mgii) [\ergscm]               & Broad \mgii\ line integrated flux \\
 9 & $\Delta^-\log L$(\mgii) [\ergs]       & Lower error on $\log L$(\mgii)\\
10 & $\Delta^+\log L$(\mgii) [\ergs]       & Upper error on $\log L$(\mgii)\\
11 & $\log \Lthree$ [\ergs]                & Monochromatic luminosity at rest-frame 3000 \AA, \lamLlam(3000 \AA) \\
12 & $\log \Fthree$ [\ergscm]                & Monochromatic flux at rest-frame 3000 \AA, \lamFlam(3000 \AA) \\
12 & $\Delta^-\log \Lthree$ [\ergs]        & Lower error on $\log\Lthree$\\
13 & $\Delta^+\log \Lthree$ [\ergs]        & Upper error on $\log\Lthree$\\
14 & EW(\mgii) [\AA]                  & Rest-frame equivalent width of the broad \mgii\ line \\
15 & $\Delta^-$EW(\mgii) [\AA]        & Lower error on EW(\mgii)\\
16 & $\Delta^+$EW(\mgii) [\AA]        & Upper error on EW(\mgii)\\
17 & $\log L_{\rm peak}$(\mgii) [\ergsA]              & Monochromatic luminosity of the broad \mgii\ line at peak \\
18 & $\log F_{\rm peak}$(\mgii) [\ergsA]              & Monochromatic flux of the broad \mgii\ line at peak \\
19 & $\Delta^- \log L_{\rm peak}$(\mgii) [\ergcmsA]     & Lower error on $\log L_{\rm peak}$(\mgii)\\
20 & $\Delta^+ \log L_{\rm peak}$(\mgii) [\ergsA]     & Upper error on $\log L_{\rm peak}$(\mgii)\\
21 & $\Delta v_{\rm peak}$(b\Hbeta) [\kms]              & Velocity shift of the broad \mgii\ line peak \\
22 & $\Delta^- \Delta v_{\rm peak}$(\mgii) [\kms]     & Lower error on $\Delta v_{\rm peak}$(\mgii)\\
23 & $\Delta^+ \Delta v_{\rm peak}$(\mgii) [\kms]     & Upper error on $\Delta v_{\rm peak}$(\mgii)\\
24 & $\Delta v_{\rm cent}$(\mgii) [\kms]              & Velocity shift of the broad \mgii\ line centroid \\
25 & $\Delta^- \Delta v_{\rm cent}$(\mgii) [\kms]     & Lower error on $\Delta v_{\rm cent}$(\mgii)\\
26 & $\Delta^+ \Delta v_{\rm cent}$(\mgii) [\kms]     & Upper error on $\Delta v_{\rm cent}$(\mgii)\\
28 & $\log\Mbh$(\mgii) [\Msun, MR16]   & Broad \mgii-based BH mass estimate, following \citet{MejiaRestrepo2016a} \\
29 & $\Delta^-\log\Mbh$(\mgii) [\Msol] &  Lower error on $\log\Mbh$(\mgii)\\
30 & $\Delta^+\log\Mbh$(\mgii) [\Msol] &  Lower error on $\log\Mbh$(\mgii)\\
31 & $z$(DR2)                    & The source redshift, as reported in the BASS/DR2 catalog \citep{Koss_DR2_catalog} \\
32 & $z_{\rm corr}$(DR2, \mgii)   & Updated redshift, based on BLR measurements of the \mgii\ emission line from this paper
\enddata
\tablecomments{All errors are $1\sigma$ equivalent, and were obtained obtained through our spectral bootstrapping  procedure. When both lower and upper errors are reported, these correspond to the 16th and 84th percentiles of the corresponding distribution. When a single error is reported, it corresponds to the standard deviation.}
\end{deluxetable}

\clearpage
\newpage

\startlongtable
\begin{deluxetable}{rll}
\tabletypesize{\small}
\label{tab:CIV}
\tablecaption{Column description for the \CIV\ measurements table.}
\tablewidth{0pt}
\tablehead{
\colhead{Column} & \colhead{Name} & \colhead{Description}}
\startdata
 1 & BAT\_ID                            & BASS identifier \\
 2 & Subsample                          & The source of the optical spectrum (facility, instrument, etc.).\\
 3 & \FQ(\civ)                          & Quality flag for the fit: 1 for a good fit and 2 for an acceptable one \\
 4 & FWHM(\civ) [\kms]                  & FWHM of the broad \civ\ emission line\\
 5 & $\Delta^-$ FWHM(\civ) [\kms]       & Lower error on FWHM(\civ)\\
 6 & $\Delta^+$ FWHM(\civ) [\kms]       & Upper error on FWHM(\civ)\\
 7 & $\log L$(\civ) [\ergs]             & Broad \civ\ line luminosity \\
 8 & $\log F$(\civ) [\ergscm]             & Broad \civ\ line integrated flux \\
 9 & $\Delta^-\log L$(\civ) [\ergs]     & Lower error on $\log L$(\civ)\\
10 & $\Delta^+\log L$(\civ) [\ergs]     & Upper error on $\log L$(\civ)\\
11 & $\log \Luv$ [\ergs]                & Monochromatic luminosity at rest-frame 1450 \AA, \lamLlam(1450 \AA) \\
12 & $\log \Fuv$ [\ergscm]                & Monochromatic flux at rest-frame 1450 \AA, \lamLlam(1450 \AA) \\
13 & $\Delta^-\log \Luv$ [\ergs]        & Lower error on $\log\Luv$\\
14 & $\Delta^+\log \Luv$ [\ergs]        & Upper error on $\log\Luv$\\
15 & EW(\civ) [\AA]                     & Rest-frame equivalent width of the broad \civ\ line \\
16 & $\Delta^-$EW(\civ) [\AA]           & Lower error on EW(\civ)\\
17 & $\Delta^+$EW(\civ) [\AA]           & Upper error on EW(\civ)\\
18 & $\log L_{\rm peak}$(\civ) [\ergsA]             & Monochromatic luminosity of the broad \civ\ line at peak \\
19 & $\log F_{\rm peak}$(\civ) [\ergcmsA]             & Monochromatic flux of the broad \civ\ line at peak \\
20 & $\Delta^- \log L_{\rm peak}$(\civ) [\ergsA]    & Lower error on $\log L_{\rm peak}$(\civ)\\
21 & $\Delta^+ \log L_{\rm peak}$(\civ) [\ergsA]    & Upper error on $\log L_{\rm peak}$(\civ)\\
22 & $\Delta v_{\rm peak}$(b\Hbeta) [\kms]          & Velocity shift of the broad \civ\ line peak \\
23 & $\Delta^- \Delta v_{\rm peak}$(\civ) [\kms]    & Lower error on $\Delta v_{\rm peak}$(\civ)\\
24 & $\Delta^+ \Delta v_{\rm peak}$(\civ) [\kms]    & Upper error on $\Delta v_{\rm peak}$(\civ)\\
25 & $\Delta v_{\rm cent}$(\civ) [\kms]             & Velocity shift of the broad \civ\ line centroid \\
26 & $\Delta^- \Delta v_{\rm cent}$(\civ) [\kms]    & Lower error on $\Delta v_{\rm cent}$(\civ)\\
27 & $\Delta^+ \Delta v_{\rm cent}$(\civ) [\kms]    & Upper error on $\Delta v_{\rm cent}$(\civ)\\
28 & $\log\Mbh$(\civ) [\Msun, MR16]   & Broad \civ-based BH mass estimate, following \citet{MejiaRestrepo2016a} \\
29 & $\Delta^-\log\Mbh$(\civ) [\Msol] &  Lower error on $\log\Mbh$(\civ)\\
30 & $\Delta^+\log\Mbh$(\civ) [\Msol] &  Lower error on $\log\Mbh$(\civ)\\
31 & $z$(ref)                    & The source redshift, as reported in the BASS/DR2 catalog \citep{Koss_DR2_catalog} \\
32 & $z_{\rm corr}$(DR2, \civ)   & Updated redshift, based on BLR measurements of the \civ\ emission line from this paper
\enddata
\tablecomments{All errors are $1\sigma$ equivalent, and were obtained obtained through our spectral bootstrapping  procedure. When both lower and upper errors are reported, these correspond to the 16th and 84th percentiles of the corresponding distribution. When a single error is reported, it corresponds to the standard deviation.}
\end{deluxetable}

\clearpage

\section{Broad H$\alpha$ vs. Mid-IR Emission}
\label{app:LHAIR_FWHM_NH}

\begin{figure*}
    \includegraphics[width=0.475\textwidth]{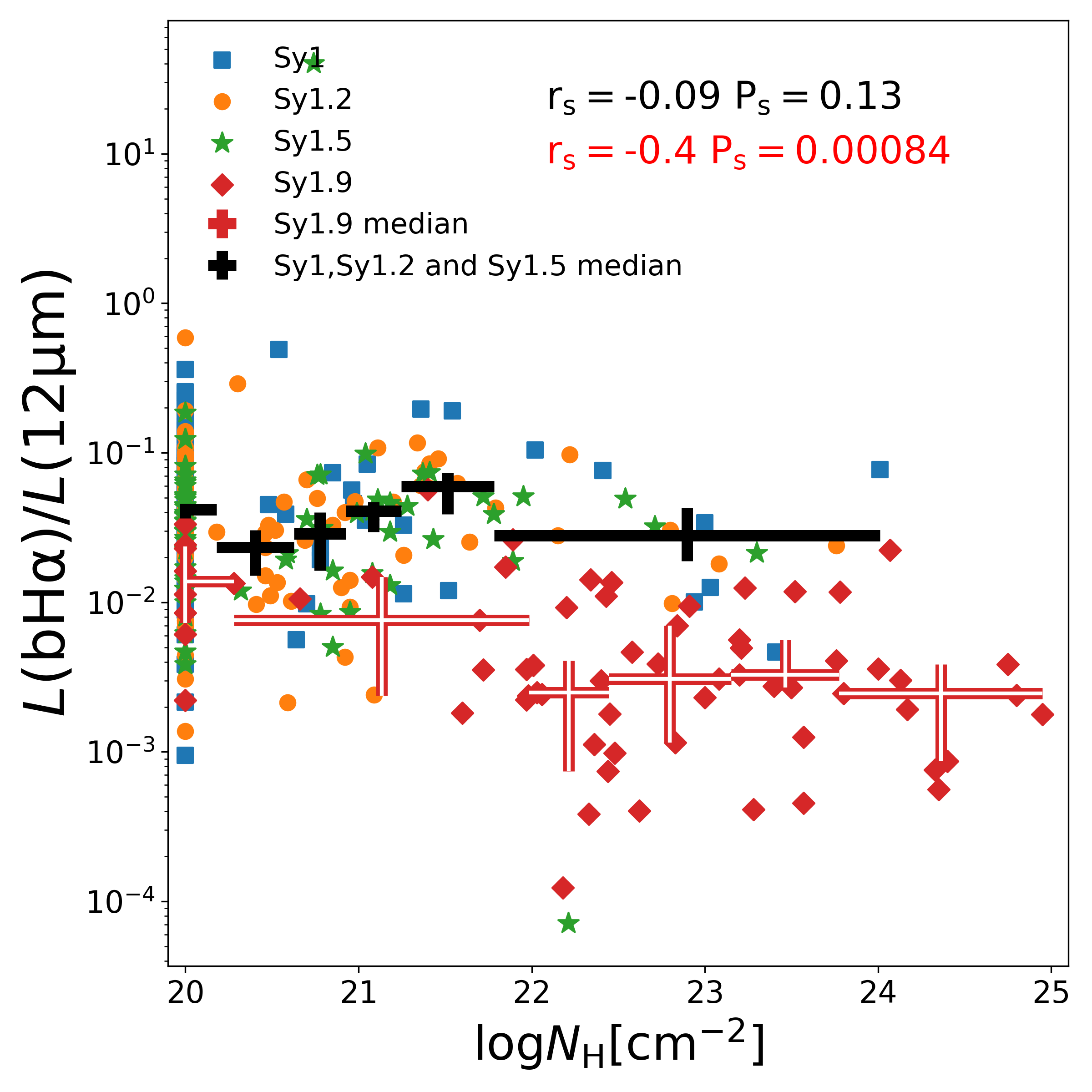}\hfill
    \includegraphics[width=0.475\textwidth]{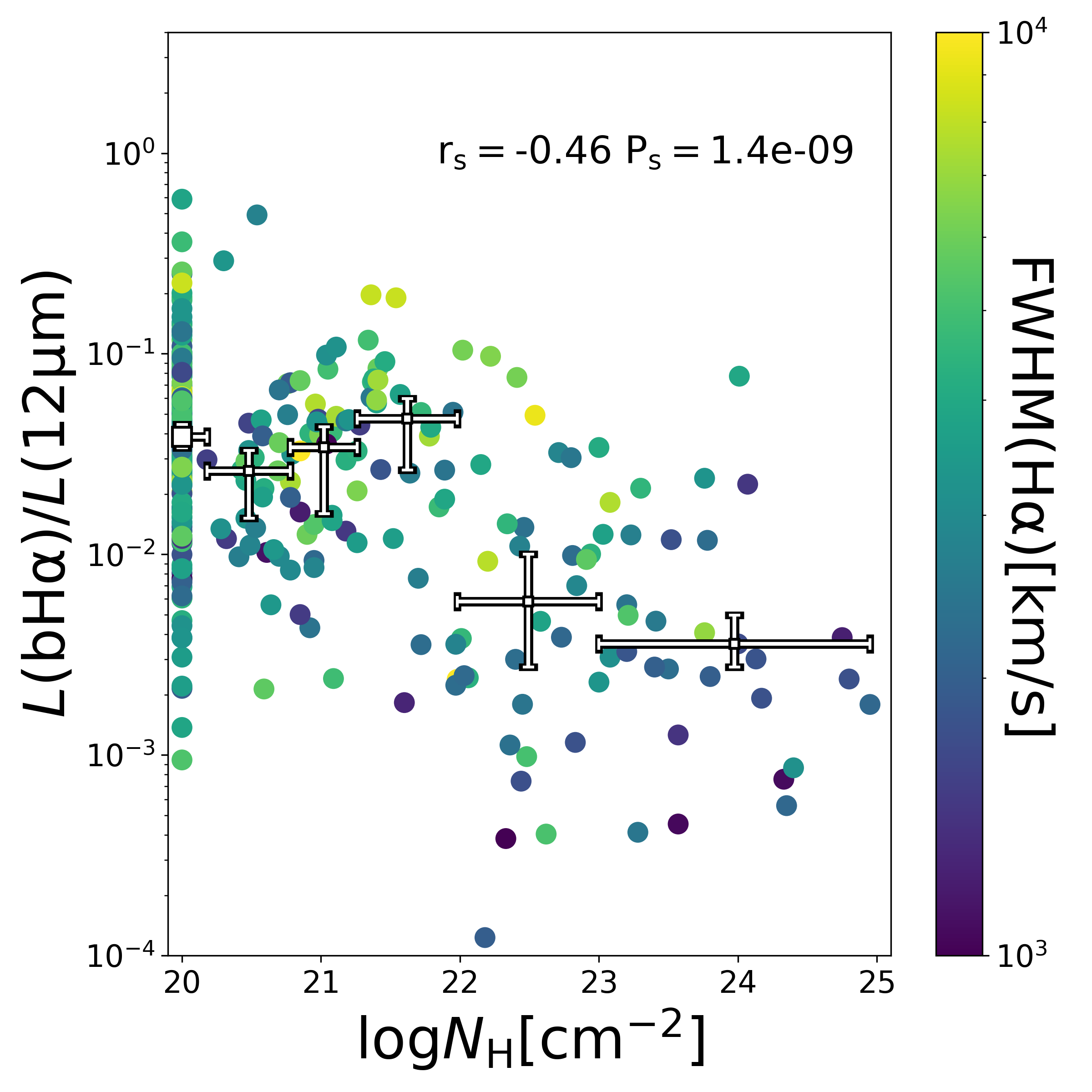}
    \caption{{\it Left:} \Lbhair\ vs \fwha. {\it Right:} \Lbhair\ vs  \logNHo\ color-coded by \fwha. 
    {\color{black} Large crosses represent} the median values of \Lbhair\ after binning in \fwha\ (left panel) and  \logNHo\ (right panel) in equally spaced quantiles.  Horizontal error-bars  represent the bin edges and vertical error bars the errors in the median
    \Lbhair\ estimated from bootstrapping.}
    \label{fig:LHAIR_FWHM_NH}
\end{figure*}

{\color{black} 
Figure~\ref{fig:LHAIR_FWHM_NH} shows the broad \Halpha to mid-IR ratios for our sample, \Lbhair, vs.\ \fwha and \logNHo.
These serve to demonstrate that the mid-IR emission can substitute the ultra-hard X-ray emission when deriving (or using) the corrections presented in Section~\ref{sec:correcting_mbh}.
}

\clearpage
\section{The broad \Hbeta\ line versus column densities}
\label{app:FWHB_NH}

\begin{figure}
\centering
	\includegraphics[width=0.7\columnwidth]{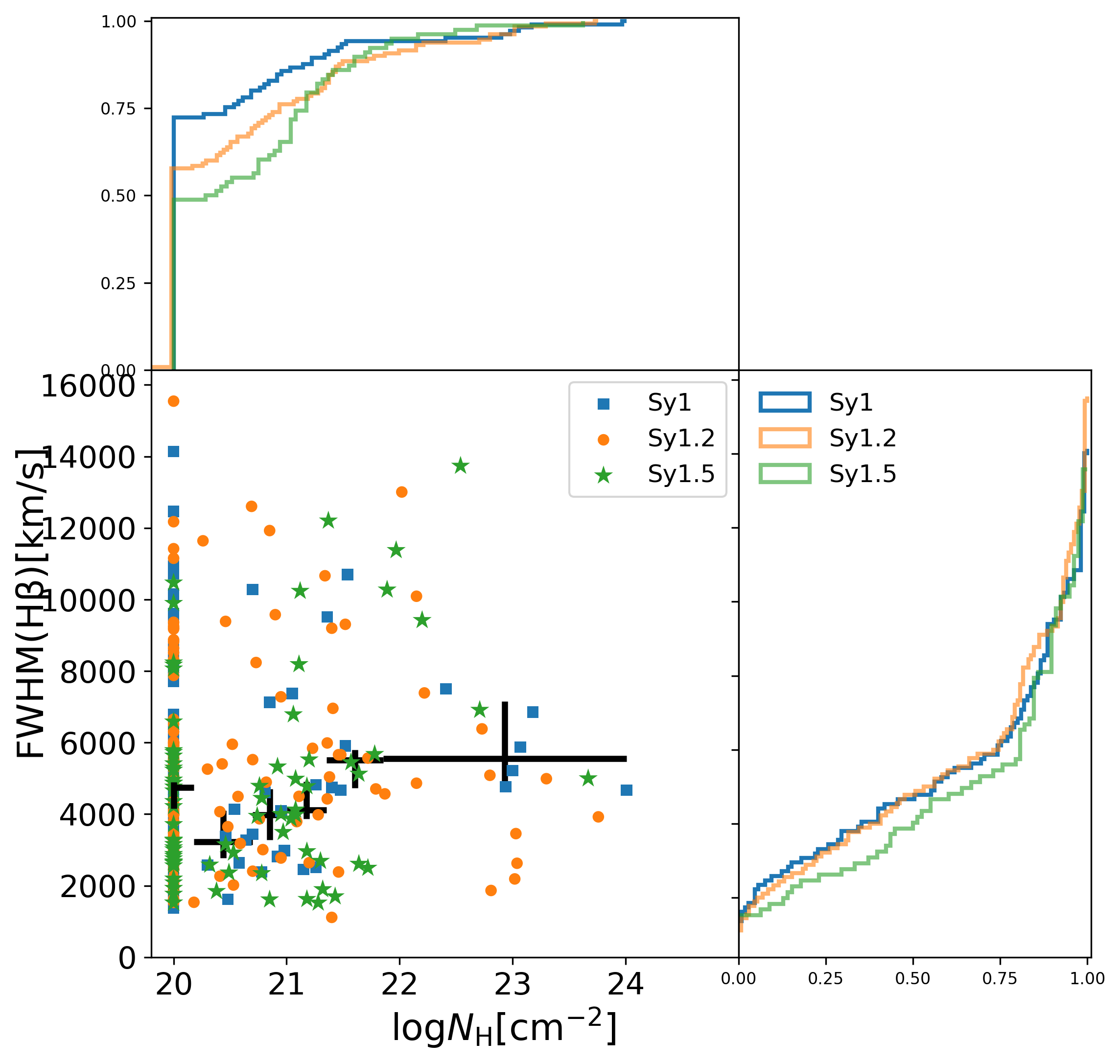}
    \caption{H$\beta$ 
    FWHMs versus \NH\ and the projected distribution of these quantities for Sy1 (light-blue), Sy1.2 (orange), Sy1.5 (light-green) and Sy1.9 (red). 
    Horizontal error-bars represent the bin edges and vertical error-bars represent the errors in the median \fwhm\ from each bin estimated from bootstrapping. 
    This figures complements Fig.~\ref{fig:FWHA_NH}, however in contrast to \fwha, here the median \fwhb\ does {\it not} vary with \logNHo\ (within the error bars).  
    } 
    \label{fig:FWHB_NH}
\end{figure}

In Figure~\ref{fig:FWHB_NH} we show the width of the broad \Hbeta\ emission line, \fwhb, vs. the line-of-sight column density, \NH. This figure complements Fig.~\ref{fig:FWHA_NH}.

\bibliography{bibliography,DR2_bib}{}
\bibliographystyle{aasjournal}



\end{document}